\documentclass[11pt]{article}
\usepackage{authblk}
\usepackage[margin=1in]{geometry}
\usepackage{amsmath}
\usepackage{graphics}
\usepackage{graphicx}
\usepackage[labelfont= bf, textfont=normalfont, font={small}]{caption}
\usepackage[labelfont= normalfont, textfont=normalfont, font={small},labelformat=simple]{subcaption}
\usepackage{enumitem}
\usepackage{float}
\usepackage{setspace}
\usepackage{color}

\title{Time lagged ordinal partition networks for capturing dynamics of continuous dynamical systems}
\date{January 16, 2015}

\author[1]{Michael McCullough}
\author[2]{Michael Small}
\author[2]{Thomas Stemler}
\author[1]{Herbert Ho-Ching Iu}
\affil[1]{School of Electrical and Electronic Engineering, The University of Western Australia, Crawley WA 6009, Australia}
\affil[2]{School of Mathematics and Statistics, The University of Western Australia, Crawley WA 6009, Australia}

\begin{document}

\maketitle

\renewcommand{\thesubfigure}{(\alph{subfigure})}

\begin{abstract}
We investigate a generalised version of the recently proposed ordinal partition time series to network transformation algorithm. Firstly we introduce a fixed time lag for the elements of each partition that is selected using techniques from traditional time delay embedding. The resulting partitions define regions in the embedding phase space that are mapped to nodes in the network space. Edges are allocated between nodes based on temporal succession thus creating a Markov chain representation of the time series. We then apply this new transformation algorithm to time series generated by the R\"ossler system and find that periodic dynamics translate to ring structures whereas chaotic time series translate to band or tube-like structures --- thereby indicating that our algorithm generates networks whose structure is sensitive to system dynamics. Furthermore we demonstrate that simple network measures including the mean out degree and variance of out degrees can track changes in the dynamical behaviour in a manner comparable to the largest Lyapunov exponent. We also apply the same analysis to experimental time series generated by a diode resonator circuit and show that the network size, mean shortest path length and network diameter are highly sensitive to the interior crisis captured in this particular data set.
\end{abstract}

\newpage

\textbf{Within the last ten years a novel approach to time series analysis has emerged whereby data is transformed into a complex network and then analysed using various measures from network science. The choice of transformation algorithm is critical in this process as different methods are inherently more effective at capturing certain aspects of dynamics and less effective at capturing others. In this paper we investigate a recently proposed algorithm known as the method of ordinal partitions. This computationally simple algorithm explicitly embeds temporal information in the network structure by partitioning the time series into a set of symbolic states which become network nodes, and then connecting these nodes based on the transition sequence present in the data. New in this work, we generalise the algorithm by introducing a time lag parameter for the elements in each partition, as is done in traditional methods of time delay embedding. Our results demonstrate that this new approach generates networks which are measurably sensitive to the dynamics present in the source time series, and has the potential to be useful as a tool for change point detection in continuous chaotic systems.}
\rule{\textwidth}{0.5pt}

\section{Introduction}
\label{sec:Intro}
Various procedures for mapping time series data to networks have attracted significant interest in recent years as a novel means of analysing complex dynamical systems. The existing gamut of time series to network transformations can be broadly classified as either proximity networks, visibility graphs or transition networks, as defined in the thorough review paper by Donner \textit{et al.}~\cite{donner_recurrence-based_2011}. While there is a significant body of work concerning proximity network methods, those which build networks based on the proximity of embedded time series points in phase space, they do not explicitly capture temporal information. The same can be said for the various visibility graph algorithms which have won favour for their computational simplicity and findings linking aspects of resulting networks' degree distributions with the Hurst exponent for time series generated by fractal Brownian motion~\cite{lacasa_time_2008}.

Several methods exist for building transition networks from time series where nodes are representative of states and edges are allocated based on temporal succession. These include coarse graining the phase space of an embedded time series~\cite{nicolis_dynamical_2005, padberg_local_2009}, partitioning a one dimensional set of observations by quartiles~\cite{campanharo_duality_2011}, or by the recently proposed method of ordinal partitions~\cite{small_complex_2013} which is an extension of the concept of permutation entropy~\cite{bandt_permutation_2002} from the field of symbolic dynamics. The last of these three methods presents the advantage that it is relatively robust to noise~\cite{amigo_true_2007} and small variations in amplitude are not discarded as a result of the coarse graining process.

In this work we define a generalised form of the ordinal partition time series to network transformation algorithm, unique at the time of this publication, and investigate its application to time series generated by chaotic flows, for both model systems and experimental data. In Section \ref{sec:Review} we present a review of the relevant existing methods that transform time series into networks. Section \ref{sec:Method} comprises the generalised definition of the ordinal partition network transformation method. We apply the method to the R\"ossler system and present results regarding network structure, discuss the selection of the embedding dimension and demonstrate the potential for tracking changes in dynamics in Section \ref{sec:Rossler}. The same tests are performed on experimental data generated by a diode resonator circuit in Section \ref{sec:Diode} before concluding statements in Section \ref{sec:Conc}.

\section{Review}
\label{sec:Review}
\subsection{Proximity Networks}
Proximity network methods include cycle networks~\cite{zhang_complex_2006}, correlation networks~\cite{yang_complex_2008} and, most prominently, networks based on proximity in phase space~\cite{marwan_complex_2009, xu_superfamily_2008}. These algorithms map states from the time series, which are commonly the states of the embedded time series but can also be cycles from pseudo-periodic time series~\cite{zhang_complex_2006} or coarse grained amplitudes~\cite{zhao_geometrical_2014}, and allocate edges between these states based on some measure of closeness or similarity.

Proximity networks therefore have the potential to capture significant information about attractor topology. For example, various measurable properties of proximity networks including node degree, clustering, betweenness centrality and network diameter have been linked to attractor density and heterogeneity or homogeneity in phase space by Xiang \textit{et al.}~\cite{xiang_multiscale_2012} and Donner \textit{et al.}~\cite{donner_recurrence_2010}. In ~\cite{marwan_complex_2009}, Marwan \textit{et al.} used the clustering coefficient to detect changes in dynamics using the proximity network paradigm as applied to the logistic map and paleo climate data. More recently degree variance was shown to track changes in dynamics along the bifurcation spectrum of the R\"ossler system~\cite{xiang_multiscale_2012}, and in~\cite{iwayama_change-point_2013} Iwayama \textit{et al.} developed a spectral clustering measure for proximity networks and used this in conjunction with surrogate data for change point detection in both R\"ossler and Lorenz systems.

Several methods have been proposed for the classification of dynamics using proximity networks. The distribution of node degrees in proximity networks was the subject of early research by Zhang and Small~\cite{zhang_complex_2006}, and Yang and Yang~\cite{yang_complex_2008} with the former reporting that cycle networks could be used to identify chaotic dynamics in flows as distinct from noisy periodic signals based on the observation of a scale free or random degree distribution respectively, and when applied to ECG data demonstrated the potential for the method to discriminate between healthy and unhealthy heart function. More recent research by Zou \textit{et al.}~\cite{zou_power-laws_2012} uncovered power law degree distributions in recurrence networks generated from a variety of deterministic test systems and experimental data, and showed that the power law exponent is directly related to the invariant density of the underlying systems. Furthermore, a cubic polynomial relationship between the Hurst exponent of a fractal Brownian motion time series and the power law exponent of the degree distribution of the corresponding recurrence network was reported by Liu \textit{et al.} in~\cite{liu_topological_2014}. Ranking the distribution of motifs in recurrence networks has also proved a robust procedure for classifying dynamical behaviour in both maps and flows as originally proposed by Xu \textit{et al.}~\cite{xu_superfamily_2008} with further contributions in~\cite{xiang_multiscale_2012} and~\cite{liu_topological_2014}. 

While the above highlights the potential of proximity networks for use in practical time series analysis, the primary drawback of most of these methods is that the resulting networks are invariant to the relabelling of nodes and hence do not explicitly preserve temporal information. Some exceptions do exist, however, including in~\cite{iwayama_change-point_2013} where the authors have deliberately added links based on temporal adjacency, in contrast to the general approach where temporally adjacent nodes are sometimes even deliberately disconnected~\cite{donner_recurrence-based_2011}. A new proximity method was recently introduced where nodes (times series points) are connected based on a threshold of relative amplitude~\cite{zhao_geometrical_2014}. It was demonstrated that this method was able to transform a time series into a network and then generate a new time series from that network such that both the original and the new time series were highly correlated, indicating that temporal information was embedded in the network structure. This result is intuitive given that for an appropriate sampling rate and threshold all temporally adjacent points will be connected, hence this algorithm straddles both the proximity network and transition network paradigms. It is important to note that in both of these cases the network edges which represent a temporal transition are not distinct from those allocated based on proximity. These exceptions aside, however, proximity networks enable us to measure aspects of attractor topology and not the evolution of trajectories. It is for this reason that we elect to investigate a transition network method in this work.

Furthermore, most proximity methods also require time delay embedding and the selection of a third parameter which is generally a proximity threshold or a fixed node degree. Proper selection of these parameters is critical to the structure of the resulting networks. For example, incorrect embedding or a poor choice of threshold can result in disconnected structures or false neighbours. A detailed discussion of these issues with respect to the various different proximity network types is presented in~\cite{donner_recurrence-based_2011}. A final practical consideration with regard to proximity networks is that the size of the networks produced scales directly with the size of the dataset. By contrast, the new method presented in this paper generates networks whose size is dependant on the attractor topology, specifically, the volume of phase space occupied by the underlying attractor, as will be discussed.

\subsection{Visibility Graphs}
The second broad class of time series to network transformation algorithms are visibility graphs. First proposed by Lacasa \textit{et al.}~\cite{lacasa_time_2008, lacasa_visibility_2009}, these are computationally simple algorithms, that map each point in the time series to a node and connect the nodes based on the convexity of successive observations. Aside from several recent extensions proposed for specific tasks~\cite{bezsudnov_time_2014, ahmadlou_visibility_2012}, visibility graph methods are parameter free. Research has demonstrated links between scale free networks and fractal time series~\cite{lacasa_time_2008}, and that the degree distributions of visibility graphs generated from fractal Brownian motion are a function of the Hurst exponent~\cite{lacasa_visibility_2009}. Visibility graphs capture some aspect of temporal information (although they do not do so explicitly). This characteristic has been exploited to test the reversibility of time series, including EEG data, without the need for surrogate data~\cite{donges_testing_2013}. The visibility graph algorithm is attractive for its simplicity and that it can be applied to data without the need for parameter selection, however it is not clear as to which aspect of the dynamics is being embedded in the networks.

\subsection{Transition Networks}
Transition networks are a simple way of conceptualising a time series based on temporal information. Algorithms of this type generally map time series data to a Markov chain by defining nodes as some set of states that spans the time series points, and allocating directed edges based on temporal succession that, when weighted, can represent transitional probabilities based on the source data. Coarse-grained phase space approaches have seen some degree of investigation by Nicolis \textit{et al.} in~\cite{nicolis_dynamical_2005} and Padberg \textit{et al.} in~\cite{padberg_local_2009}. Alternatively the time series can be partitioned into quartiles of amplitude as was investigated by Campanharo \textit{et al.}~\cite{campanharo_duality_2011}. The authors demonstrated that when considering time series generated by a simple map with a variable noise parameter, their method could be inverted using a random walk to generate time series from networks with autocorrelation, power spectrums and probability densities all similar to the original data. While coarse-graining approaches provide some degree of robustness to noise they also discard small amplitude information. Selecting how the data or the embedded phase space of the data is coarse grained will affect both the time and amplitude scale of information embedded in the network. As such, we note here that a multiscale coarse-graining time series to network analysis method could be an interesting avenue for future research.

An alternative strategy for defining states for a transition network is the method of ordinal partitions, as recently proposed by Small~\cite{small_complex_2013}, which arises naturally as an extension to the concept of permutation entropy originally developed by Bandt and Pompe~\cite{bandt_permutation_2002} and then sub-sequentially extended by Cao \textit{et al.}~\cite{cao_detecting_2004}. Computing the permutation entropy of a time series involves first embedding the data in \(D\)-dimensional space with embedding lag \(\tau\), then converting each embedded vector into a symbol which is the rank order of the elements in that vector, hence only two parameters are required. The time series therefore become a symbolic series comprising some distinct set of symbols, and the permutation entropy is defined as the Shannon entropy of this set. It has been demonstrated that this metric can track dynamical changes in the logistic map in a manner similar to the largest Lyapunov exponent~\cite{bandt_permutation_2002} and identify change points in the transient Lorenz system and in EEG data~\cite{cao_detecting_2004}. More recent investigations have developed the method into a multiscale approach~\cite{zunino_distinguishing_2012} and the permutation spectrum test~\cite{kulp_discriminating_2014} --- two distinct visual analysis methods shown to be effective in discriminating between a range of chaotic and stochastic dynamics.

Additional information can be extracted from the set of rank order permutation symbols by mapping each distinct symbol to a node in a complex network and allocating edges based on temporal succession. In~\cite{small_complex_2013}, Small used a non-overlapping window and fixed lag \(\tau=1\) for embedding, and presented numerical results for a range of chaotic maps, flows and stochastic dynamics, demonstrating the potential for ordinal partition networks to discriminate between these different systems. One distinct advantage of permutation symbols over coarse-grained states for time series to network transformation algorithms is that they capture relative amplitude differences on all scales while remaining robust to noise. However, information about absolute amplitude is not captured in the symbol, and hence it is impossible to discern where a particular sequence occurred with respect to the phase space based on its permutation symbol alone. To overcome this, a dual symbol method was proposed by Sun \textit{et al.} in~\cite{sun_characterizing_2014} where the second symbol was a coarse grained state of the signal amplitude. The authors used an overlapping window and time lag \(\tau=1\). In the resulting networks, the mean degree, degree variance, density and mean shortest path length were able to track changes in dynamical behaviour along the bifurcation spectrum of the R\"ossler system in a way that was visually reminiscent of the largest Lyapunov exponent. Other network methods and standard permutation entropy techniques have shown similar tracking but this has generally only been reported for simple maps, not chaotic flows, hence the significance of these results. However, the use of a second symbol imposes the selection of a third parameter. The primary motivation for this work was to eliminate the need for this additional complexity yet still achieve similar results.

When using the ordinal partition method the choice of the embedding dimension (sometimes referred to as the window size) is critical as this determines the maximum time scale on which dynamical information is captured. As noted by Sun \textit{et al.}~\cite{sun_characterizing_2014}, this parameter must be chosen in relation to the sampling rate for continuous systems, but simply using a large embedding dimension to capture a longer time span will make the method less robust to noise. The authors proposed using a time delay \(\tau>1\) as a solution to this problem, but did not attempt such a scheme. In the following section we define and investigate a generalised version of the ordinal partition time series to network transformation algorithm where \(\tau\) is selected by traditional methods used for time series embedding, and the embedding dimension \(D\) is selected based on simple network measures. As will be shown, this allows the permutation symbols to capture absolute amplitude information with respect to phase space without the need for a second symbol and enables simple tracking of dynamical change in chaotic flows.

We note here that our method is not the first proposed for constructing a symbolic sequence and a corresponding transition network from an embedded time series. In~\cite{kennel_estimating_2003} and~\cite{hirata_characterizing_2005} embedded time series points are mapped to partitions symbolised by the unit square and a binary sequence respectively. Like these methods, our algorithm maps embedded time series points from continuous phase space into disjoint sets whose union comprises the complete phase space. However, the primary motivation in both~\cite{kennel_estimating_2003} and~\cite{hirata_characterizing_2005} was to find generating partitions --- partitions for which trajectories are uniquely mapped to symbol sequences and therefore provide a complete representation of the system dynamics --- rather than constructing networks from time series and then using measures from network science to analyse the data, as is the case in this work and the aforementioned time series to network literature. Furthermore, the symbolic mapping procedure used in this paper is also fundamentally different. For example, the ordinal partition method is only guaranteed to provide an injective mapping in the trivial case where \(D\) is chosen sufficiently large, and thus permutation symbols made sufficiently long, that each data point corresponds to its own symbol.

\section{Method}
\label{sec:Method}
\begin{figure}
	\centering
	\includegraphics [width=11cm] {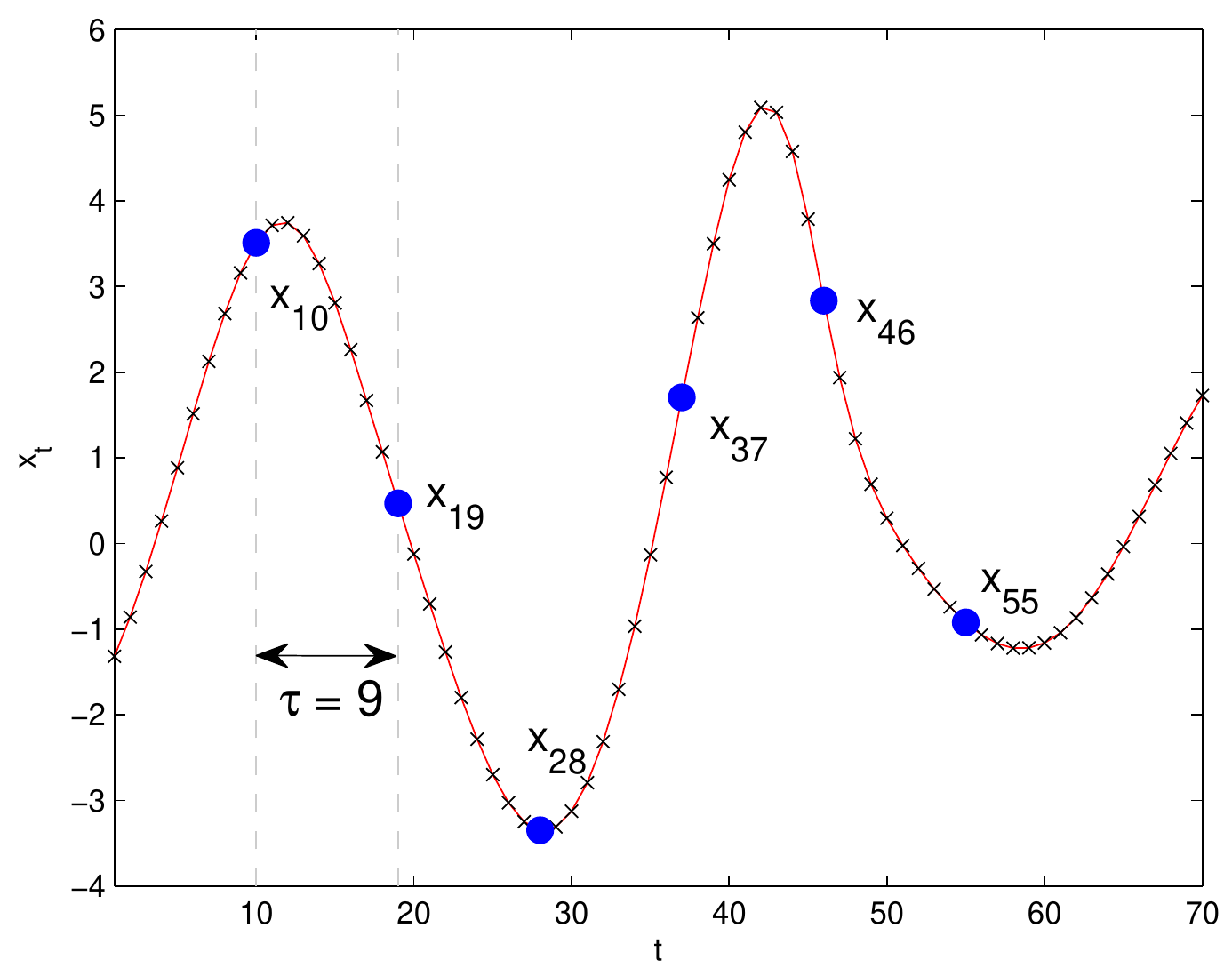}
	\caption{This figure illustrates the process of defining permutation symbols from a discrete sampled time series where sampled points are marked with a cross. Assume \(\tau=9\) and \(D=6\). Starting at \(x_{10}\) the embedded vector is \(v_{10}=\{x_{10},x_{19},x_{28},x_{37},x_{46},x_{55}\}\) (blue dots) and the corresponding symbol defined by the rank ordering is \(s_{10} = \{1,4,6,3,2,5\}\). Likewise, \(s_{11} = \{1,4,6,{\bf2,3},5\}\) and so on.}
	\label{fig:meth}
\end{figure}

The generalised method of ordinal partitions is defined here. The time series, a set of discrete observations \(x=\{x_1,x_2,x_3,...,x_n\}\), is first embedded in \(D\)-dimensional space with embedding lag \(\tau\). The elements in each of the resulting state vectors \(v_i = (x_i,x_{i+\tau},x_{i+2\tau},...x_{i+(D-1)\tau})\) are assigned an ordinal rank in descending order to form symbols \(s_i=(\pi_1,\pi_2,...,\pi_D)\) where \(\pi_j\in \{1,2,...,D\}\) and \(\pi_i \neq \pi_j \iff i \neq j\) (see Figure \ref{fig:meth}). If two or more elements of \(v_i\) are tied then rank for those elements is assigned based on order of appearance in the vector. The unique set of symbols from \(s\) are mapped to nodes in a network represented by the adjacency matrix \(A\). Weighted and directed edges are allocated between nodes based on the transition sequence between permutation symbols, that is, \(s = \{s_1,s_2,s_3,...,s_{(n-D+1)}\}\) corresponds to the embedded state vectors \(v = \{v_1,v_2,v_3,...,v_{(n-D+1)}\}\). Hence, in terms of the existing literature, we have partitioned the time series with an overlapping window. The weight of an edge \(a_{i,j}\) is equal to the number of times that a dynamical transition occurs from node \(i\) to node \(j\).

The selection of \(\tau\) is based on traditional criteria used for time delay embedding. In this study we have used the first zero of the autocorrelation of the time series because it provides a sufficiently good phase space reconstruction for both the R\"ossler time series (Section \ref{sec:Rossler}) and the experimental data (Section \ref{sec:Diode}). However, other methods such as mutual information could foreseeably be used where they provide a better reconstruction for a given time series. By choosing \(\tau\) in this manner, each permutation symbol can be interpreted as a set of inequalities with respect to the coordinates of the embedding phase space. For example, consider the case where \(D=3\) and the time series point \(x_i\) has the symbol \(s_i={1,2,3}\). In terms of the \(3\)-dimensional embedding phase space \(\{x,y,z\}\) we can say that \(v_i\), the embedded vector corresponding to \(x_i\), must have coordinates such that \(x>y\), \(x>z\) and \(y>z\). More generally the symbol \(s_i\) of length \(D\), corresponding to the time series point \(x_i\) and embedded vector \(v_i\), comprises \({D}\choose{2}\) inequalities, each of which defines a subspace of the \(D\)-dimensional embedding phase space. We can therefore say that, based on the information contained in the symbolic sequence alone, \(v_i\) lies within the intersect of said subspaces. These intersect regions are the states which define the nodes in ordinal partition networks. The number of nodes in an ordinal partition network, \(N\), is bounded: \(N\leq D!\) for an infinite time series, and \(N\leq n-D+1\) for a finite time series of length \(n\). Moreover, the effective volume of each state reduces with increasing \(D\) if we only consider systems which are bounded in phase space of some fixed dimension less than \(D\). While there does not yet exist a robust metric for determining the correct choice of \(D\), in the cases examined as part of this study we found that networks with the most visually intuitive structure often corresponded with peak values of degree variance with respect to \(D\). We also found the range \(6\leq D\leq10\) to be the most useful when using simple network measures to track changes in dynamics.

We can also say that \(N\) is dependent on the volume of phase space occupied by the embedded time series, given that each node corresponds to a disjoint region of phase space and assuming that \(D\) is large enough that these regions are sufficiently restricted. This is essentially an equivalent statement to the findings presented by Amigo \textit{et al.} in~\cite{amigo_true_2007} regarding the possible number of permutation patterns present in a time series. Considering only the permutation symbols (not any interpretation of embedded phase space as they did not use a time lag) the authors found that whilst an infinite random sequence will contain all possible permutation patterns, there will always exist patterns which do not occur in periodic and chaotic time series, and furthermore, periodic dynamics will always be characterised by more forbidden patterns than for chaos. Returning to our phase space interpretation, periodic limit cycles will occupy less volume in phase space than a chaotic attractor and, in turn, embedded noise should occupy more phase space than a chaotic attractor given that both time series have the same length and variance.

\section{Application to the R\"ossler system}
\label{sec:Rossler}
We use the R\"ossler equations as a test system:

\begin{equation}
\begin{aligned} \frac{dx}{dt} &= -y - z \\ \frac{dy}{dt} &= x + \alpha y \\ \frac{dz}{dt} &= \beta + z(x-\gamma) \end{aligned}
\label{eq:Rossler}
\end{equation}

\begin{figure}[]
	\centering
		\centering
	\begin{subfigure} {0.24\textwidth}
		\centering
		\includegraphics [width={1\textwidth}] {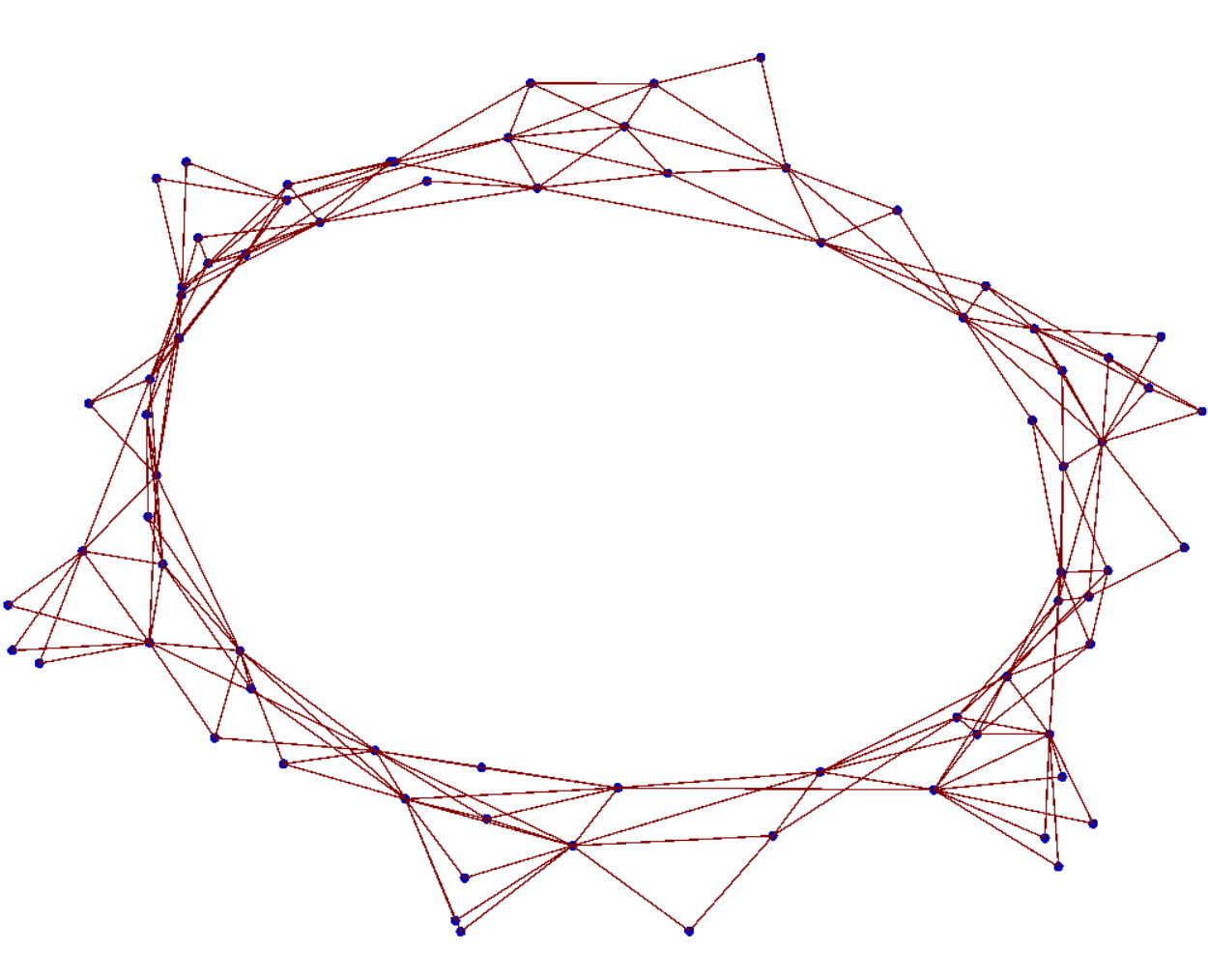}
		\subcaption{}
		\label{fig:NET0601-6}
	\end{subfigure}
	\begin{subfigure} {0.24\textwidth}
		\centering
		\includegraphics [width={1\textwidth}] {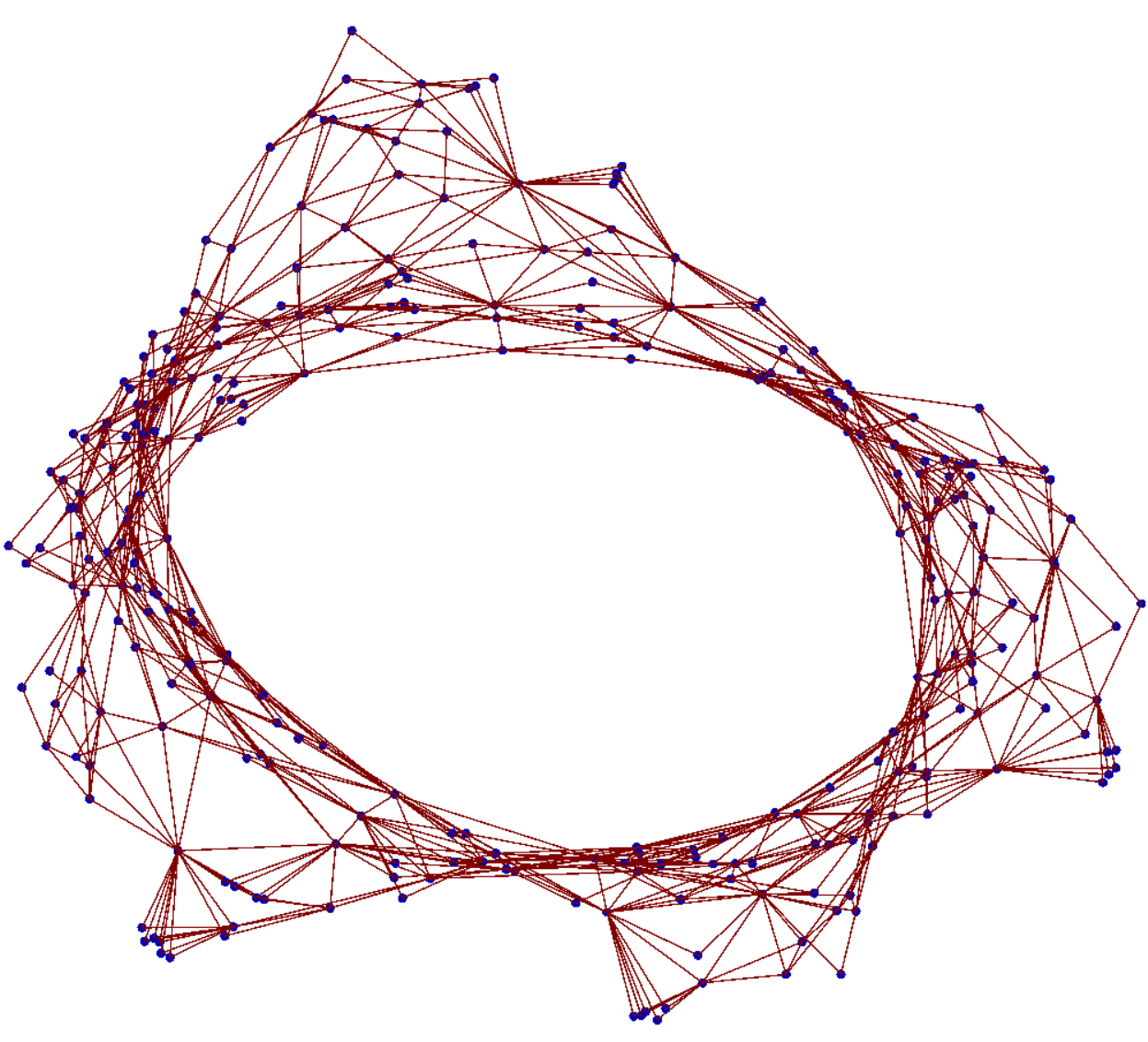}
		\subcaption{}
		\label{fig:NET0601-8}
	\end{subfigure}
	\begin{subfigure} {0.24\textwidth}
		\centering
		\includegraphics [width={1\textwidth}] {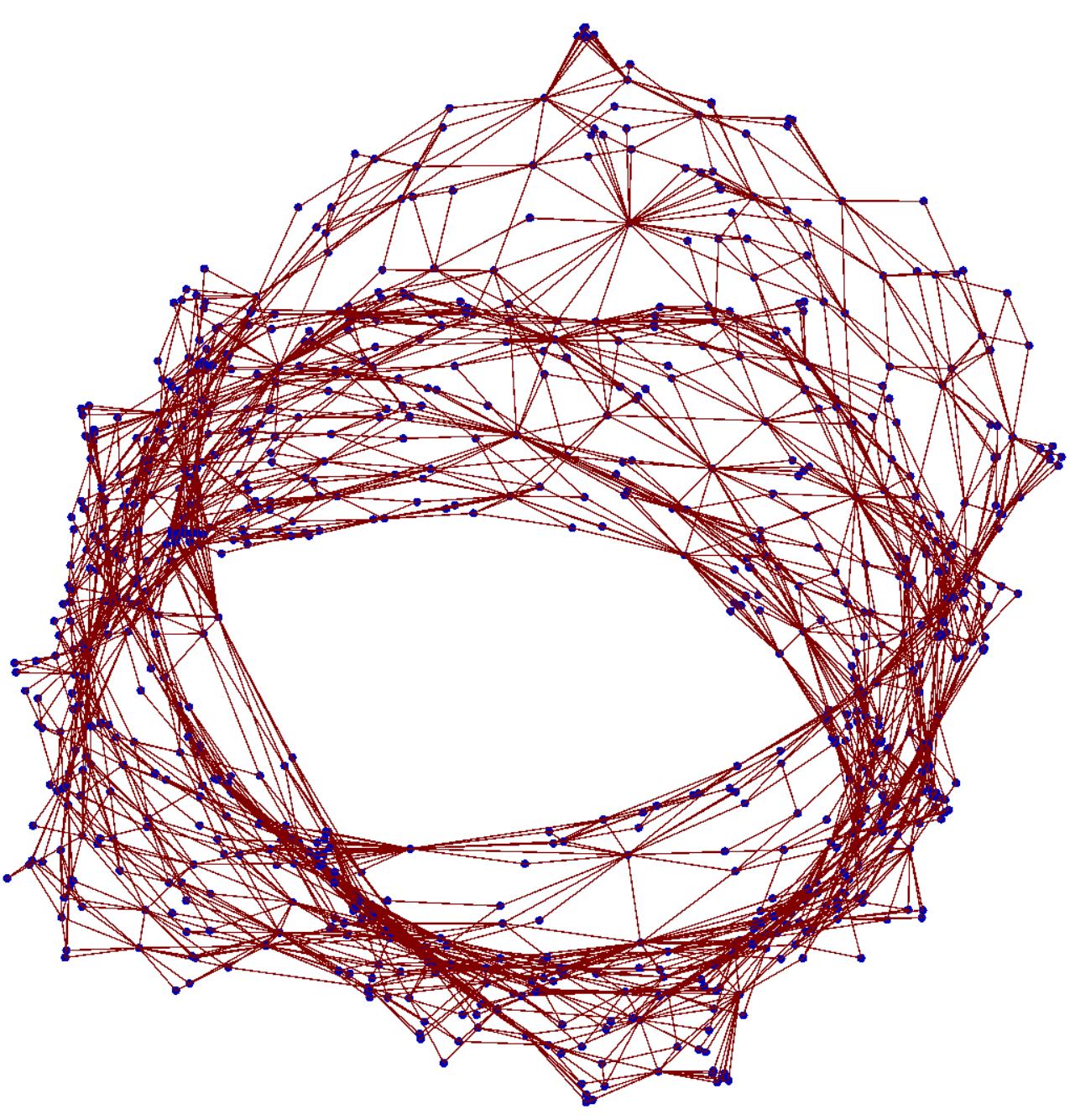}
		\subcaption{}
		\label{fig:NET0601-10}
	\end{subfigure}
	\begin{subfigure} {0.24\textwidth}
		\centering
		\includegraphics [width={1\textwidth}] {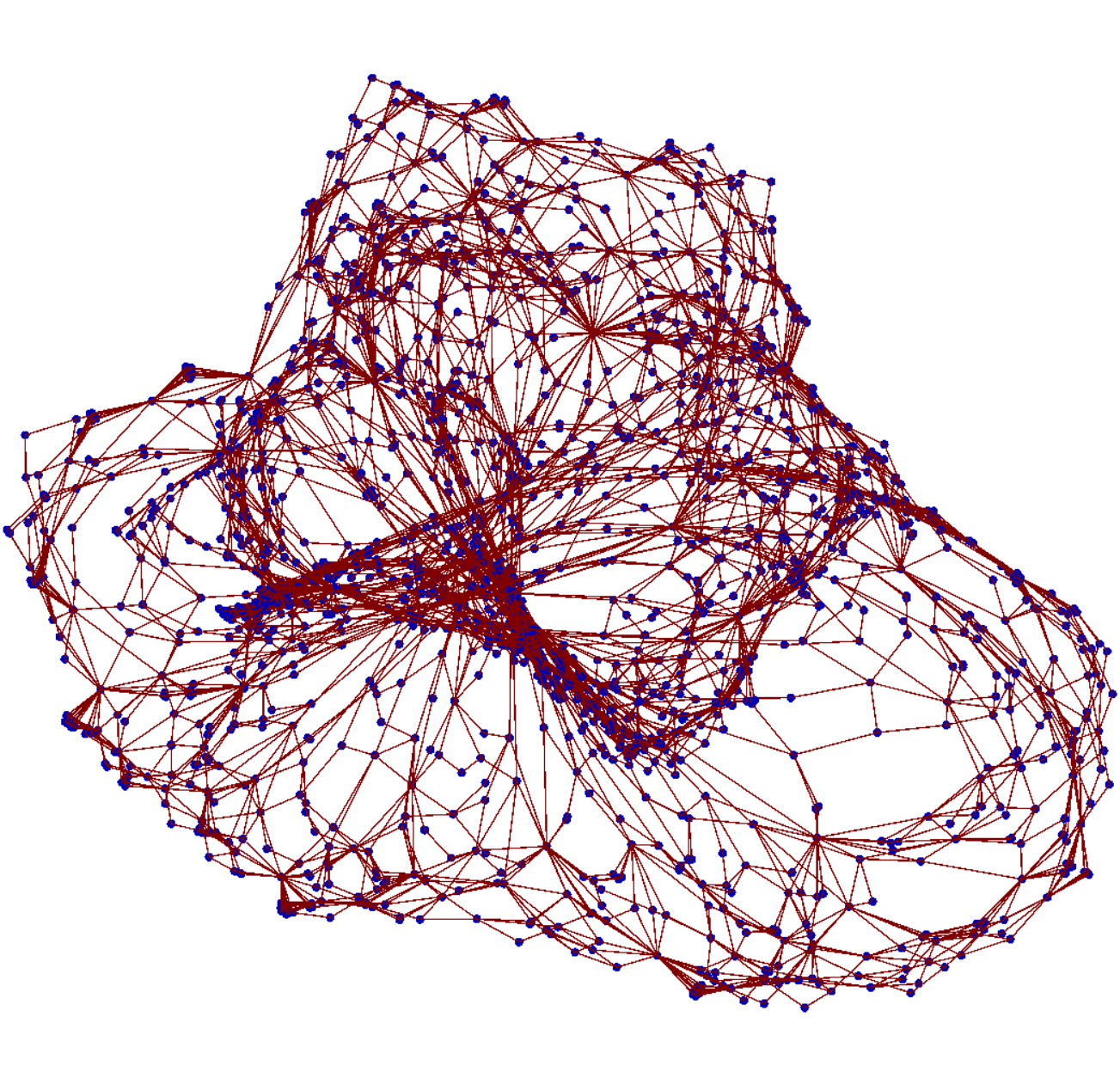}
		\subcaption{}
		\label{fig:NET0601-12}
	\end{subfigure}
	\caption{Networks generated from a chaotic R\"ossler time series as shown in Figure \ref{fig:RosAtr0601} (\(\alpha=0.4\)) using the ordinal partition method with \(\tau=9\) and \subref{fig:NET0601-6} \(D=6\), \subref{fig:NET0601-8} \(D=8\), \subref{fig:NET0601-10} \(D=10\) and \subref{fig:NET0601-12} \(D=12\).}
	\label{fig:RosNet0601}
\end{figure}

\begin{figure}[]
	\centering
		\centering
	\begin{subfigure} {0.24\textwidth}
		\centering
		\includegraphics [width={1\textwidth}] {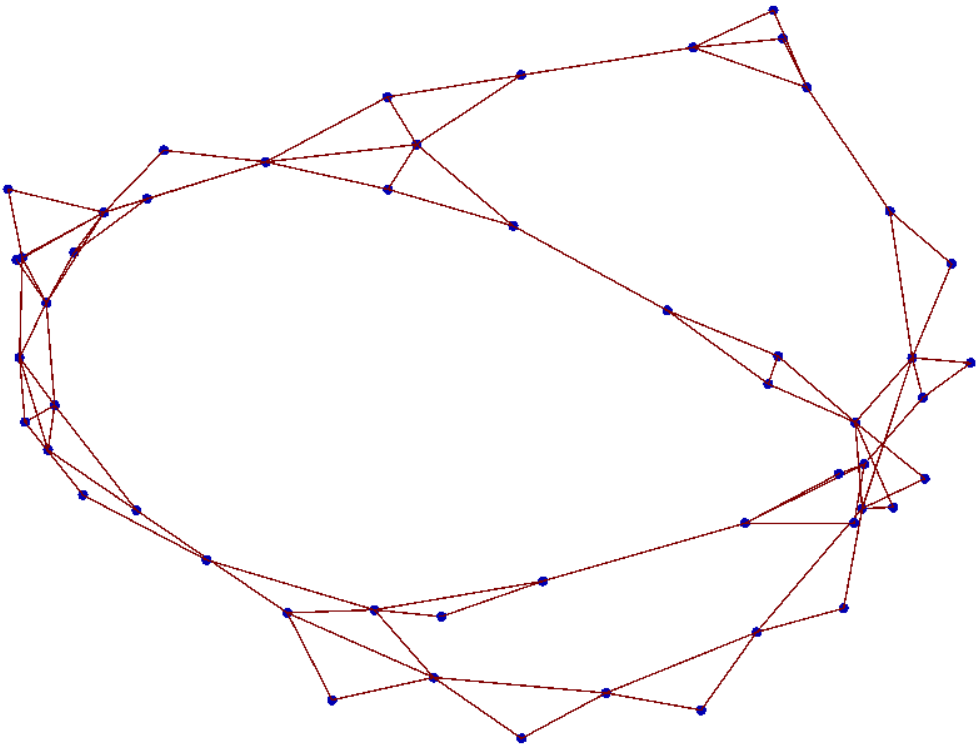}
		\subcaption{}
		\label{fig:NET0001-6}
	\end{subfigure}
	\begin{subfigure} {0.24\textwidth}
		\centering
		\includegraphics [width={1\textwidth}] {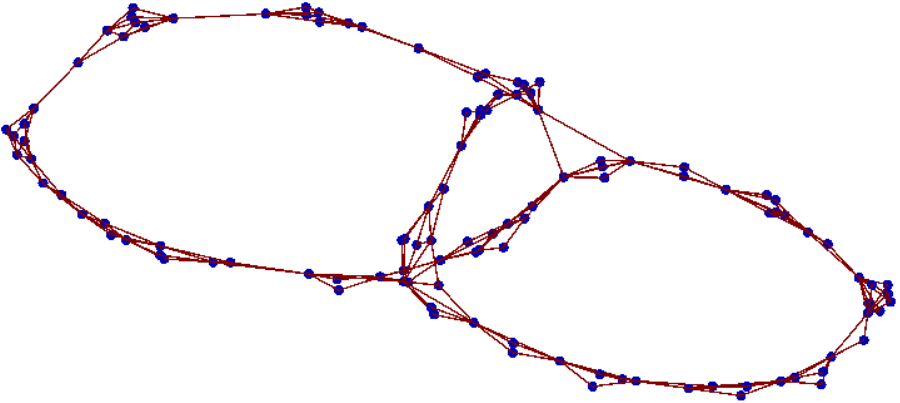}
		\subcaption{}
		\label{fig:NET0001-8}
	\end{subfigure}
	\begin{subfigure} {0.24\textwidth}
		\centering
		\includegraphics [width={1\textwidth}] {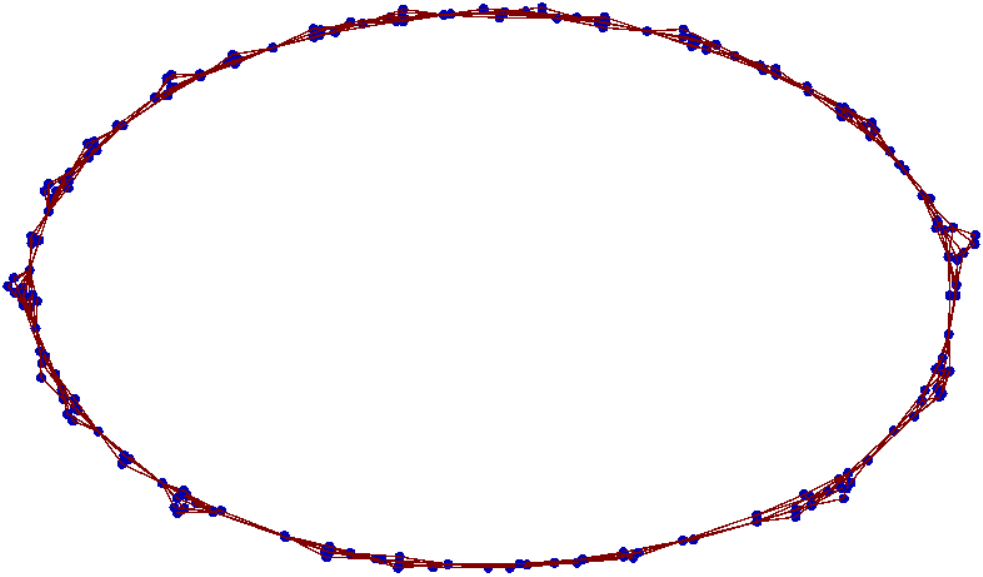}
		\subcaption{}
		\label{fig:NET0001-10}
	\end{subfigure}
	\begin{subfigure} {0.24\textwidth}
		\centering
		\includegraphics [width={1\textwidth}] {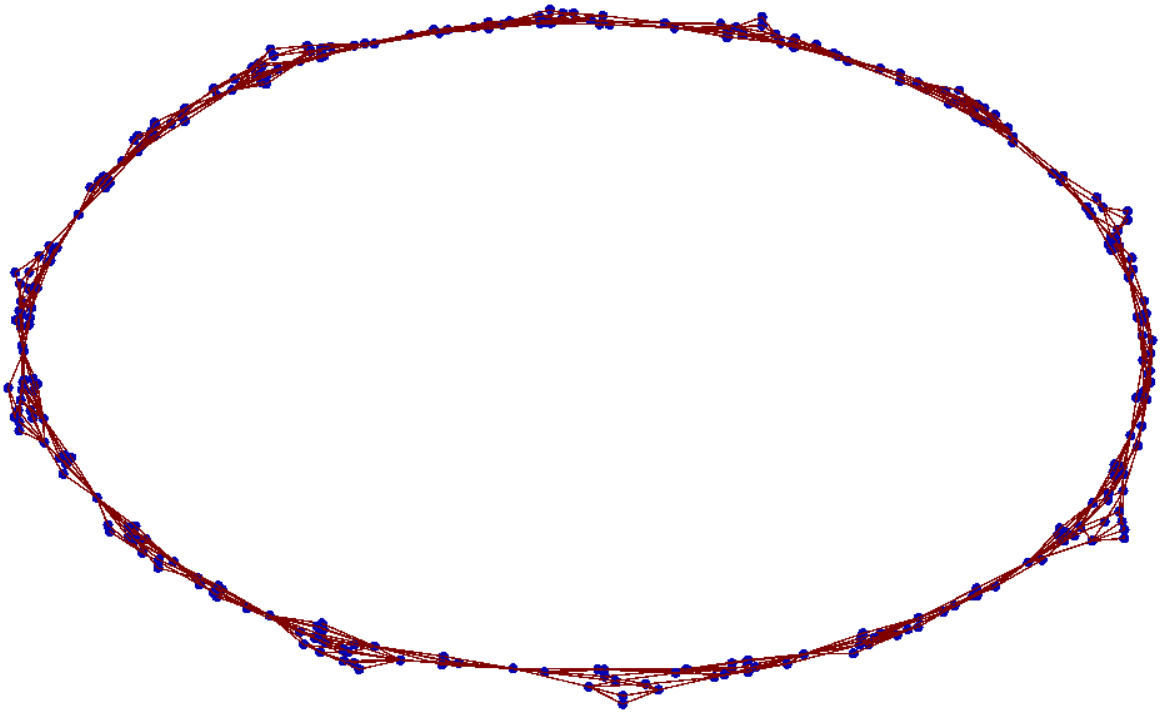}
		\subcaption{}
		\label{fig:NET0001-12}
	\end{subfigure}
	\caption{Networks generated from a period-2 R\"ossler time series as shown in Figure \ref{fig:RosAtr0001} (\(\alpha=0.37\)) using the ordinal partition method with \(\tau=9\) and \subref{fig:NET0001-6} \(D=6\), \subref{fig:NET0001-8} \(D=8\), \subref{fig:NET0001-10} \(D=10\) and \subref{fig:NET0001-12} \(D=12\).}
	\label{fig:RosNet0001}
\end{figure}

\begin{figure}[]
	\centering
		\centering
	\begin{subfigure} {0.24\textwidth}
		\centering
		\includegraphics [width={1.15\textwidth}] {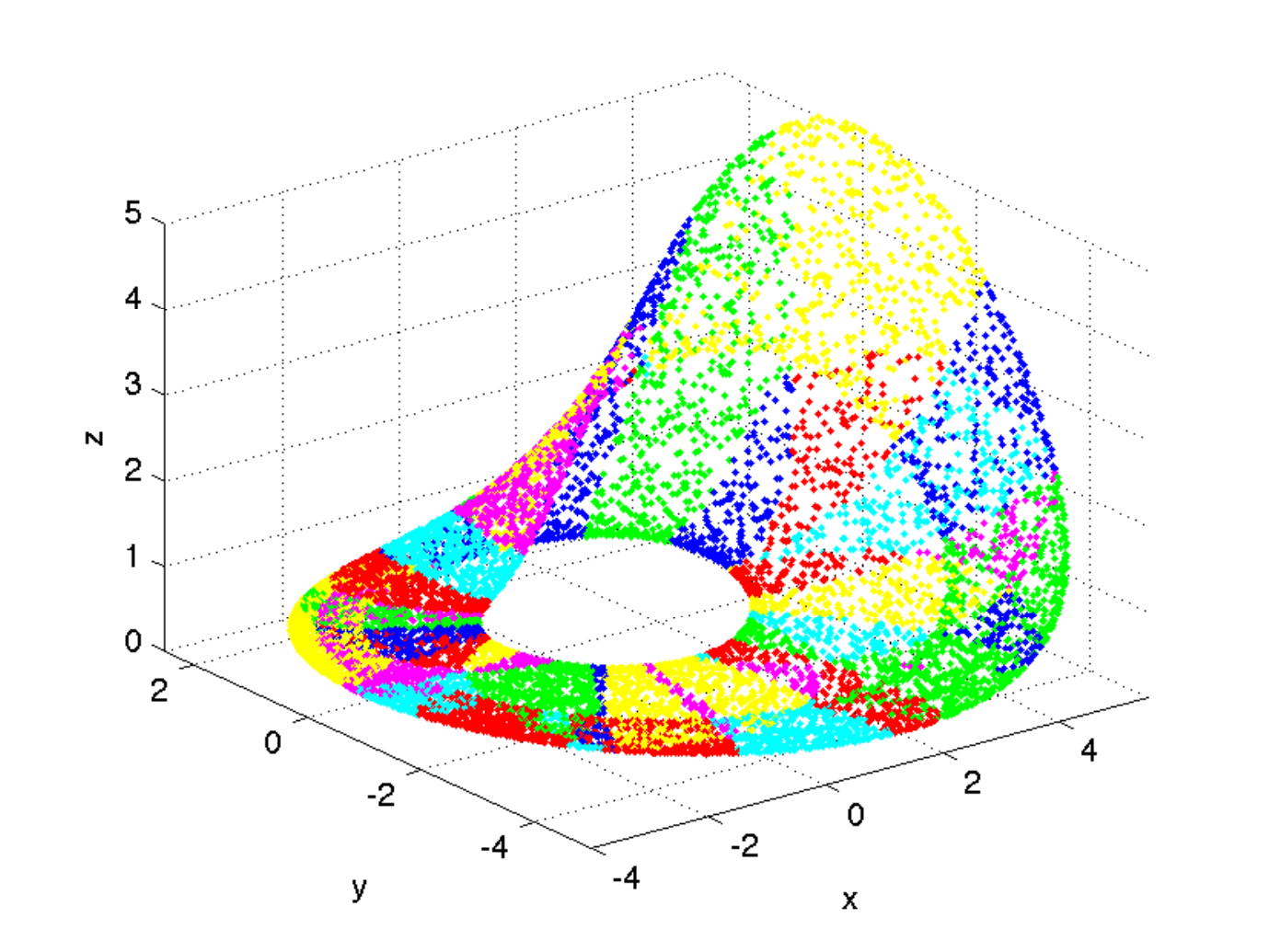}
		\subcaption{}
		\label{fig:0601-6}
	\end{subfigure}
	\begin{subfigure} {0.24\textwidth}
		\centering
		\includegraphics [width={1.15\textwidth}] {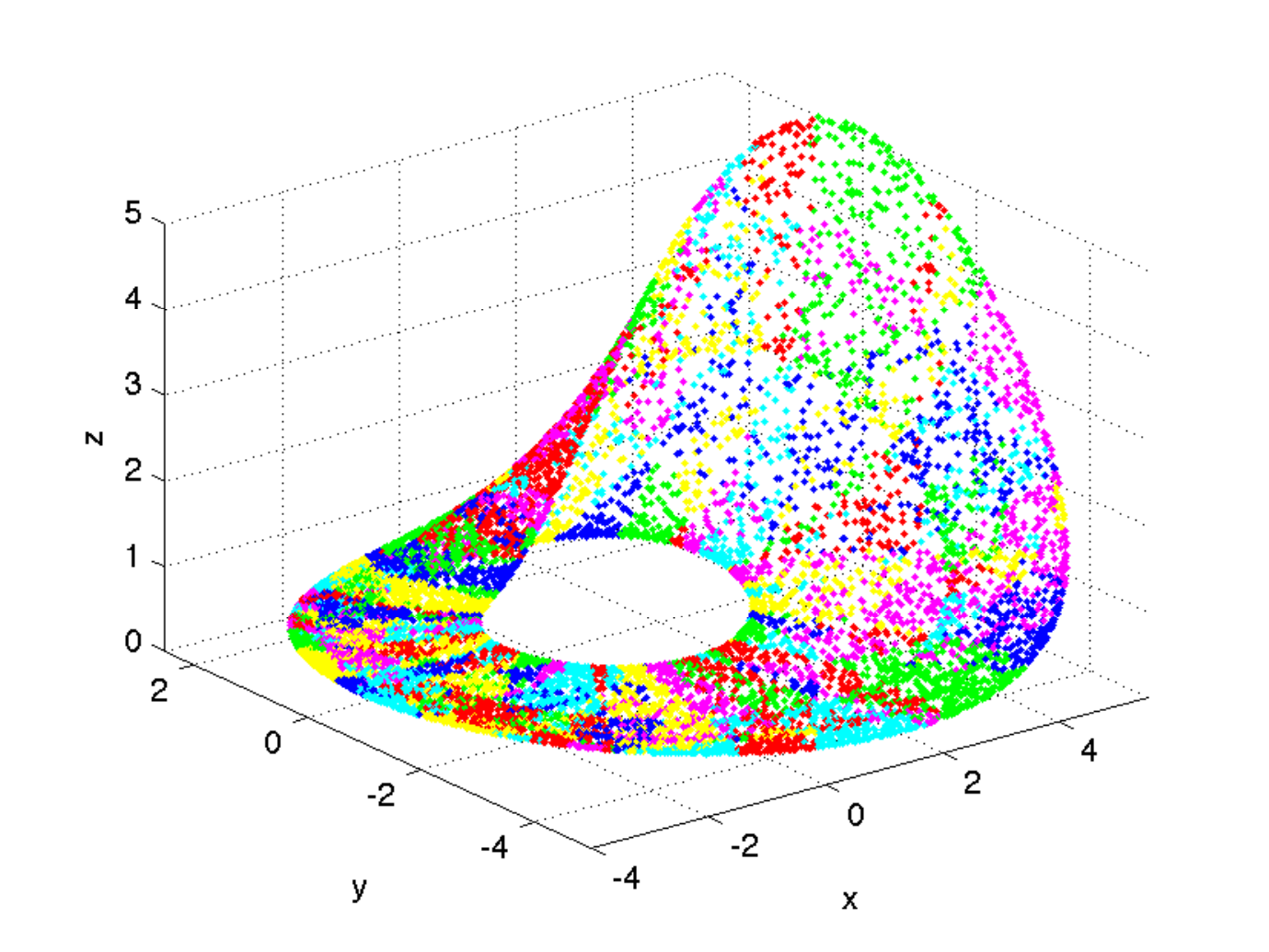}
		\subcaption{}
		\label{fig:0601-8}
	\end{subfigure}
	\begin{subfigure} {0.24\textwidth}
		\centering
		\includegraphics [width={1.15\textwidth}] {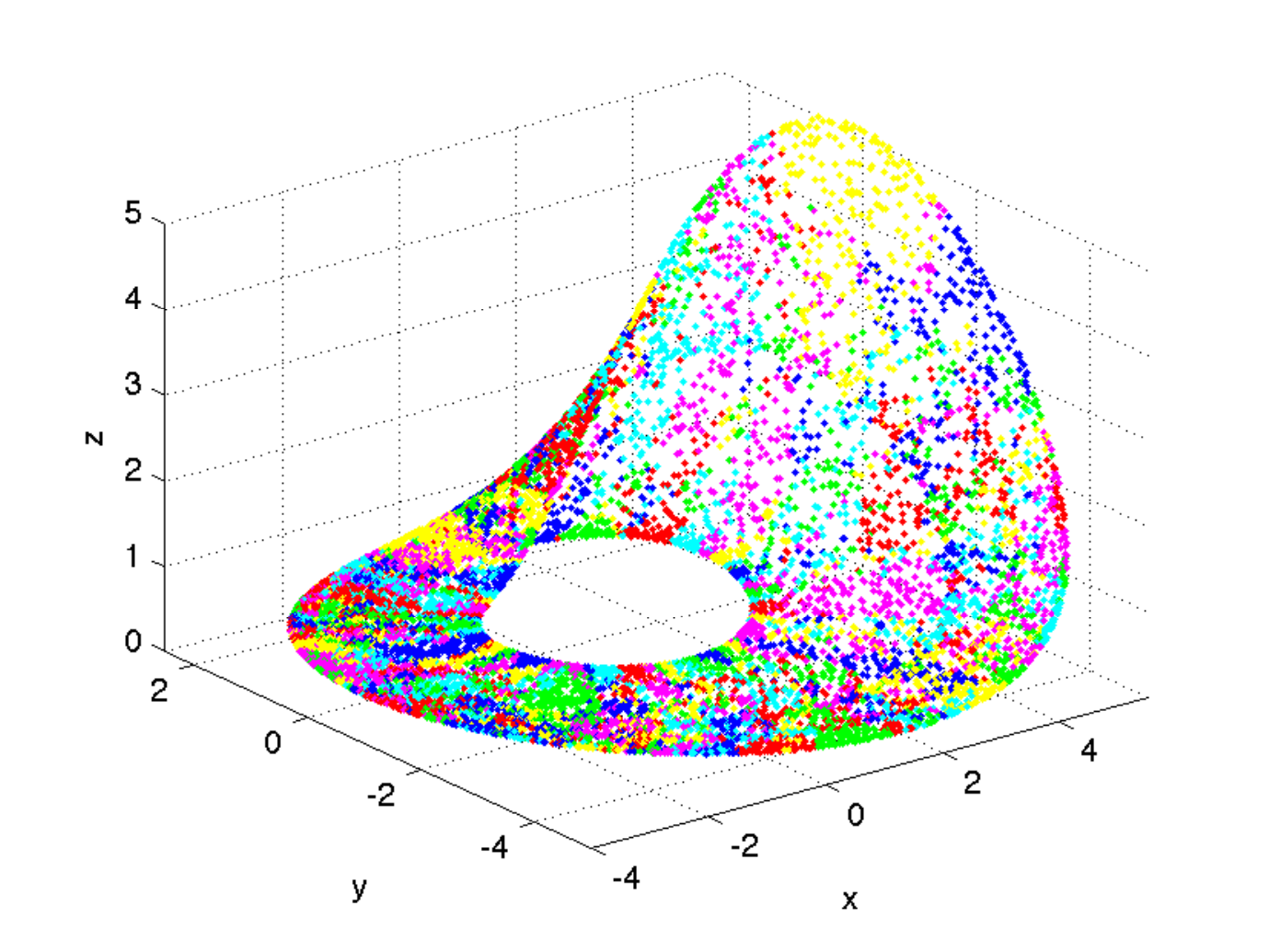}
		\subcaption{}
		\label{fig:0601-10}
	\end{subfigure}
	\begin{subfigure} {0.24\textwidth}
		\centering
		\includegraphics [width={1.15\textwidth}] {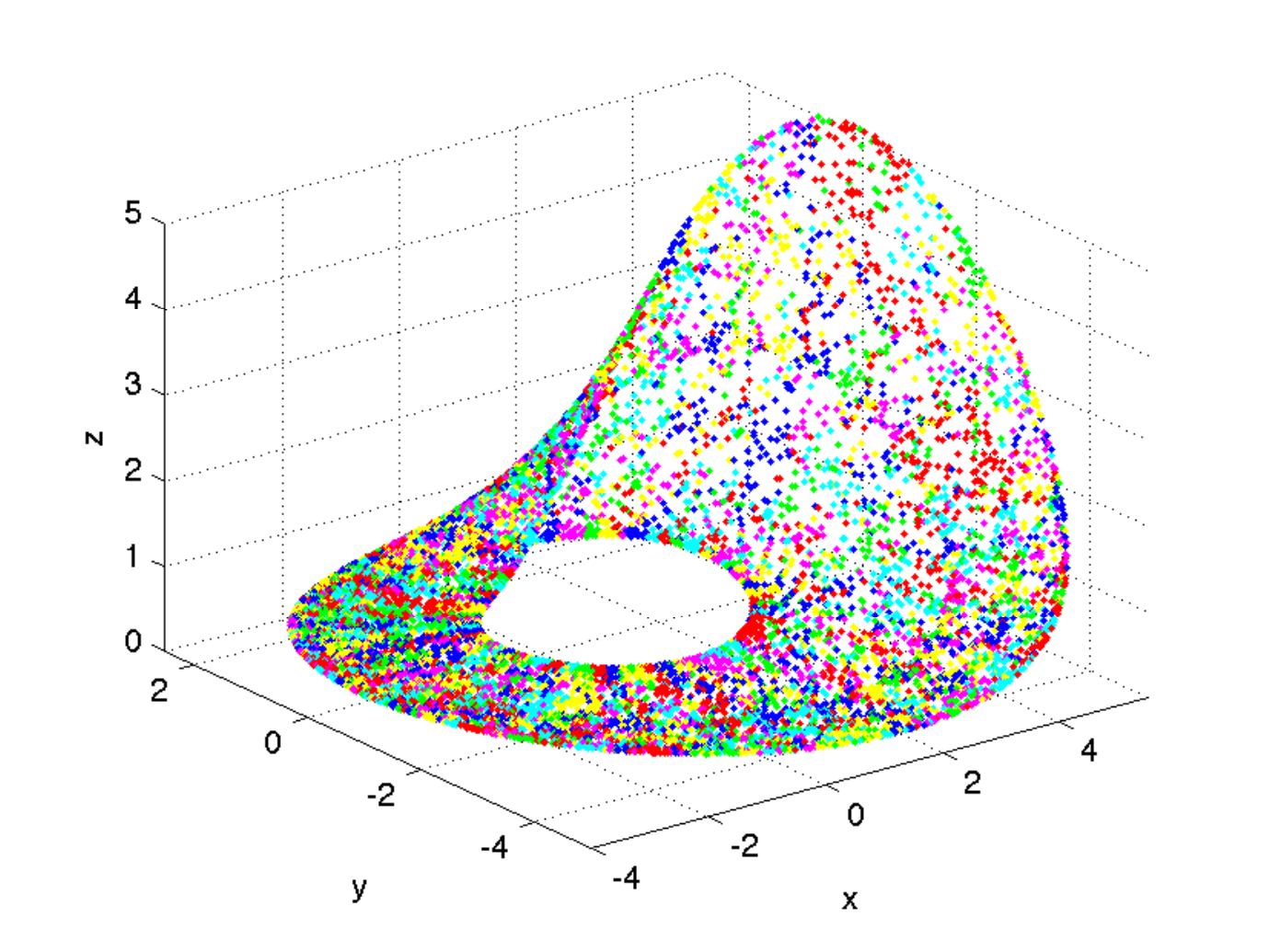}
		\subcaption{}
		\label{fig:0601-12}
	\end{subfigure}
	\caption{Time series points in phase space for a chaotic R\"ossler system (\(\alpha=0.4\)). The nodes from networks generated using this time series with \(D=6\) , \(D=8\) , \(D=10\) and \(D=12\) (see Figure \ref{fig:RosNet0601}) have been colour mapped back to the attractor using one of seven colours in subfigures \subref{fig:0601-6}, \subref{fig:0601-8}, \subref{fig:0601-10} and \subref{fig:0601-12} respectively. Temporally adjacent nodes are assigned different colours such that distinctly coloured regions on the attractor are representative of the states defined by the permutation symbols in phase space.}
	\label{fig:RosAtr0601}
\end{figure}

\begin{figure}[]
	\centering
		\centering
	\begin{subfigure} {0.24\textwidth}
		\centering
		\includegraphics [width={1.15\textwidth}] {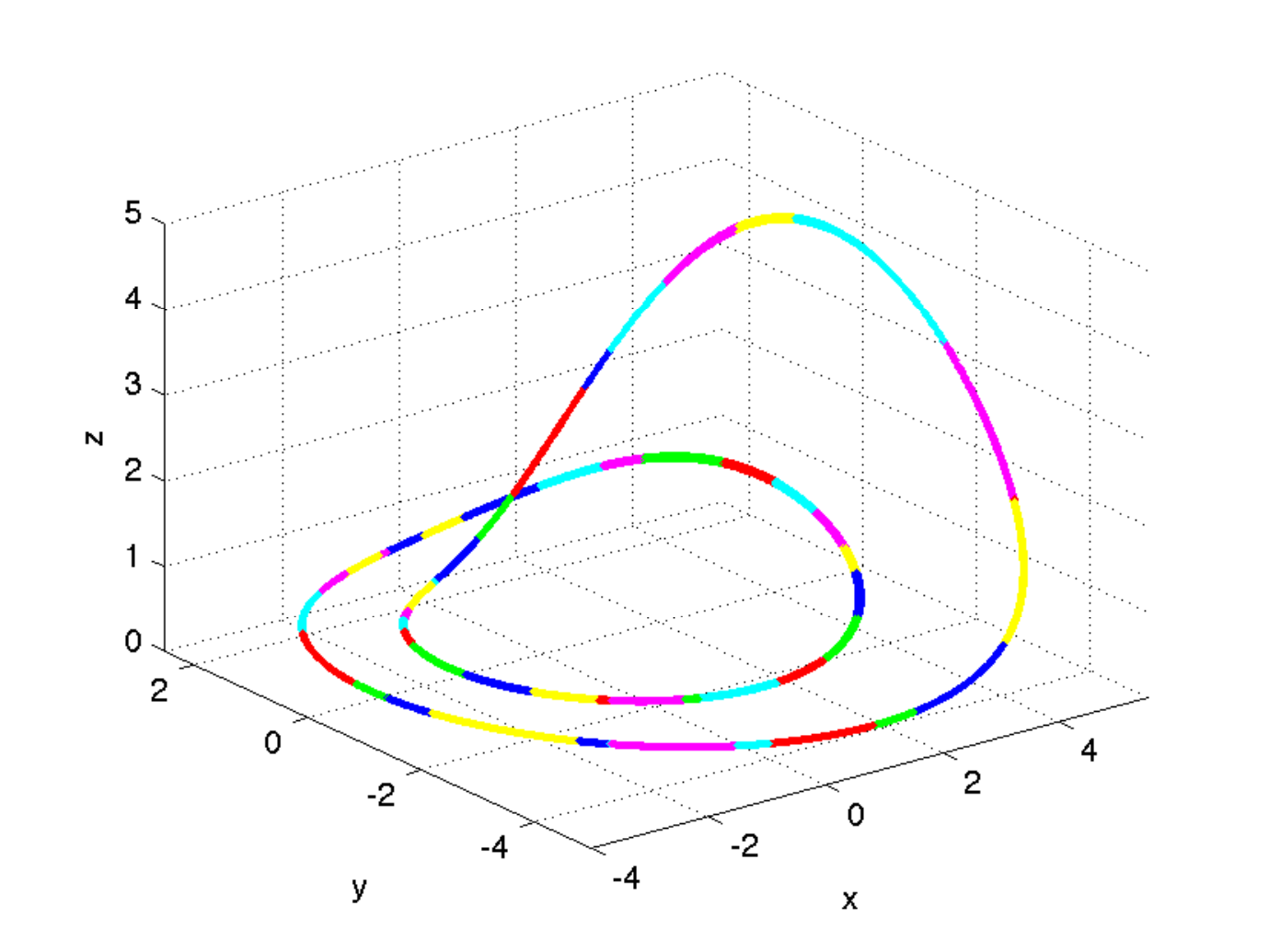}
		\subcaption{}
		\label{fig:0001-6}
	\end{subfigure}
	\begin{subfigure} {0.24\textwidth}
		\centering
		\includegraphics [width={1.15\textwidth}] {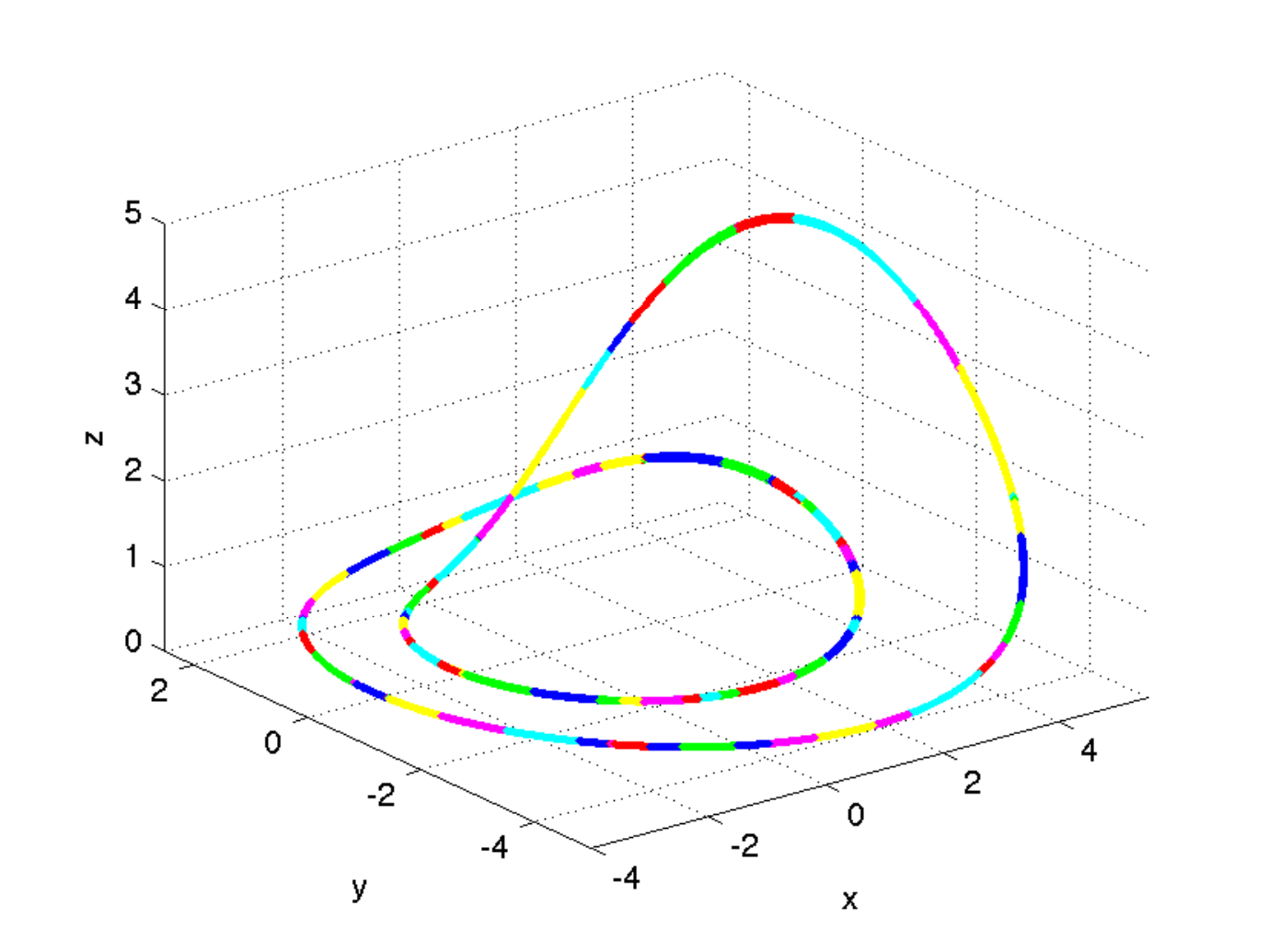}
		\subcaption{}
		\label{fig:0001-8}
	\end{subfigure}
	\begin{subfigure} {0.24\textwidth}
		\centering
		\includegraphics [width={1.15\textwidth}] {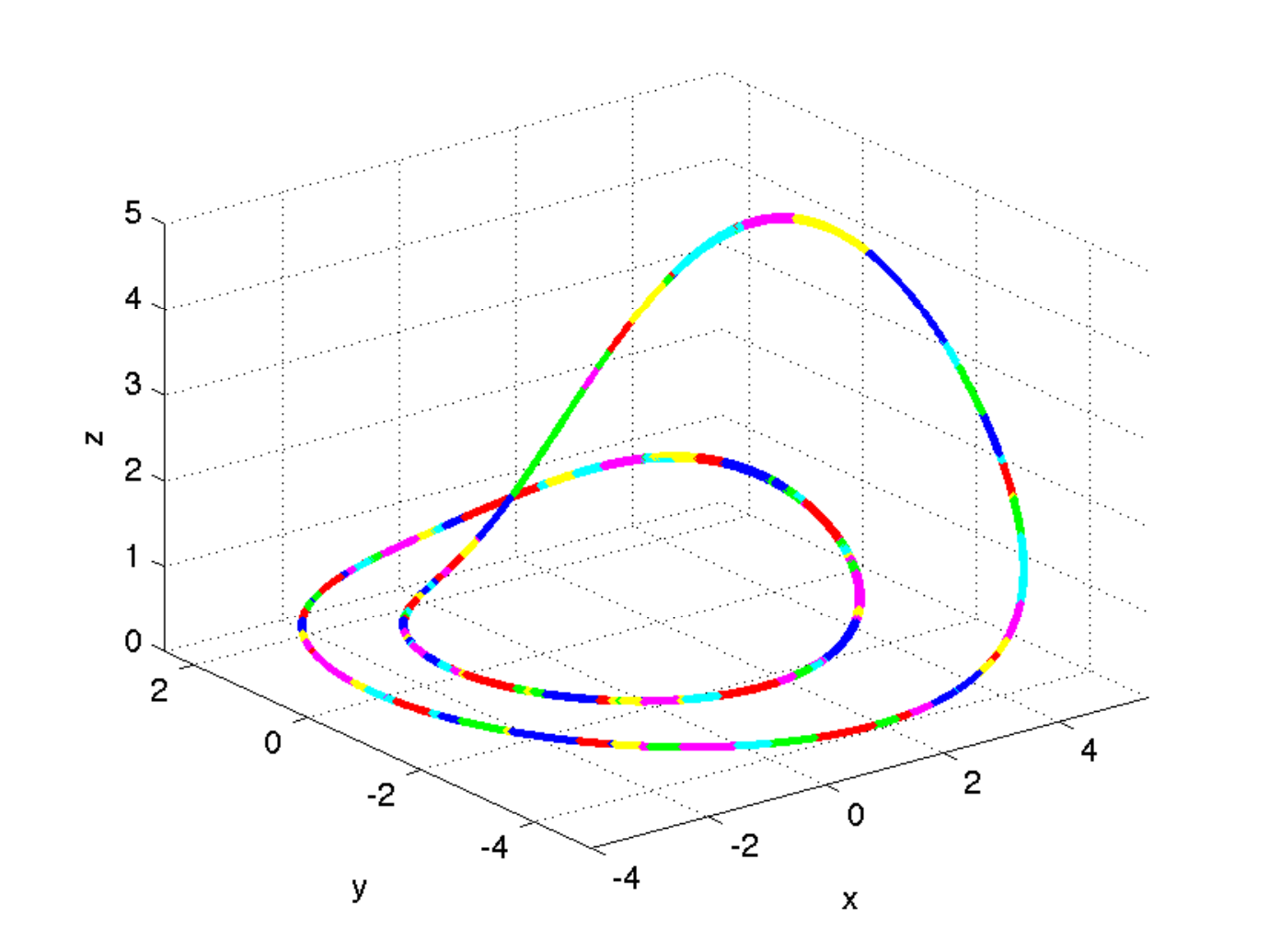}
		\subcaption{}
		\label{fig:0001-10}
	\end{subfigure}
	\begin{subfigure} {0.24\textwidth}
		\centering
		\includegraphics [width={1.15\textwidth}] {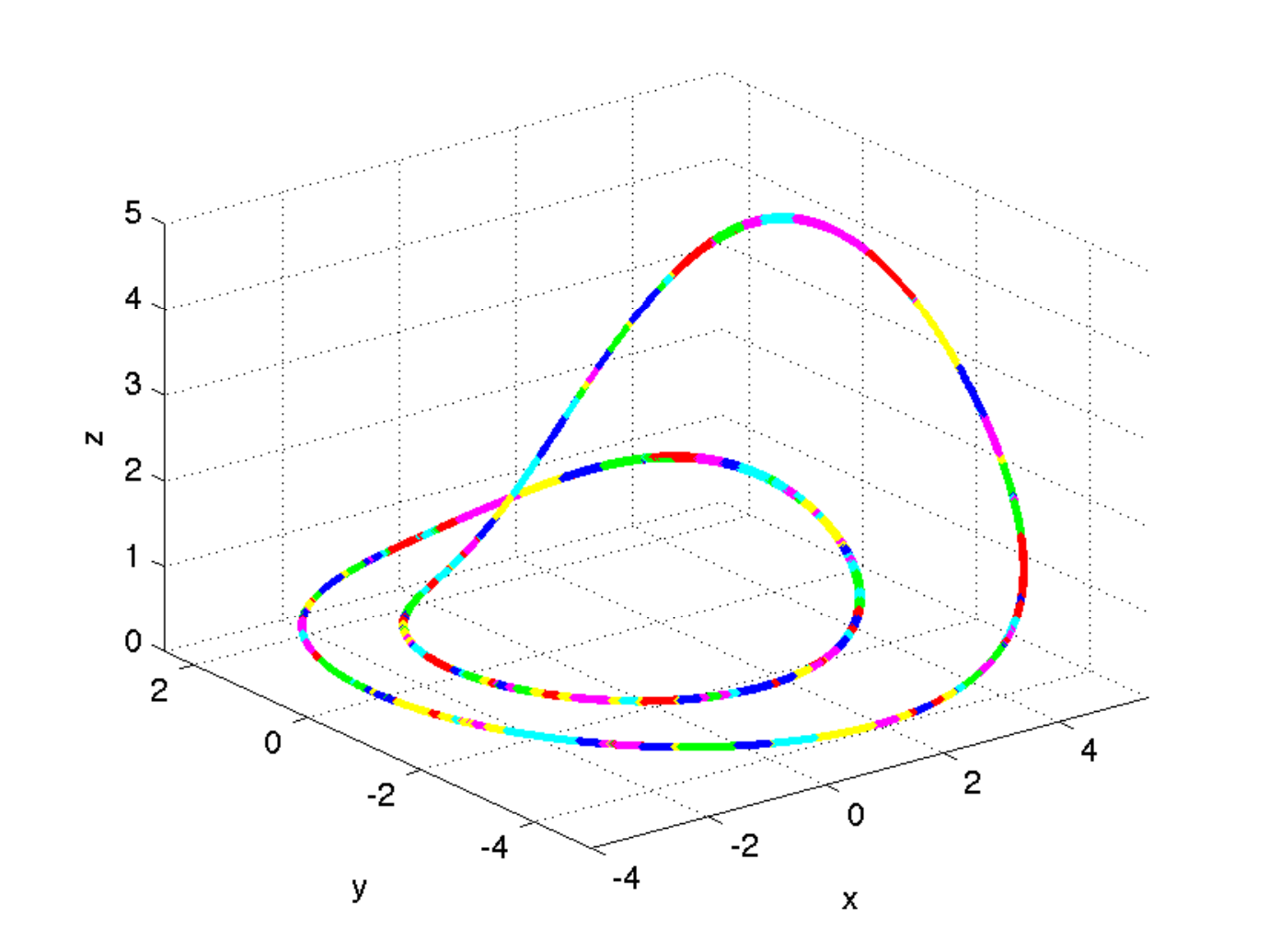}
		\subcaption{}
		\label{fig:0001-12}
	\end{subfigure}
	\caption{Time series points in phase space for a period-2 R\"ossler system (\(\alpha=0.37\)). Colour mapping procedure is the same is in Figure \ref{fig:RosAtr0601} but for the networks shown in Figure \ref{fig:RosNet0001}.}
	\label{fig:RosAtr0001}
\end{figure}

\noindent We solve the equations using a fourth/fifth order Runge-Kutta method for \(\beta=2\), \(\gamma=4\) and 1201 evenly spaced values of the bifurcation parameter in the range \(0.37\leq \alpha \leq 0.43\). Each time series was solved with random initial conditions in the range 0 to 1 for \(2\times10^4\) points at a time step of 0.2, the first \(10^4\) of which were discarded to remove transients. In the following analysis we use the remaining \(10^4\) time series points from the \(x\) component of the system to generate networks. The first zero of the autocorrelation of the ensemble of time series is consistently \(\tau=9\) so this value is used to generate all the networks in this section. Figures \ref{fig:RosNet0601} and \ref{fig:RosNet0001} show networks generated for different embedding dimensions from a time series exhibiting broadband chaos (\(\alpha=0.4\)) and from a period-2 oscillation (\(\alpha=0.37\)) respectively. Corresponding attractors are shown in Figures \ref{fig:RosAtr0601} and \ref{fig:RosAtr0001} (the colour map shall be explained in the following analysis). The network images are produced using a spring electrical embedding algorithm as implemented in Wolfram Mathematica 9~\cite{wolfram_documentation_2014}. This algorithm positions vertices at an equilibrium point in the network space as if the network were a physical system where nodes experience an attractive spring force proportional to their separation in Euclidian space if they are adjacent, and a repulsive force that is inversely proportional to Euclidian separation regardless of connectivity.

Firstly we consider the chaotic case (Figures \ref{fig:RosNet0601} and \ref{fig:RosAtr0601}). For \(D=6,8,10\) the network maintains a tight band structure which is most reminiscent of the attractor when \(D=10\), but this structure breaks down at \(D=12\). Figure \ref{fig:RosAtr0601} is a colour map of the network nodes back onto the original attractor (not the embedding) in three dimensional phase space to illustrate how the attractor has been partitioned into states. When \(D=6\) the states are clearly observable. As expected, when the embedding dimension increases the states on the attractor become more numerous and smaller until they are difficult to distinguish when \(D=12\).

\begin{figure}[]
	\centering
		\centering
	\begin{subfigure} {0.32\textwidth}
		\centering
		\includegraphics [width={1.1\textwidth}] {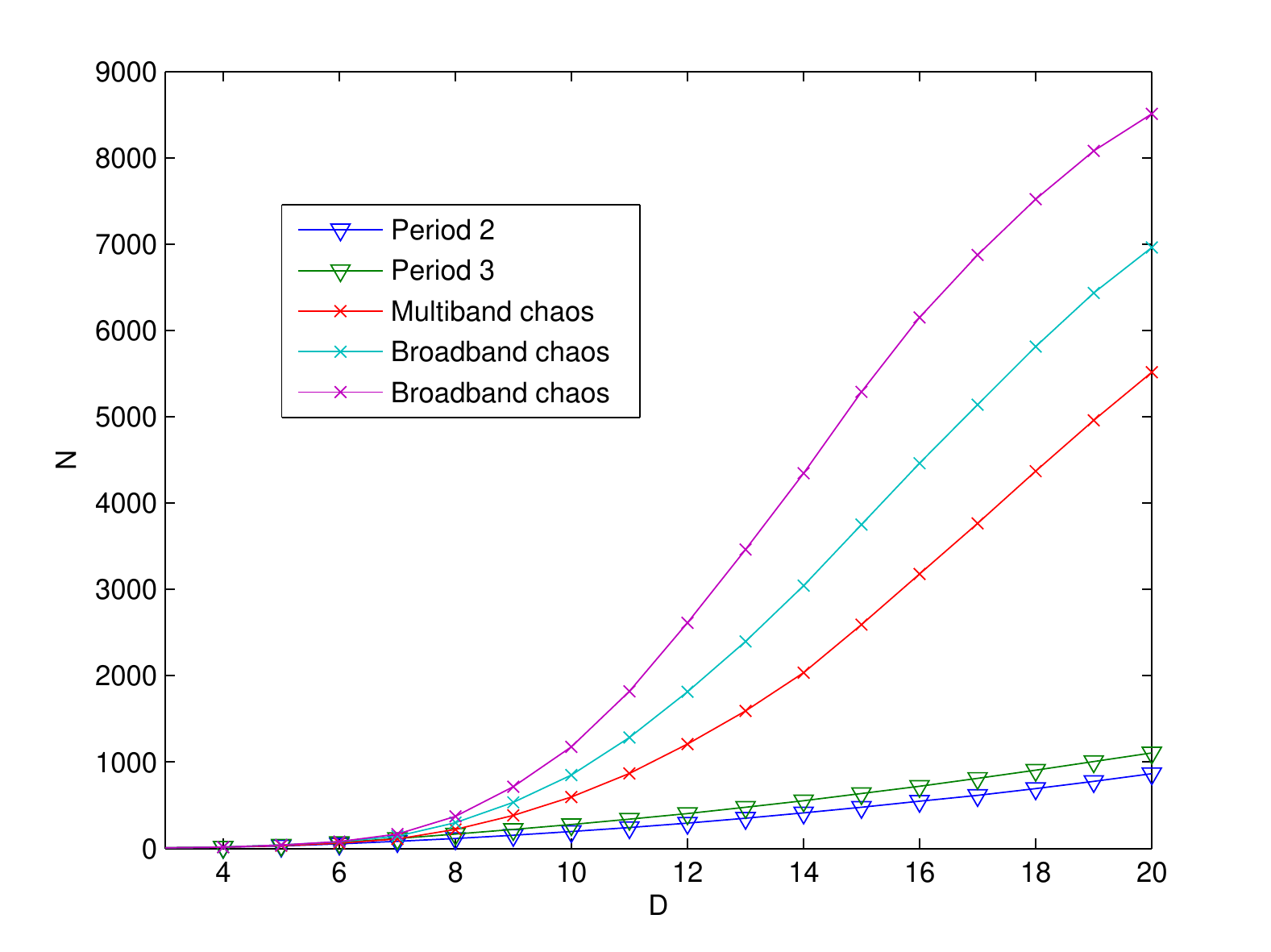}
		\subcaption{}
		\label{fig:NVsD}
	\end{subfigure}
		\begin{subfigure} {0.32\textwidth}
		\centering
		\includegraphics [width={1.1\textwidth}] {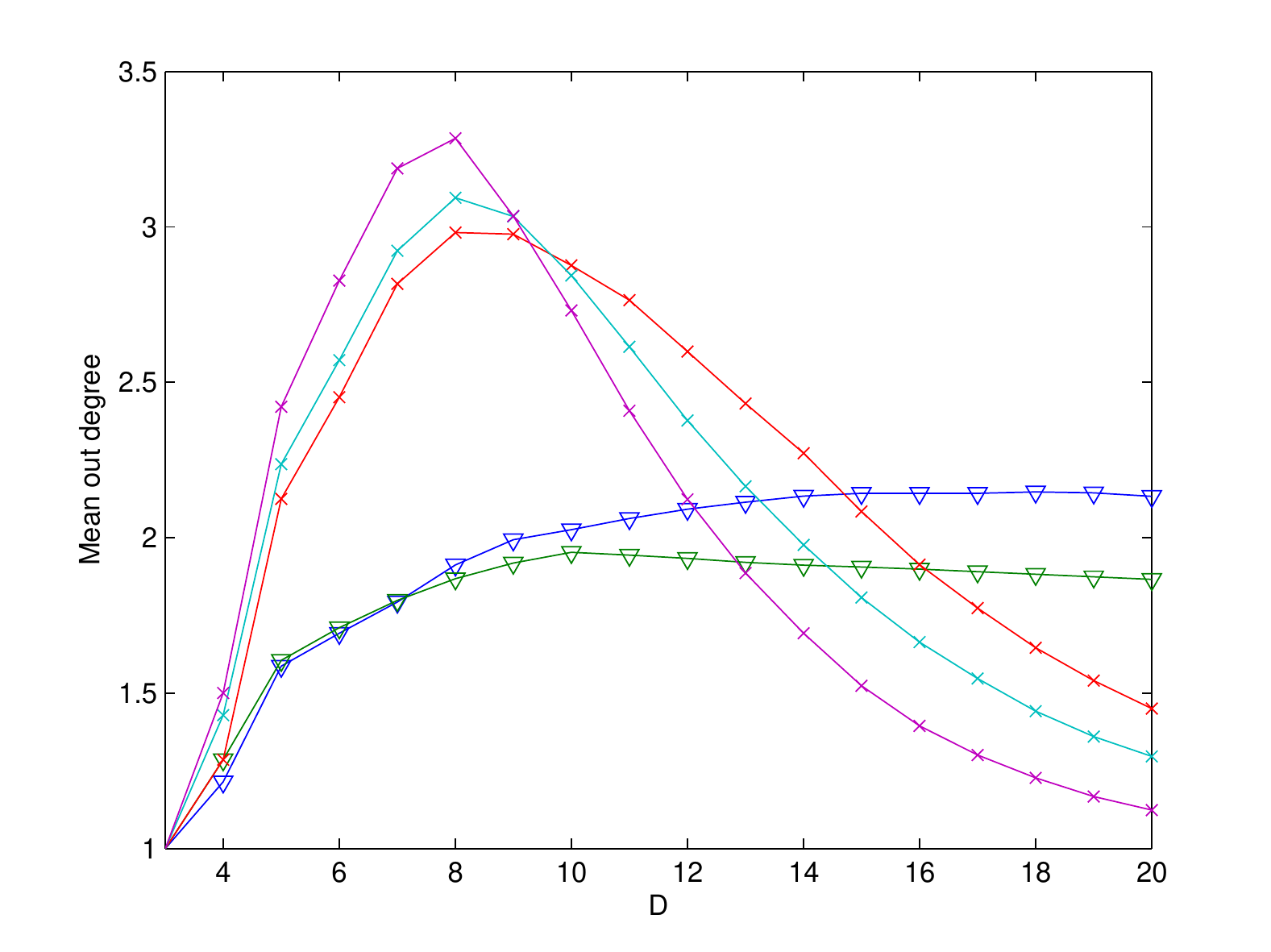}
		\subcaption{}
		\label{fig:MeanDegVsD}
	\end{subfigure}
	\begin{subfigure} {0.32\textwidth}
		\centering
		\includegraphics [width={1.1\textwidth}] {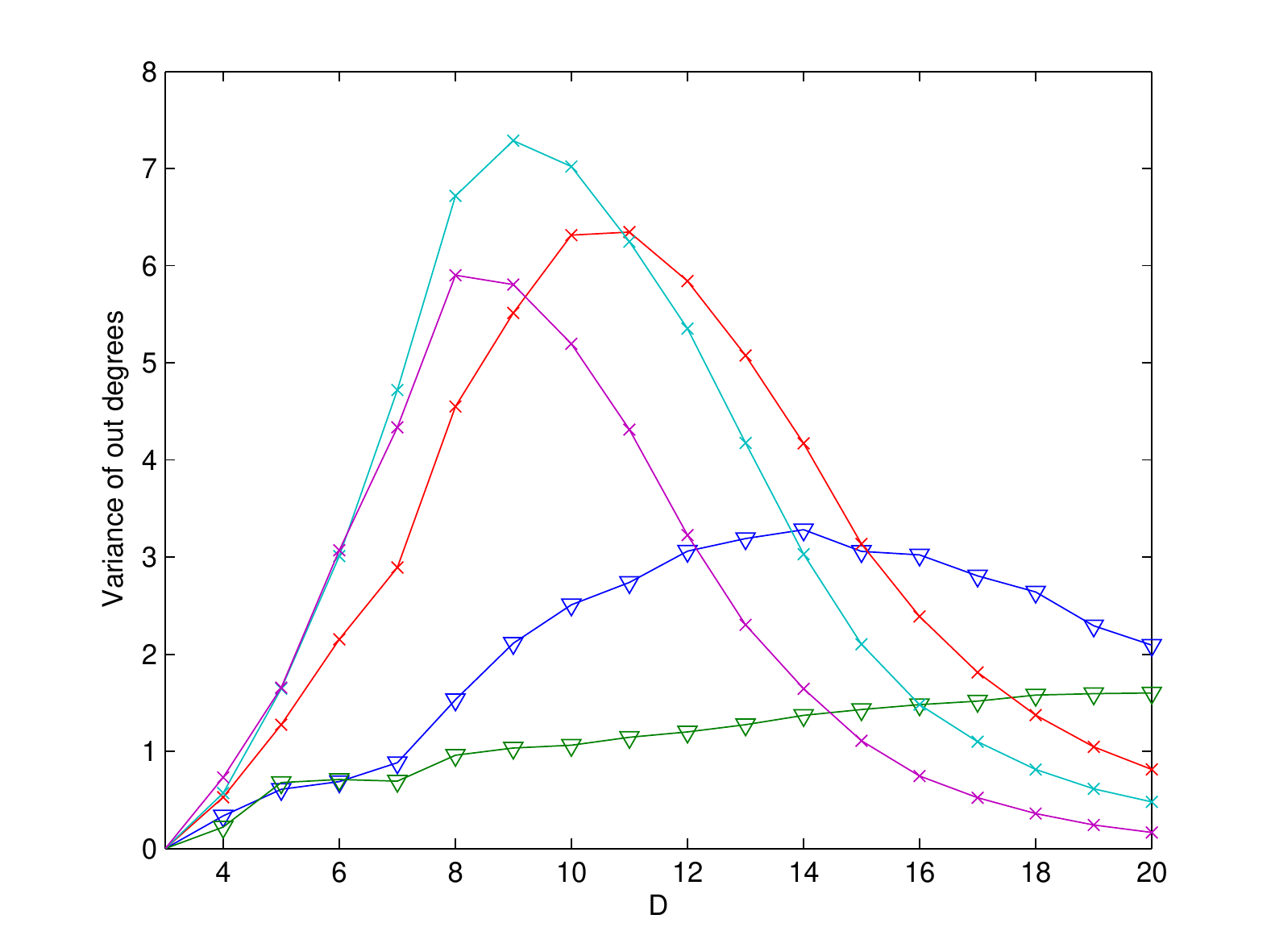}
		\subcaption{}
		\label{fig:DegVarVsD}
	\end{subfigure}
	\caption{Network measures plotted against the embedding dimension for the R\"ossler system in different dynamical regimes: period 2, \(\alpha=0.37\) (blue line); period 3, \(\alpha=0.41\) (green line); multiband chaos, \(\alpha=0.39\) (red line); broadband chaos, \(\alpha=0.4\) (cyan line); and broadband chaos, \(\alpha=0.42\) (magenta line). Network measures shown are \subref{fig:NVsD} number of nodes, \subref{fig:MeanDegVsD} mean out degree, and \subref{fig:DegVarVsD} variance of out degrees.}
	\label{fig:RosDim}
\end{figure}

We now discuss the selection of \(D\) with respect to the results presented in Figure \ref{fig:RosDim}. Transforming the time series into a transition network is a process of mapping the temporal information into a Markov chain to obtain a compressed or simplified representation of the dynamics. The level of simplification is governed by \(D\). In Figure \ref{fig:NVsD}, observe that \(N\) begins to rapidly approach a value close to the length of the data for \(D>9\) in networks from chaotic time series (red, cyan and magenta lines). In this range the phase space has been partitioned into too many states and the network is no longer a useful simplification of the system dynamics. We propose that the variance of out degrees and mean node out degree can provide an indication of an appropriate range for \(D\) for reasons as follows.

In transition networks, node degree is proportional to the uncertainty about where the system will be after the next time step. Consider that a periodic system perfectly embedded in a directed transition network will have mean node out degree \(\langle k_{out}\rangle=1\) with variance \(\sigma=0\). In practice these values will be larger due to the shape of the trajectories and sampling time step which both impact on the capacity for any phase space partitioning method to accurately represent the dynamics (e.g. if the length of a segment of a trajectory for a single time step skips over a state in the transition network because \(\Delta t\) is too large). For chaotic dynamics both \(\langle k_{out}\rangle\) and \(\sigma\) will be nonzero due to the stretching and folding of the trajectories. If \(\sigma\) is maximised then we might assume that information about transitional probabilities is being captured on as wider scale as possible for the given dataset. Figure \ref{fig:DegVarVsD} shows clear peaks in \(\sigma\) against \(D\) for the chaotic R\"ossler time series. The cyan line, which corresponds to the networks in Figure \ref{fig:RosNet0601}, peaks at \(D=9\) then falls away rapidly. Generally, when \(D\) was selected close to the peak value for \(\sigma\) we found the networks for chaotic systems most closely resembled the attractor, but for larger \(D\) the visually coherent network structure begins to break down.

Peaks also occur in the plots for mean degree against \(D\) (Figure \ref{fig:MeanDegVsD}) providing a further indication for parameter selection. To explain this consider an alternative interpretation of the permutation symbols as information about where a time series point is with respect to a history of \(D\) steps of length \(\tau\). Due to the exponential divergence of trajectories in chaotic attractors, if this history is too long then each state will correspond to fewer and longer sections of trajectory that will have a limited number of transitional possibilities (low node degree) such that the network is no longer a useful simplification of the dynamics.

Networks have ring structures for periodic regimes as shown in Figure \ref{fig:NET0001-10} and \ref{fig:NET0001-12}. The corresponding period-2 limit cycle is shown in Figure \ref{fig:RosAtr0001}. The colour map again demonstrates the addition and constriction of states in phase space as \(D\) increases, however the number of nodes in the network increases at a far slower rate than in the chaotic case (Figure \ref{fig:NVsD}). Large peaks in \(\sigma\) and \(\langle k_{out} \rangle \) are not present as they were for chaotic regimes. Note that the embedding dimension must be large enough to ensure that two or more independent segments of trajectory do not both pass through a single state, otherwise the network will include a false transitional edge --- an edge connecting two or more states that are not temporally adjacent. A false transitional edge is clearly observable in a network generated from a periodic time series because the ring structure will fold onto itself as shown in Figures \ref{fig:NET0001-6} and \ref{fig:NET0001-8}. 

\begin{figure}[]
	\centering
		\centering
	\begin{subfigure} {0.32\textwidth}
		\centering
		\includegraphics [width={1.1\textwidth}] {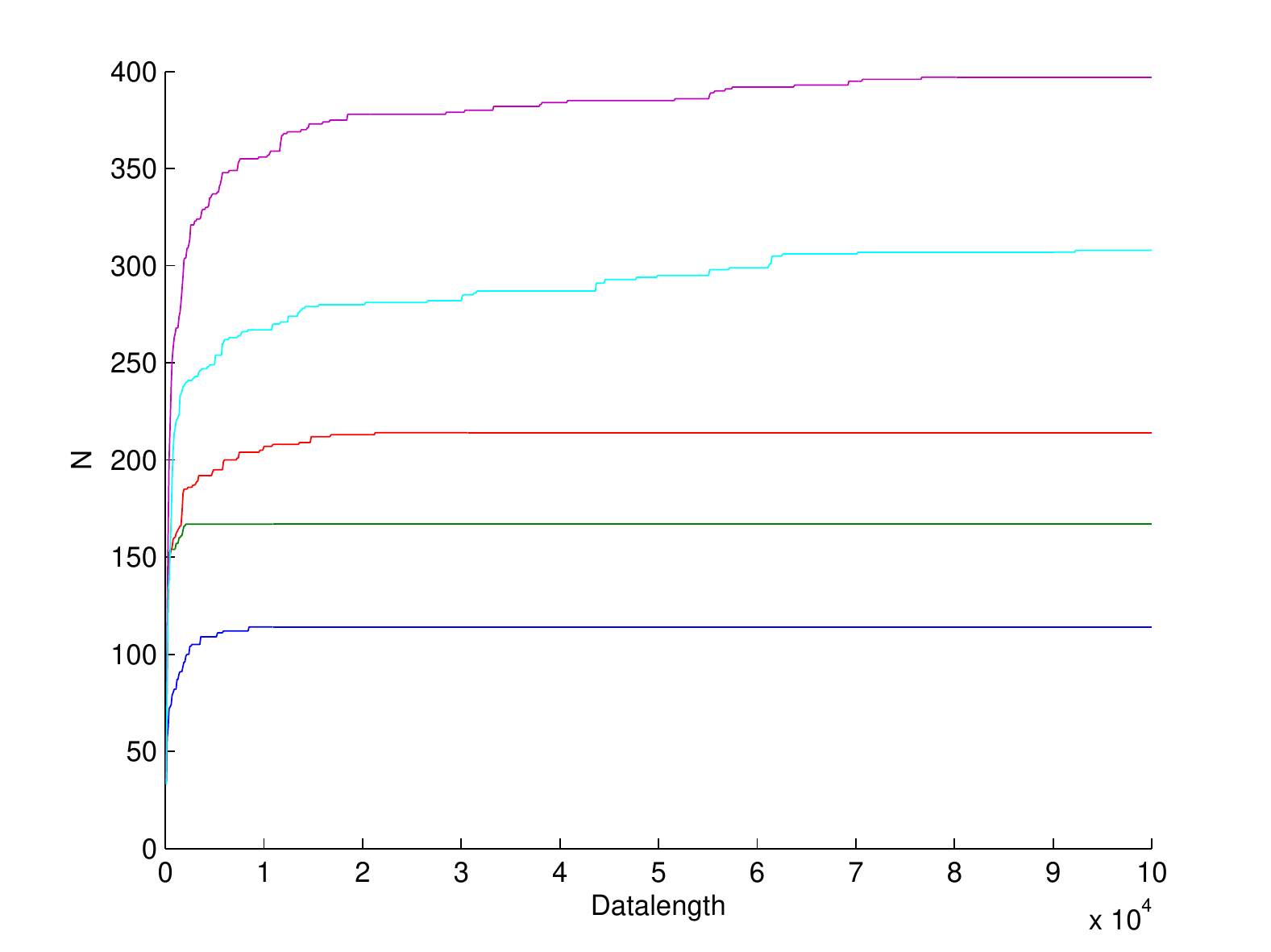}
		\subcaption{}
		\label{fig:NVsn}
	\end{subfigure}
		\begin{subfigure} {0.32\textwidth}
		\centering
		\includegraphics [width={1.1\textwidth}] {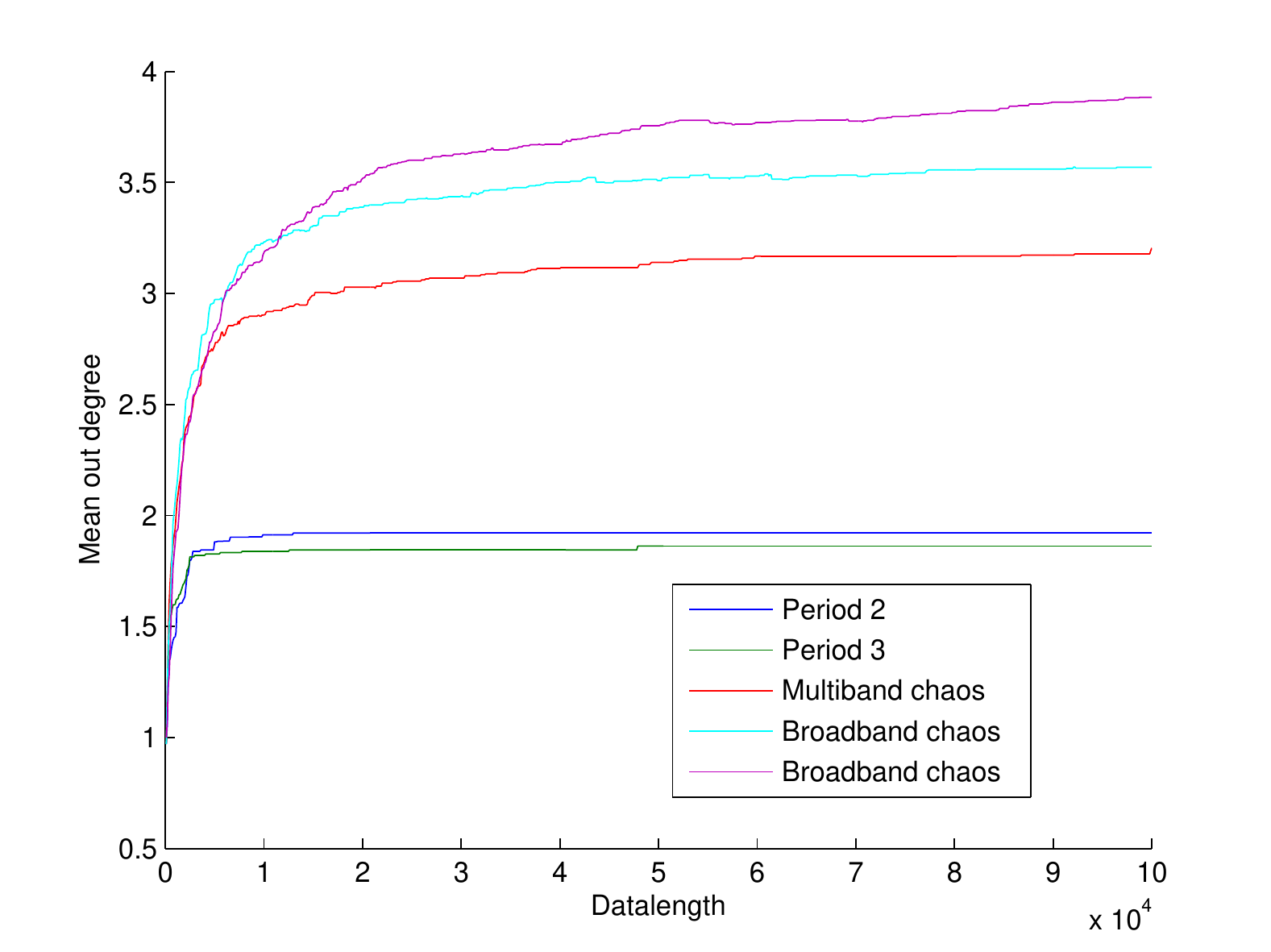}
		\subcaption{}
		\label{fig:MeanDegVsn}
	\end{subfigure}
	\begin{subfigure} {0.32\textwidth}
		\centering
		\includegraphics [width={1.1\textwidth}] {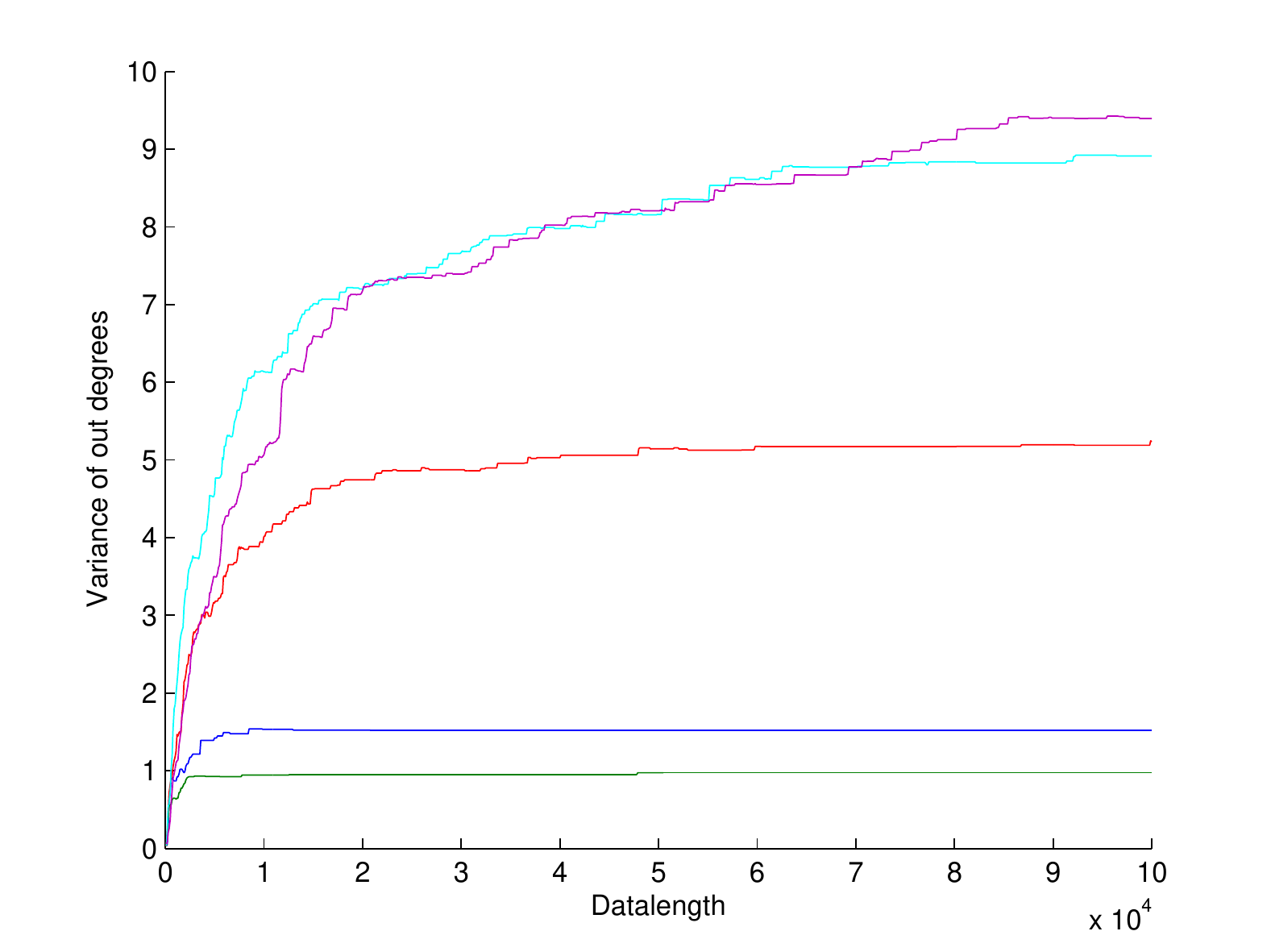}
		\subcaption{}
		\label{fig:DegVarVsn}
	\end{subfigure}
	\caption{Network measures plotted against the length of the dataset used to generate an ordinal partition network for the R\"ossler system in different dynamical regimes: period 2, \(\alpha=0.37\) (blue line); period 3, \(\alpha=0.41\) (green line); multiband chaos, \(\alpha=0.39\) (red line); broadband chaos, \(\alpha=0.4\) (cyan line); and broadband chaos, \(\alpha=0.42\) (magenta line). Network measures shown are \subref{fig:NVsn} number of nodes, \subref{fig:MeanDegVsn} mean out degree, and \subref{fig:DegVarVsn} variance of out degrees.}
	\label{fig:RosDatalength}
\end{figure}

To test the dependence of our method on time series length we generated five additional time series from different dynamical regimes each with \(10^5\) points with transients removed, and mapped these to networks using varying amounts of the data. Ideally there should be enough points in the time series such that \(N\ll n\) for an appropriate value of \(D\). This is to say that the dataset should contain a sufficiently complete representation of the dynamics. Figure \ref{fig:RosDatalength} shows the number of nodes \(N\), the variance of out degrees \(\sigma\) and the mean out degree \(\langle k_{out}\rangle\) plotted against the length of the time series used. In the periodic case these properties converge for relatively short time series (less than \(10^4\) data points). As expected, networks generated from chatoic time series take far longer to converge because more time series data is required to capture a sufficient portion of the attractor. However, \(\sigma\) and \(\langle k_{out}\rangle\) still provide good discrimination between the different dynamical regimes for \(n=10^4\).

\begin{figure}[]
	\centering
		\centering
	\begin{subfigure} {0.32\textwidth}
		\centering
		\includegraphics [width={1.1\textwidth}] {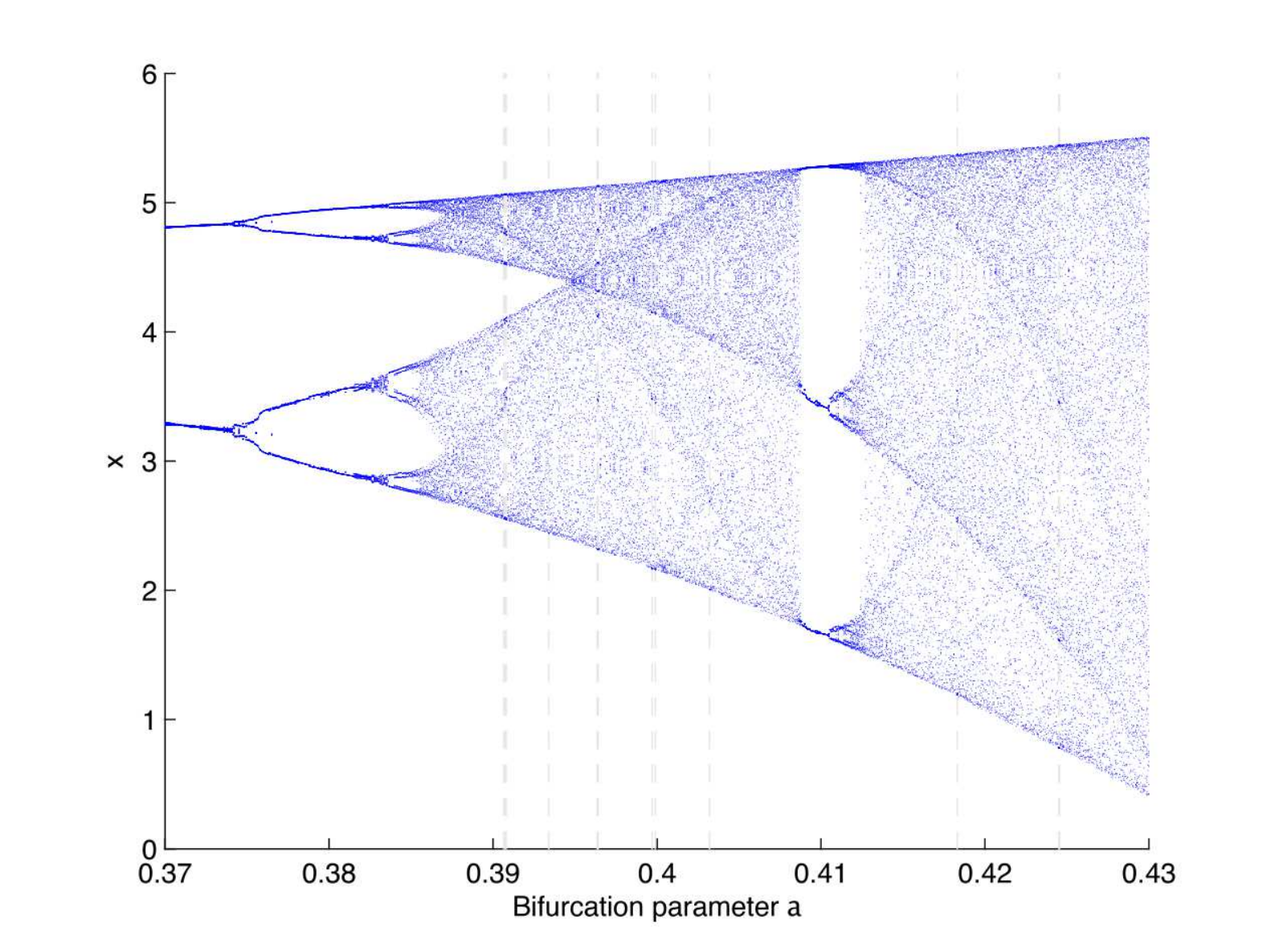}
		\subcaption{}
		\label{fig:RosBif}
	\end{subfigure}
	\begin{subfigure} {0.32\textwidth}
		\centering
		\includegraphics [width={1.1\textwidth}] {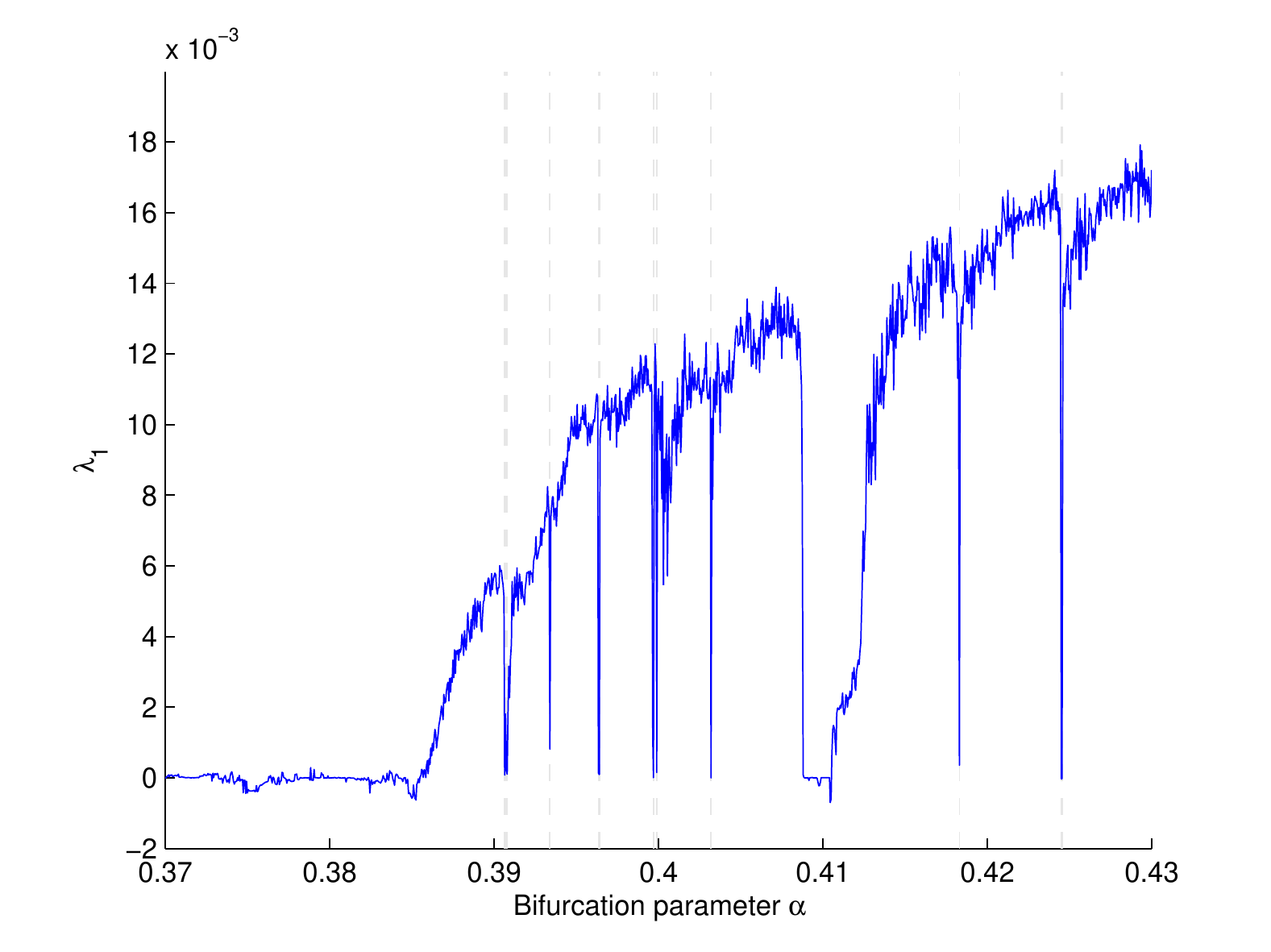}
		\subcaption{}
		\label{fig:LyapVsBif}
	\end{subfigure}
	\begin{subfigure} {0.32\textwidth}
		\centering
		\includegraphics [width={1.1\textwidth}] {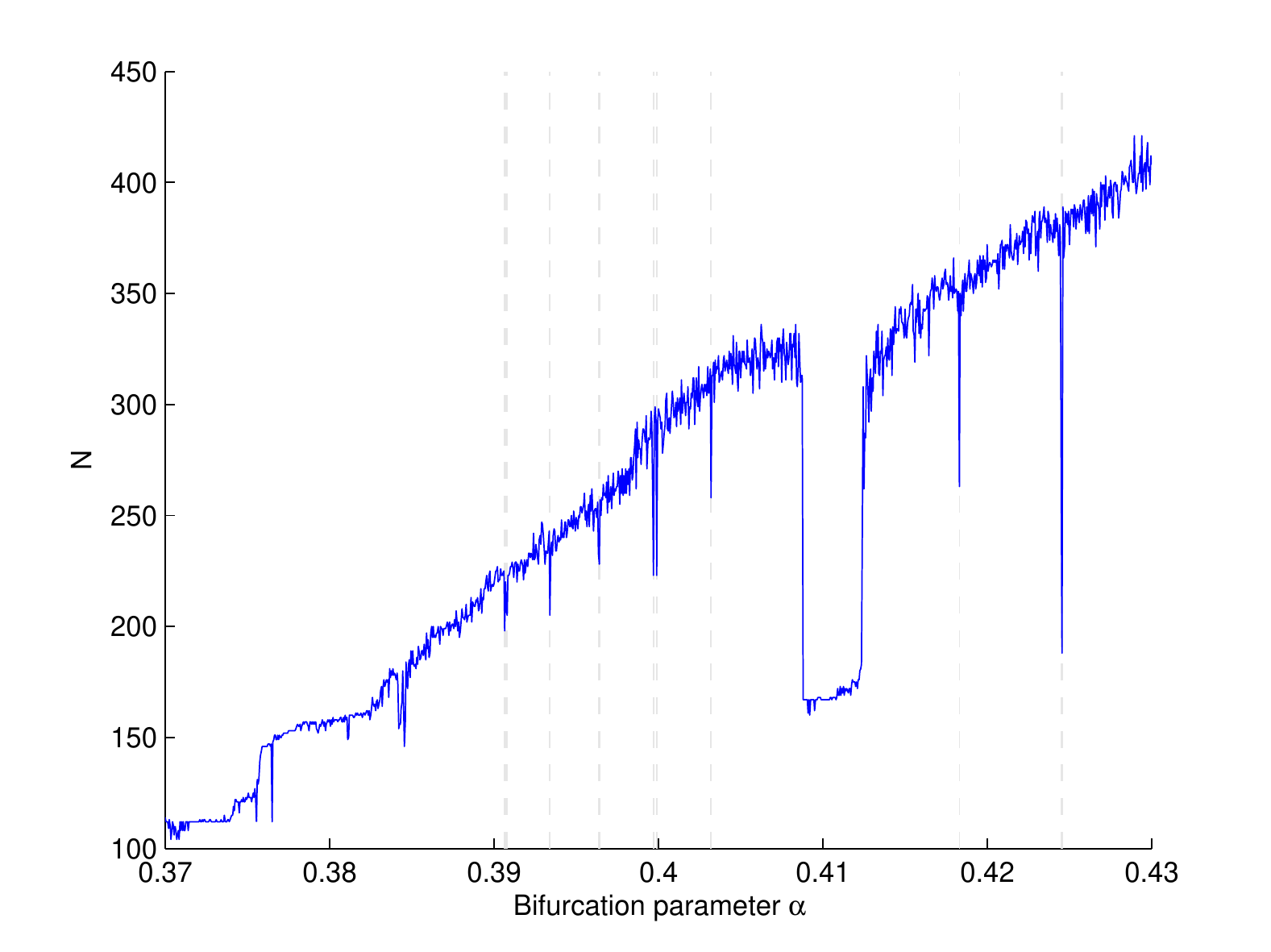}
		\subcaption{}
		\label{fig:NVsBif}
	\end{subfigure}
	\begin{subfigure} {0.32\textwidth}
		\centering
		\includegraphics [width={1.1\textwidth}] {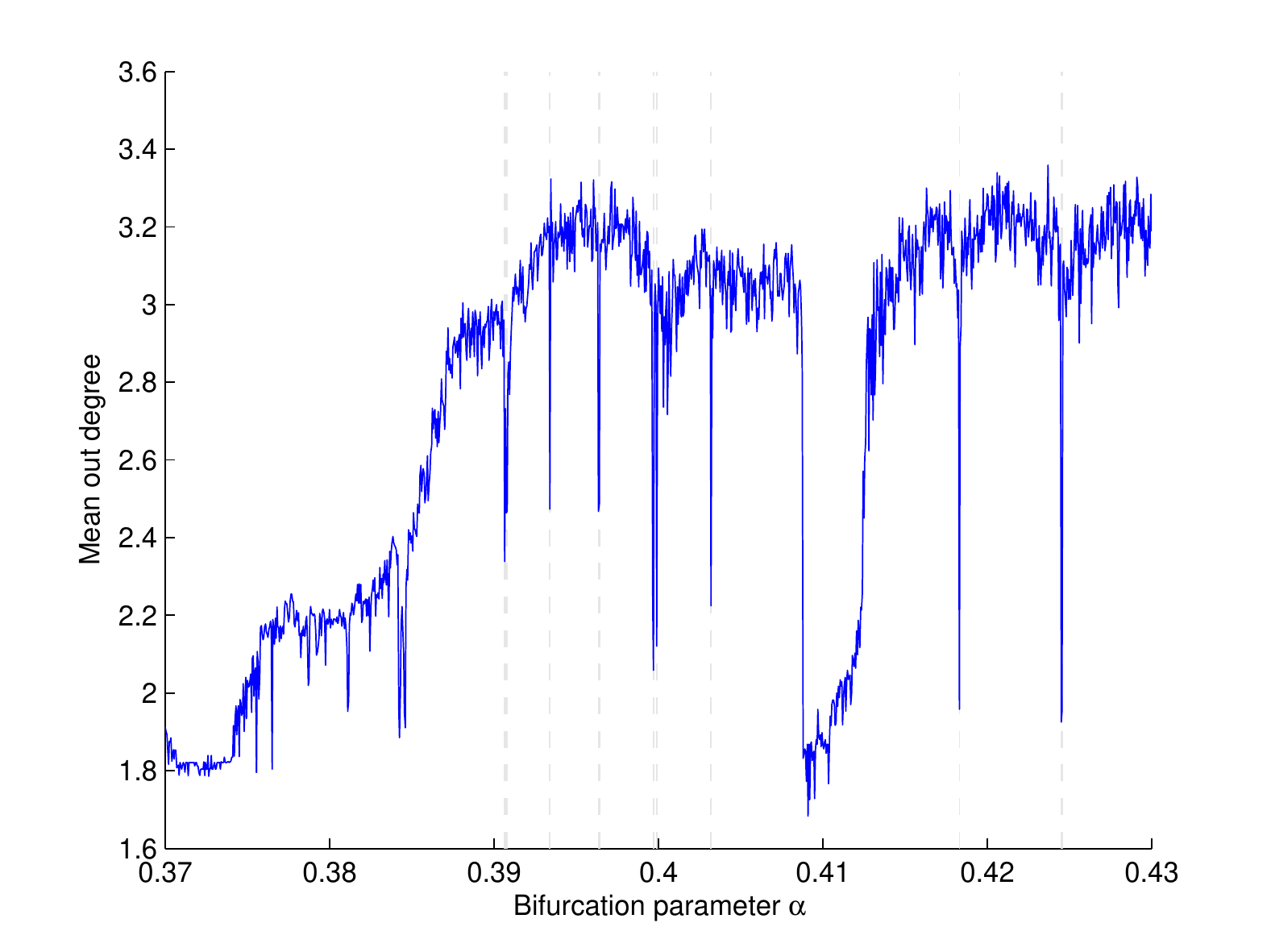}
		\subcaption{}
		\label{fig:MeanDegVsBif}
	\end{subfigure}
	\begin{subfigure} {0.32\textwidth}
		\centering
		\includegraphics [width={1.1\textwidth}] {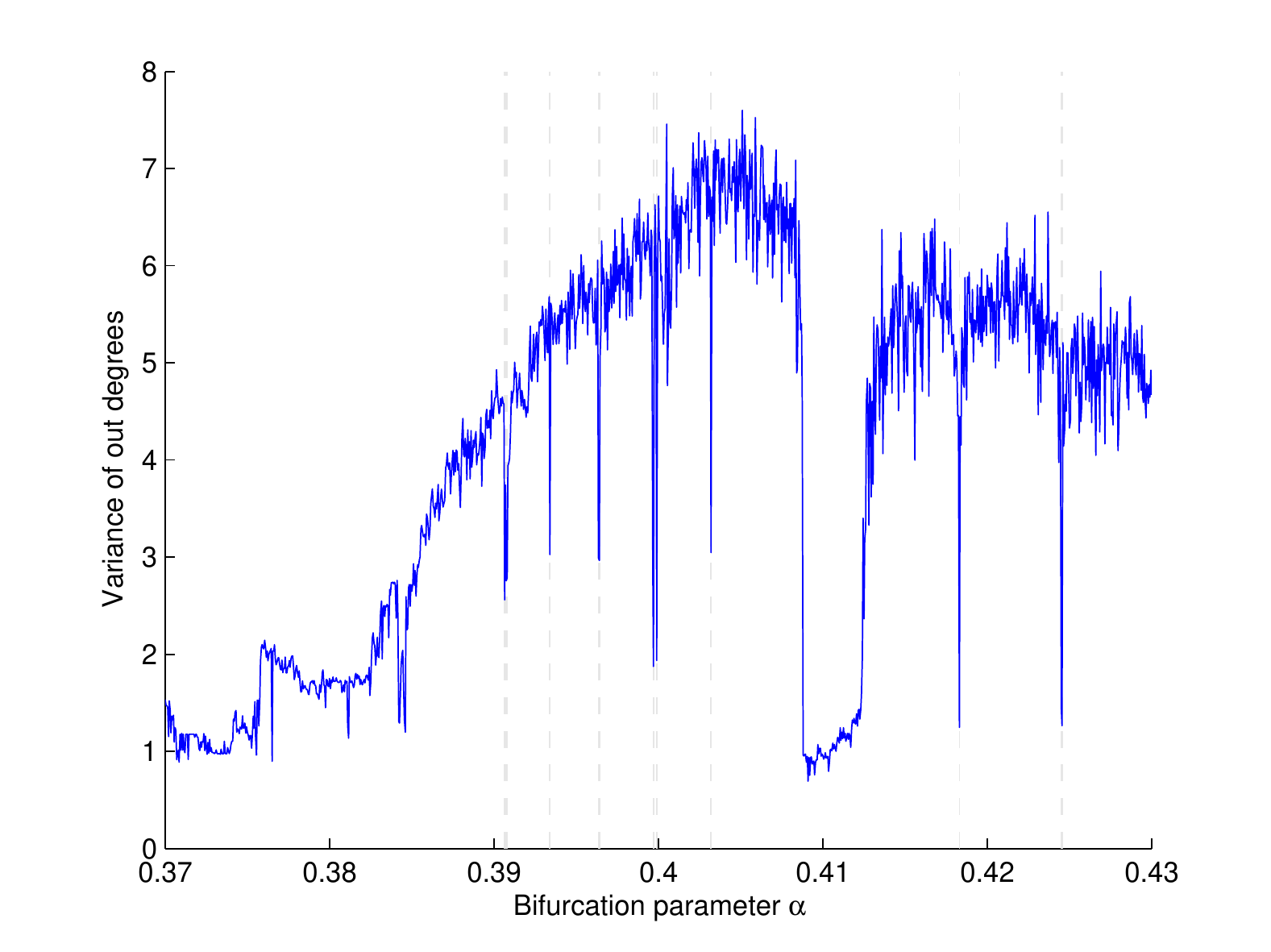}
		\subcaption{}
		\label{fig:DegVarVsBif}
	\end{subfigure}
	\begin{subfigure} {0.32\textwidth}
		\centering
		\includegraphics [width={1.1\textwidth}] {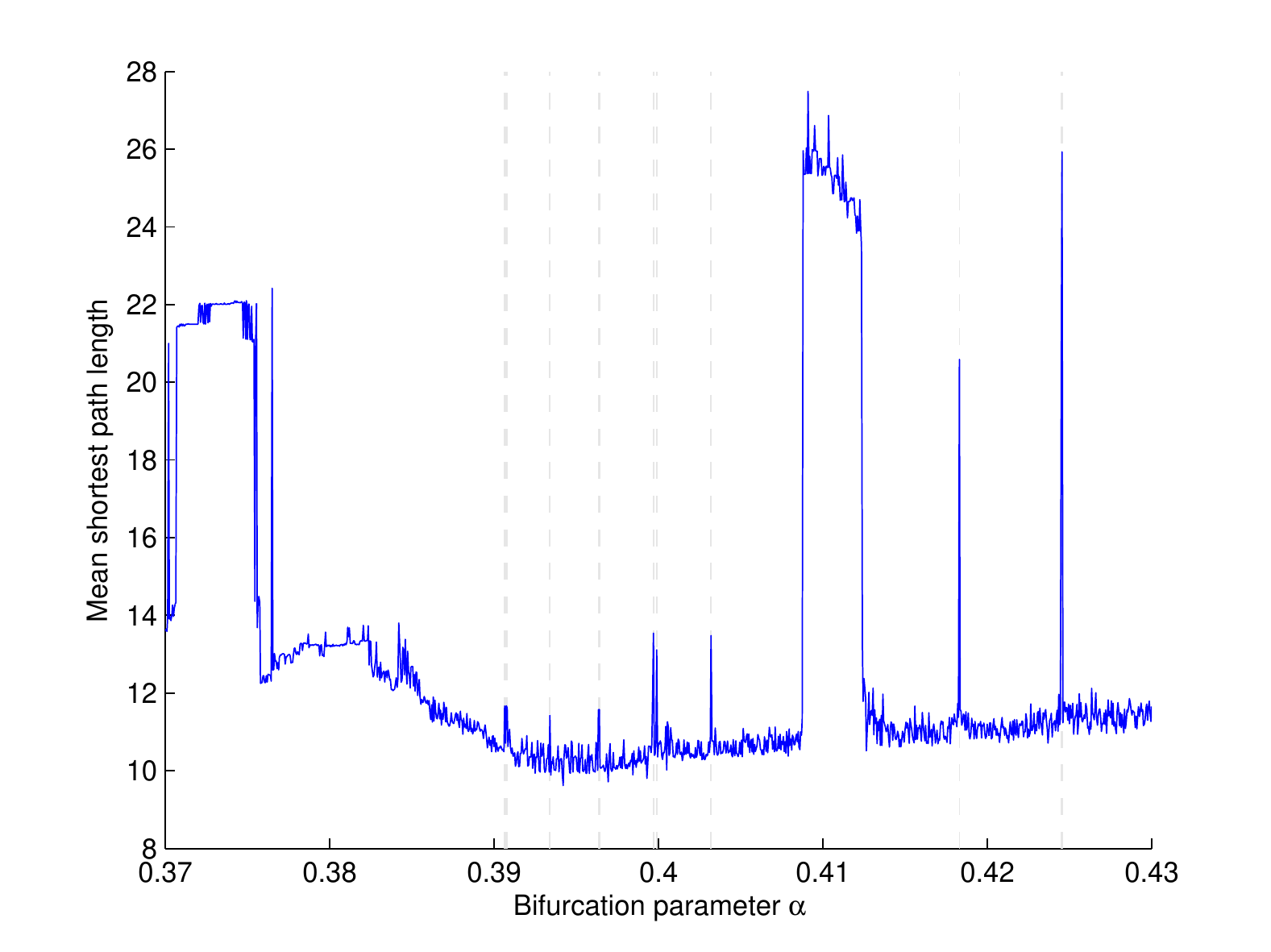}
		\subcaption{}
		\label{fig:MeanSPLVsBif}
	\end{subfigure}
	\begin{subfigure} {0.32\textwidth}
		\centering
		\includegraphics [width={1.1\textwidth}] {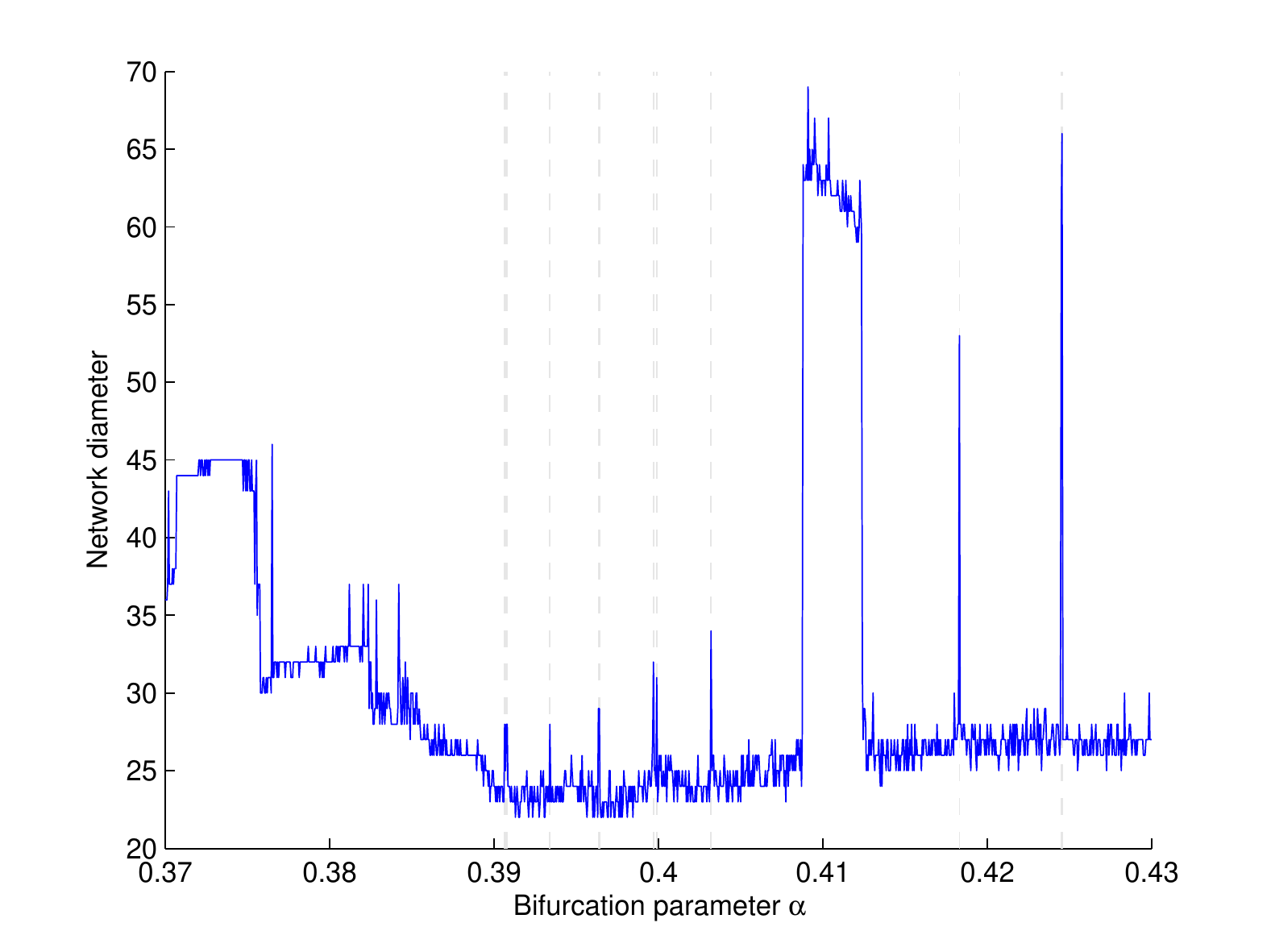}
		\subcaption{}
		\label{fig:DiamVsBif}
	\end{subfigure}
	\caption{\subref{fig:RosBif} The bifurcation diagram for the R\"ossler system and \subref{fig:LyapVsBif} the largest Lyapunov exponent for the range \(0.37\leq\alpha\leq0.43\). The vertical dotted lines correspond to periodic windows. Networks were generated for each time series with \(\tau=9\) and \(D=8\). Plotting \subref{fig:NVsBif} the number of nodes, \subref{fig:MeanDegVsBif} the mean out degree, \subref{fig:DegVarVsBif} variance of out degrees, \subref{fig:MeanSPLVsBif} mean shortest path length, \subref{fig:DiamVsBif} and network diameter over the same range demonstrates that the dynamical information is embedded in the network structure.}
	\label{fig:RosslerMeas}
\end{figure}

We have plotted the bifurcation diagram(Figure \ref{fig:RosBif}) and the largest Lyapunov exponent (Figure \ref{fig:LyapVsBif}) of the R\"ossler system from the generated time series using the \textit{extrema} and \textit{lyapk} functions from the TISEAN software package~\cite{hegger_practical_1999} respectively. The system undergoes a period doubling route to chaos then passes through several periodic windows including the large period-3 window at \(\alpha \approx 0.41\). Other periodic windows are denoted by the dotted grey lines. Networks have been generated for each time series with \(D=8\). This value was selected based on Figures \ref{fig:MeanDegVsD} and \ref{fig:DegVarVsD} from which it appears that there should be good discrimination between different dynamical regimes based on \(\langle k_{out} \rangle \) and \(\sigma\) for this choice of embedding dimension, although it may not be the optimal choice for individual time series in the dataset (i.e. not all of the periodic ring networks have unfolded at \(D=8\)). Figures \ref{fig:MeanDegVsBif} and \ref{fig:DegVarVsBif} show that \(\langle k_{out} \rangle\) and \(\sigma\) both exhibit sensitivity to system dynamics, as they appear to track the relative change in the largest Lyapunov exponent and detect all of the periodic windows. Similar but less effective tracking is apparent from \(N\), the number of nodes in the network, however this is only a measure of how many different permutation symbols are present in the times series and therefore is related to the volume of phase space occupied by the embedded attractor rather than characteristics of temporal succession. The mean shortest path length and network diameter (maximum shortest path length) both display sensitivity to dynamical changes. For example, a pronounced step change is evident in both of these measures at the point of the first period doubling bifurcation and there are peaks at each of the periodic windows. In summary, this set of results demonstrates that while \(\langle k_{out} \rangle \), \(\sigma\), mean shortest path length and diameter all share the deficiency that they do not provide an absolute criteria for discriminating between periodic and chaotic dynamics, they have the potential to be useful as an indicator for dynamical discrimination in a relative sense, and for detecting change points.

\begin{figure}[]
	\centering
		\centering
	\begin{subfigure} {0.24\textwidth}
		\centering
		\includegraphics [width={0.9\textwidth}] {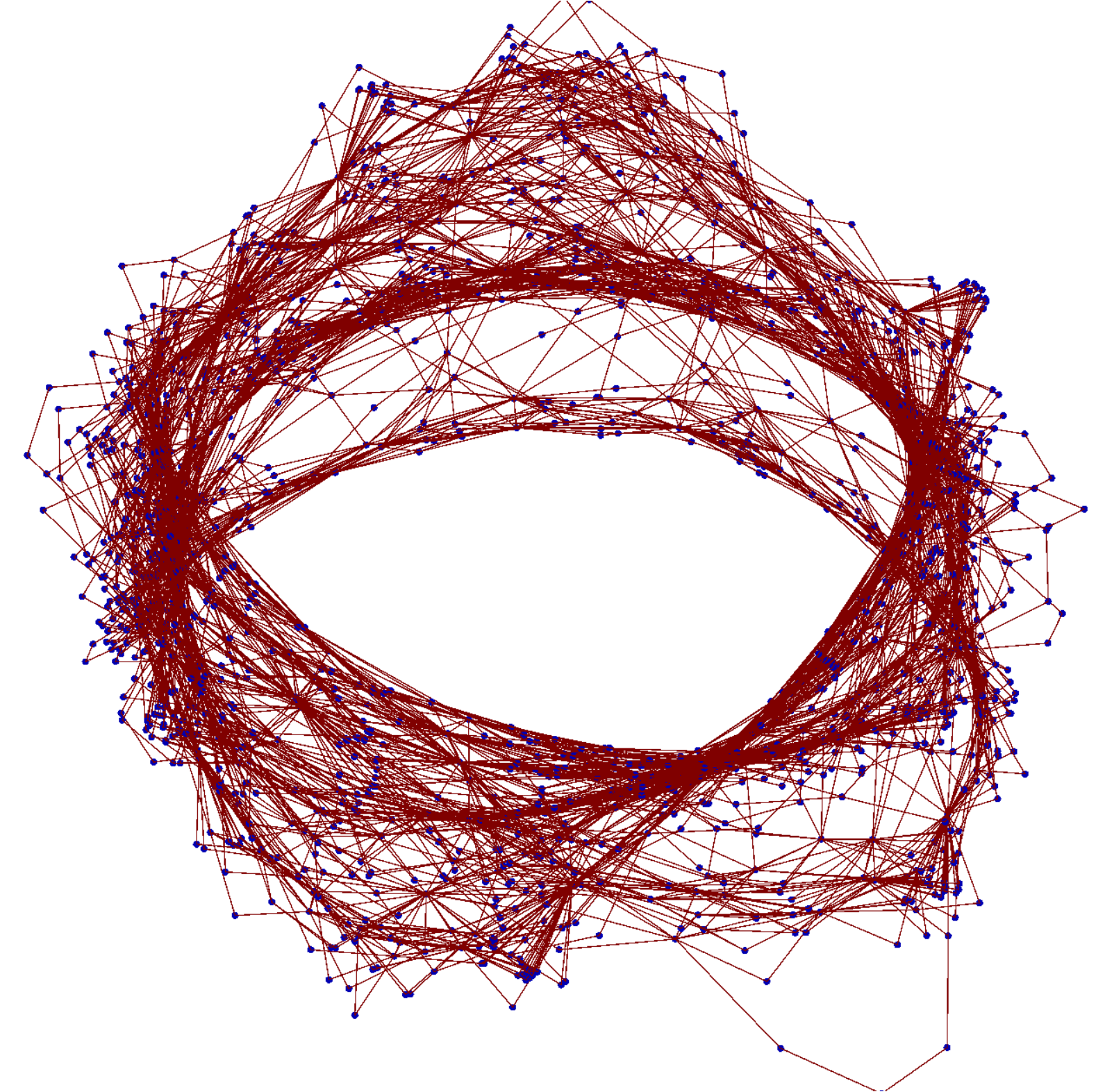}
		\subcaption{}
		\label{fig:0601-10-05}
	\end{subfigure}
	\begin{subfigure} {0.24\textwidth}
		\centering
		\includegraphics [width={0.9\textwidth}] {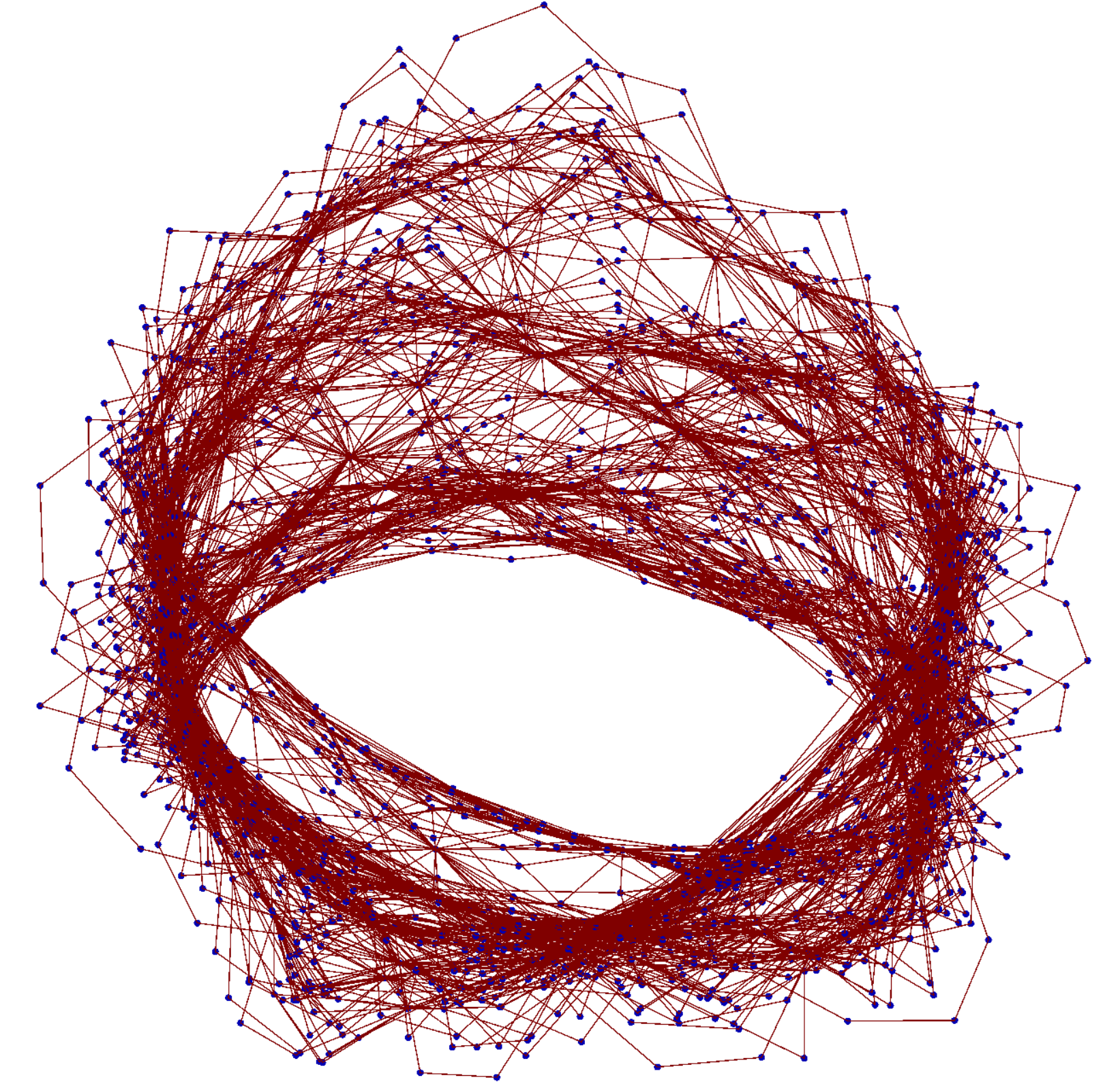}
		\subcaption{}
		\label{fig:0601-10-1}
	\end{subfigure}
	\begin{subfigure} {0.24\textwidth}
		\centering
		\includegraphics [width={0.9\textwidth}] {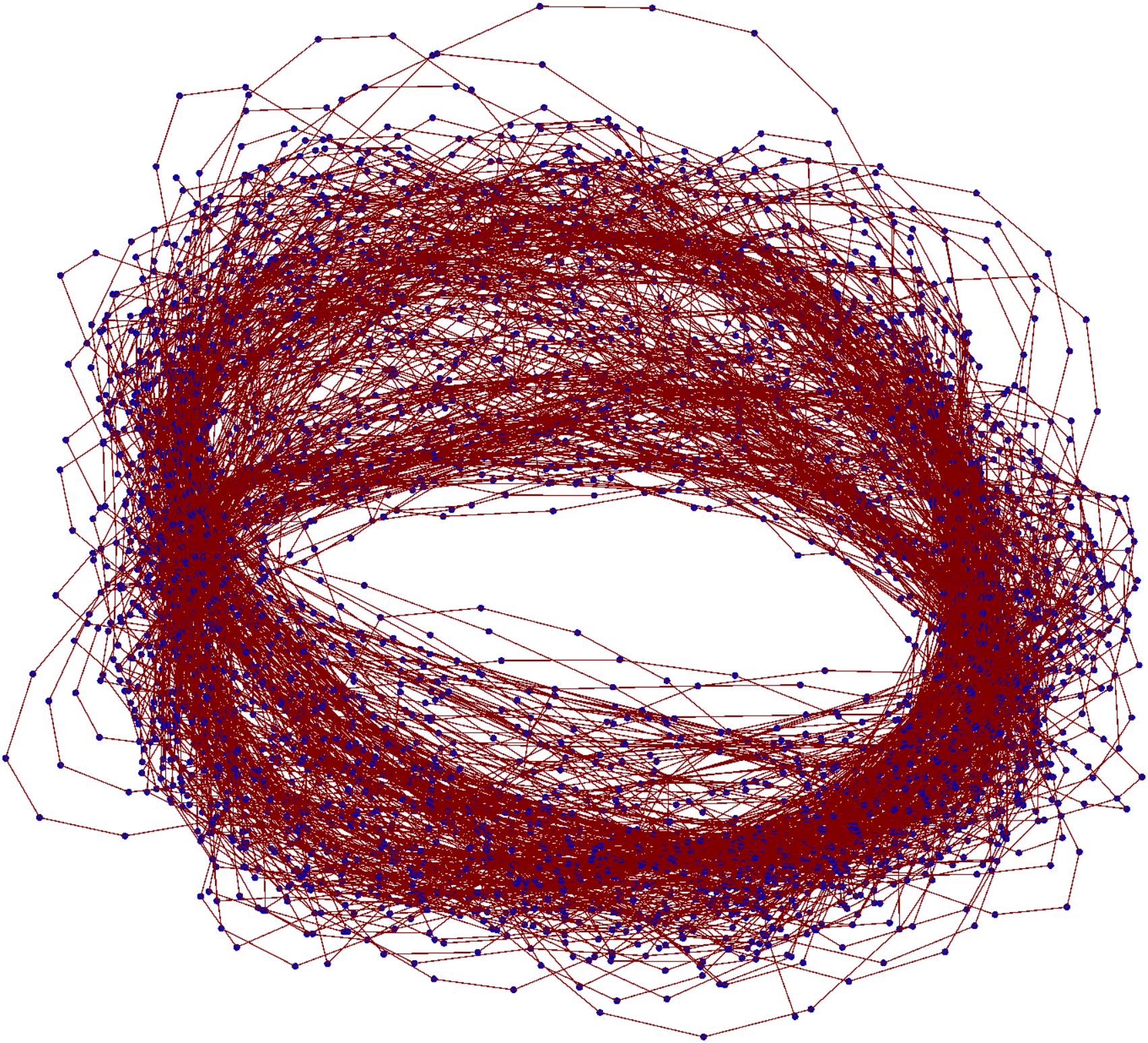}
		\subcaption{}
		\label{fig:0601-10-5}
	\end{subfigure}
	\begin{subfigure} {0.24\textwidth}
		\centering
		\includegraphics [width={1.3\textwidth}] {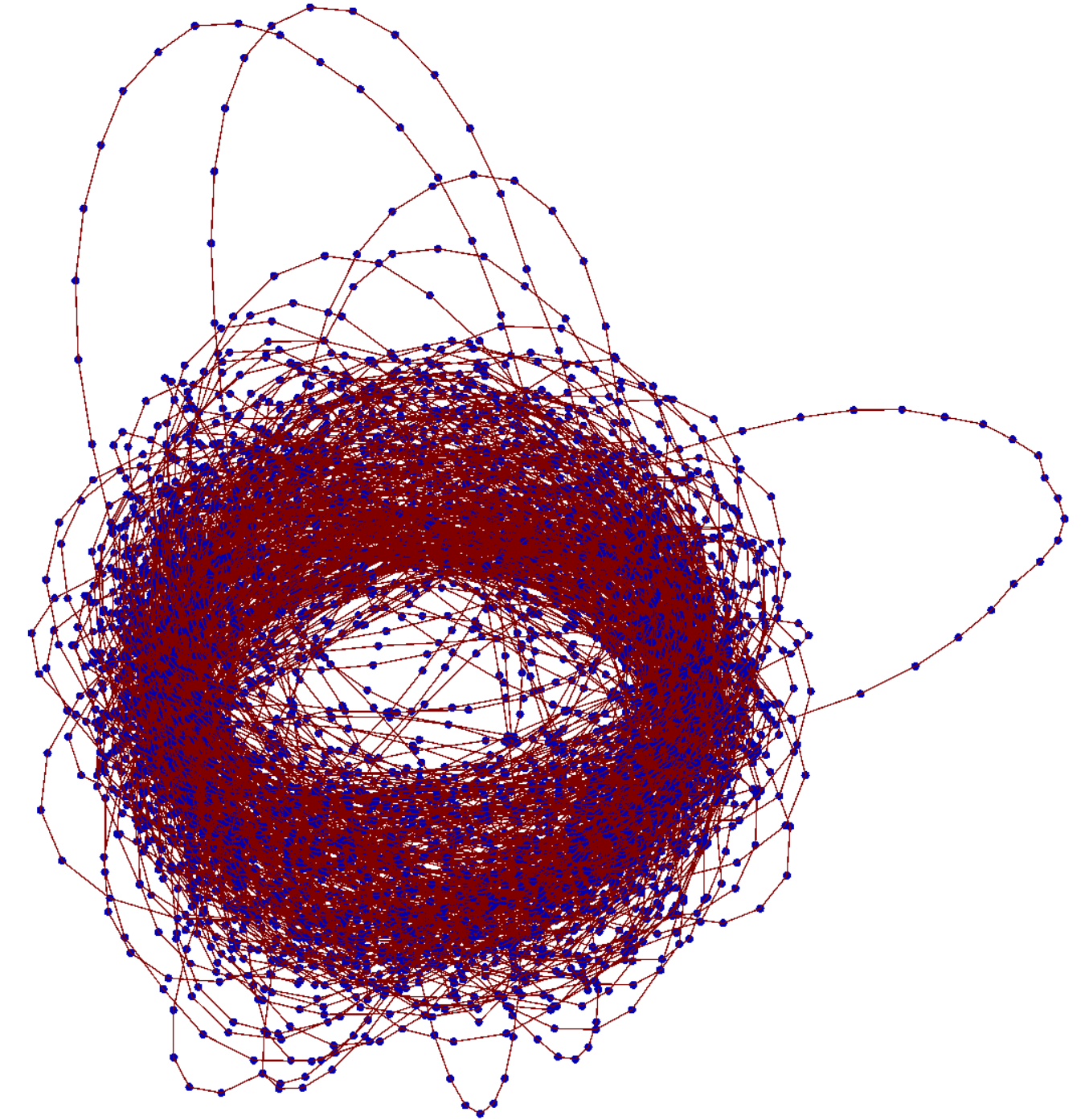}
		\subcaption{}
		\label{fig:0601-10-10}
	\end{subfigure}
	\caption{Networks generated from a chaotic R\"ossler time series (\(\alpha=0.4\)) with \(\tau=9\) and \(D=10\) for increasing levels of additive white gaussian noise: \subref{fig:0601-10-05} 0.5\%, \subref{fig:0601-10-1} 1\%, \subref{fig:0601-10-5} 5\%, and \subref{fig:0601-10-10} 10\%. Noise variance is expressed as a percentage of the variance of the original signal.}
	\label{fig:Noise0601}
\end{figure}

\begin{figure}[]
	\centering
		\centering
	\begin{subfigure} {0.24\textwidth}
		\centering
		\includegraphics [width={1\textwidth}] {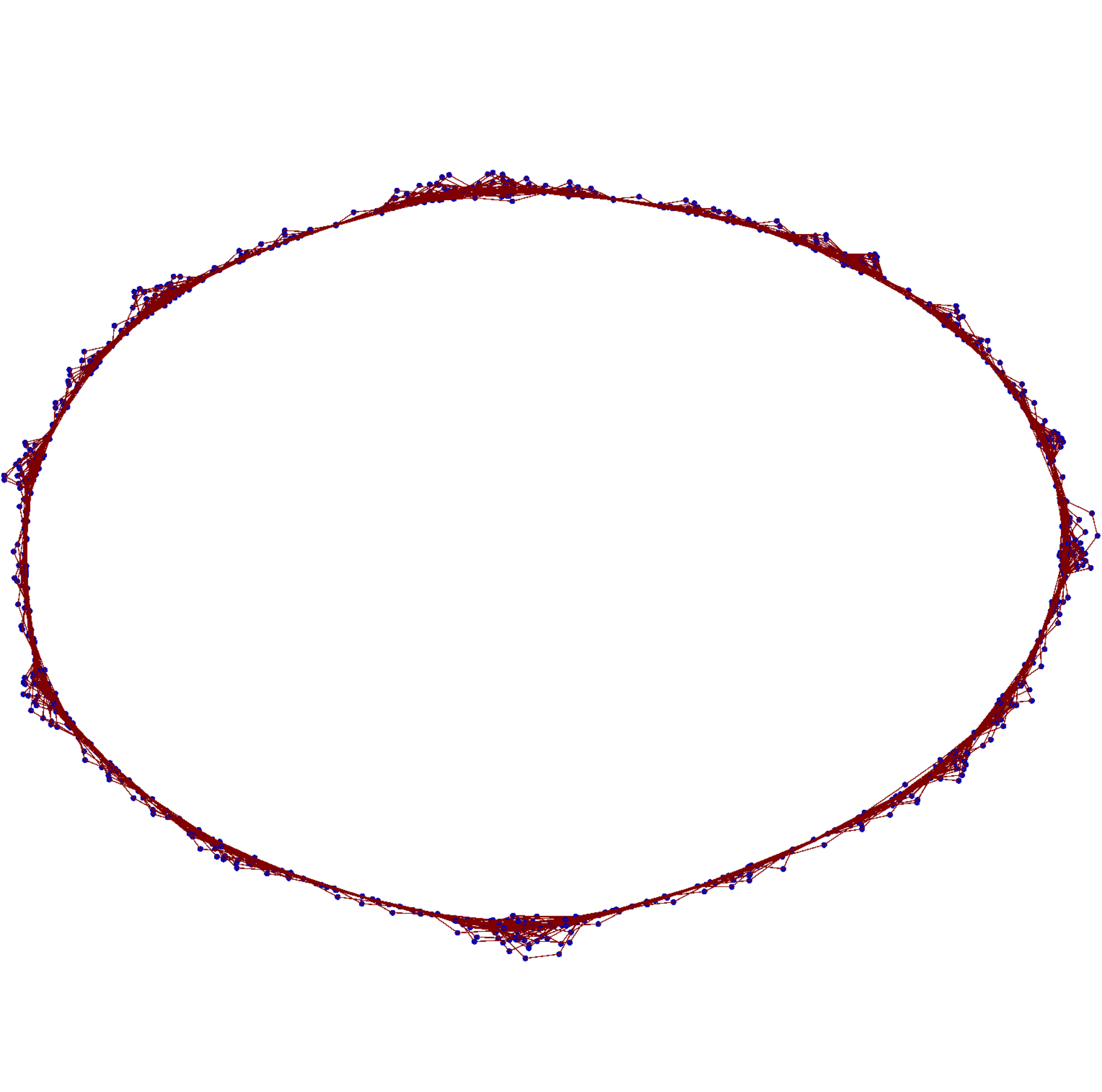}
		\subcaption{}
		\label{fig:0801-10-05}
	\end{subfigure}
	\begin{subfigure} {0.24\textwidth}
		\centering
		\includegraphics [width={1\textwidth}] {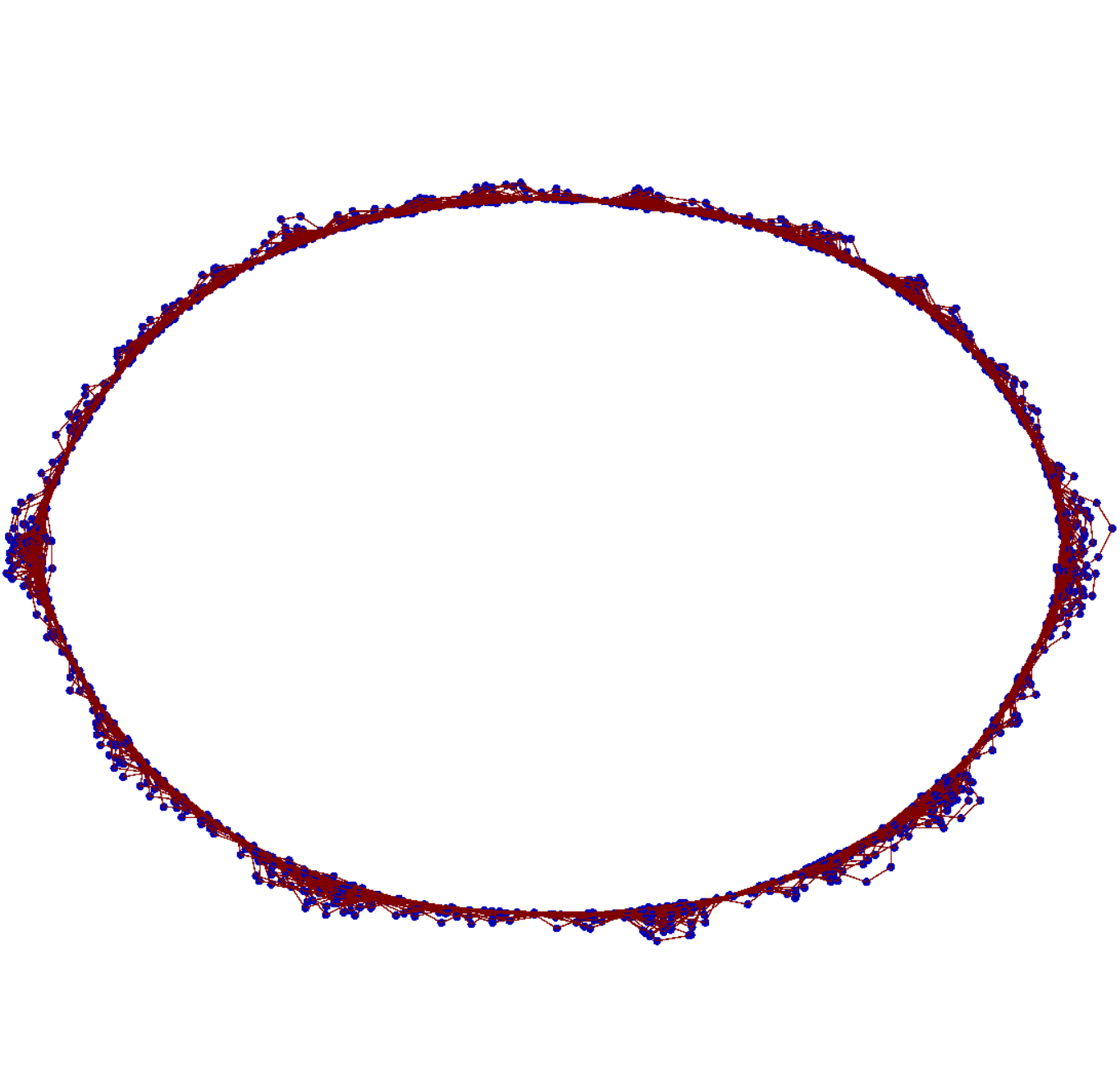}
		\subcaption{}
		\label{fig:0801-10-1}
	\end{subfigure}
	\begin{subfigure} {0.24\textwidth}
		\centering
		\includegraphics [width={1\textwidth}] {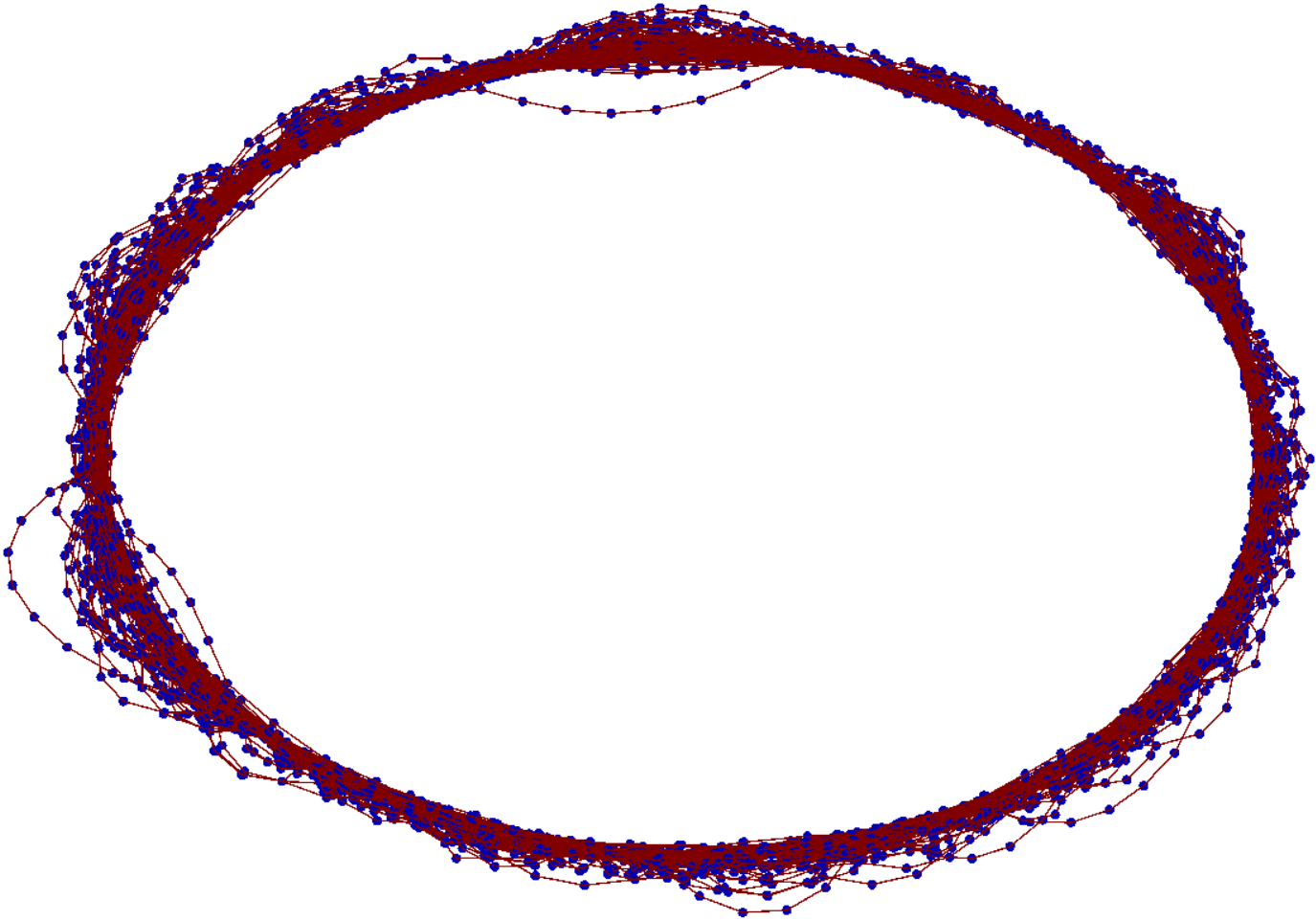}
		\subcaption{}
		\label{fig:0801-10-5}
	\end{subfigure}
	\begin{subfigure} {0.24\textwidth}
		\centering
		\includegraphics [width={1\textwidth}] {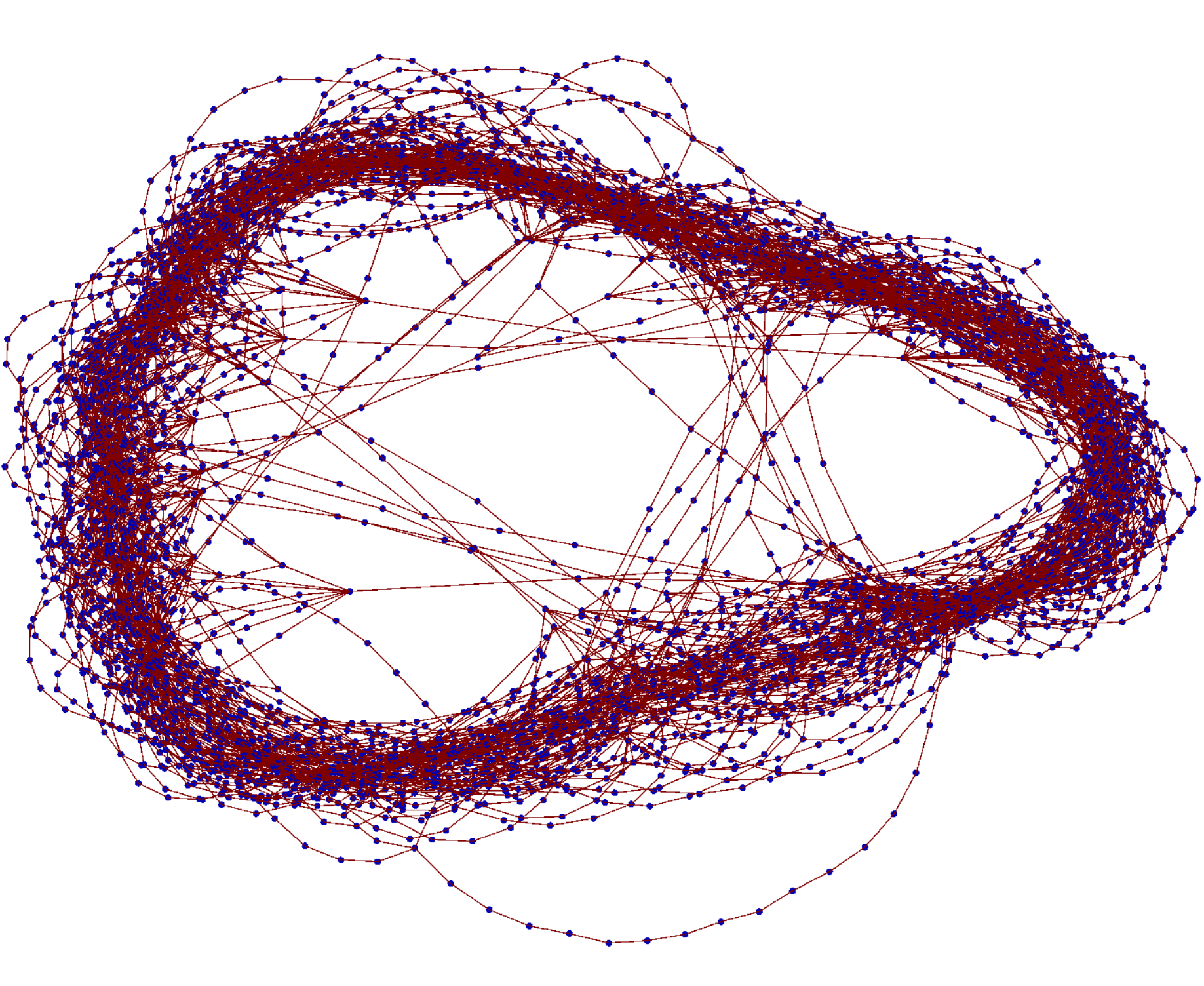}
		\subcaption{}
		\label{fig:0801-10-10}
	\end{subfigure}
		\caption{Networks generated from a period-3 R\"ossler time series (\(\alpha=0.41\)) with \(\tau=9\) and \(D=10\) for increasing levels of additive white gaussian noise: \subref{fig:0801-10-05} 0.5\%, \subref{fig:0801-10-1} 1\%, \subref{fig:0801-10-5} 5\%, and \subref{fig:0801-10-10} 10\%. Noise variance is expressed as a percentage of the variance of the original signal.}
	\label{fig:Noise0801}
\end{figure}

When the time series is corrupted with additive white Gaussian noise, \(\langle k_{out} \rangle\) and \(\sigma\) become far less effective in discriminating different dynamical behaviour. These simple network properties are global averages of local node properties which represent transitional possibilities for a single time step, hence these measures will be sensitive to noise. However, from a qualitative perspective, Figures \ref{fig:Noise0601} and \ref{fig:Noise0801} show that the distinct ring and band structures which we observed in the noiseless case for periodic and chaotic dynamics respectively are fairly well maintained even in the presence of significant additional noise. We postulate that network measures which operate on an intermediate scale such as motif distributions~\cite{xu_superfamily_2008} or subgraph expansion rates~\cite{padberg_local_2009} will be far more effective at quantifying these observable differences in network structure.

\section{Application to experimental data}
\label{sec:Diode}
\begin{figure}
	\centering
	\includegraphics [width=10cm] {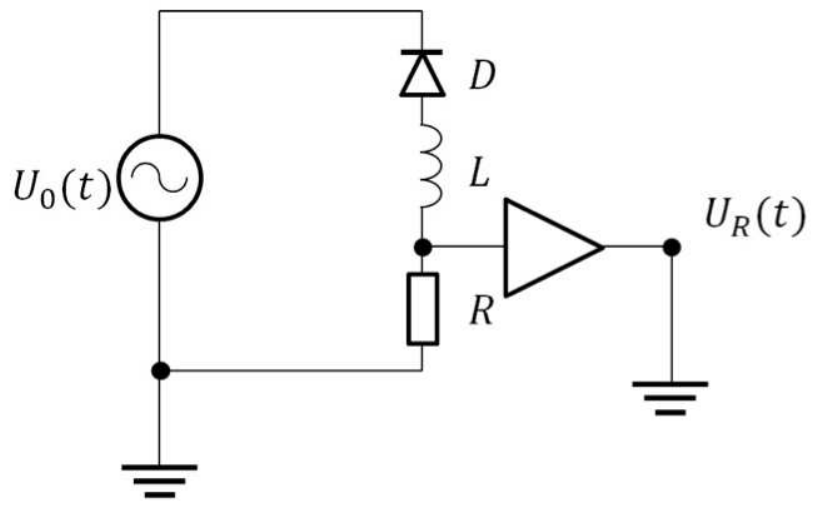}
	\caption{The diode resonator circuit.}
	\label{fig:Circuit}
\end{figure}

\begin{figure}[]
	\centering
		\centering
	\begin{subfigure} {0.24\textwidth}
		\centering
		\includegraphics [width={1.15\textwidth}] {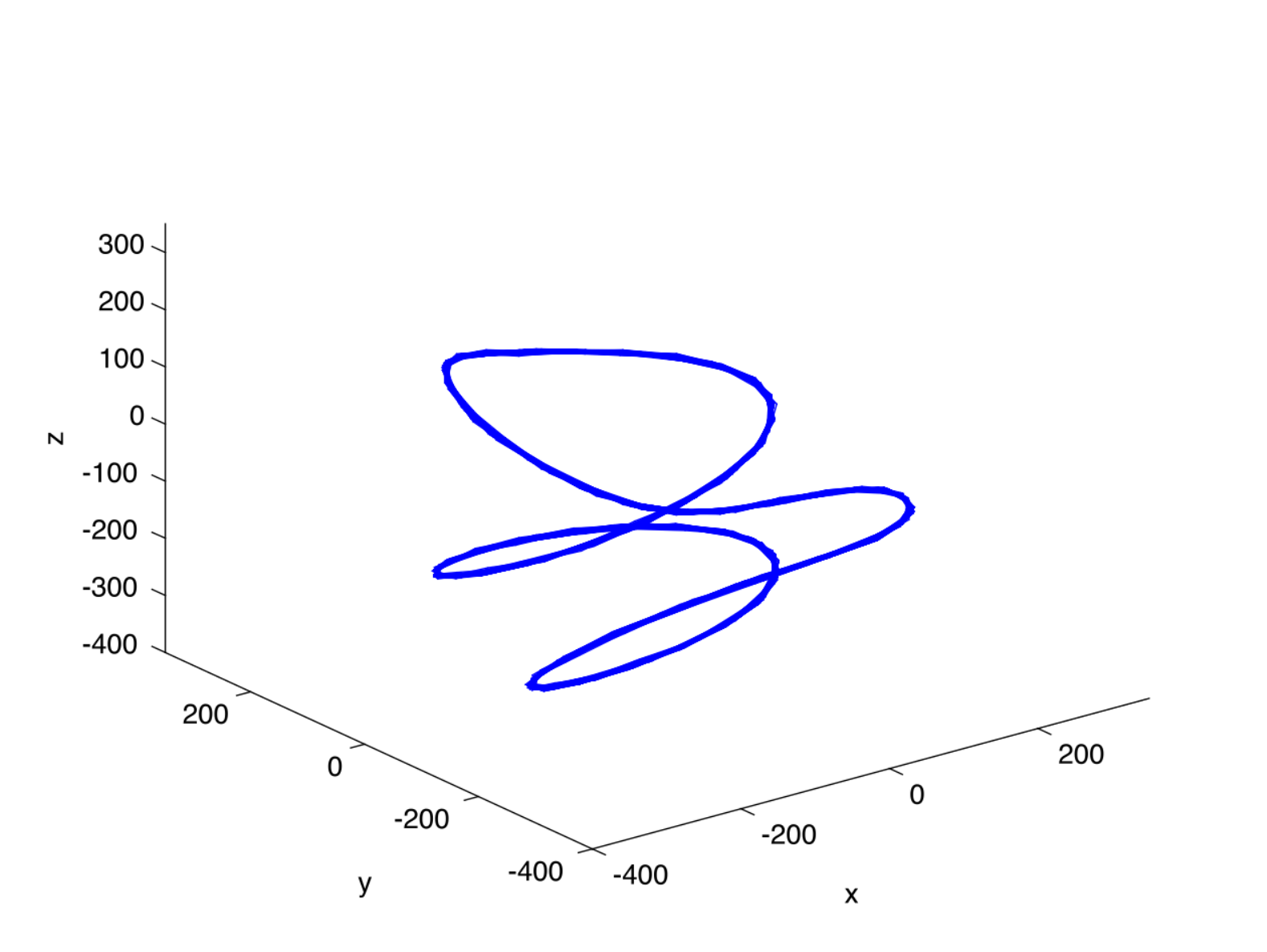}
		\subcaption{}
		\label{fig:Diode0000}
	\end{subfigure}
	\begin{subfigure} {0.24\textwidth}
		\centering
		\includegraphics [width={1.15\textwidth}] {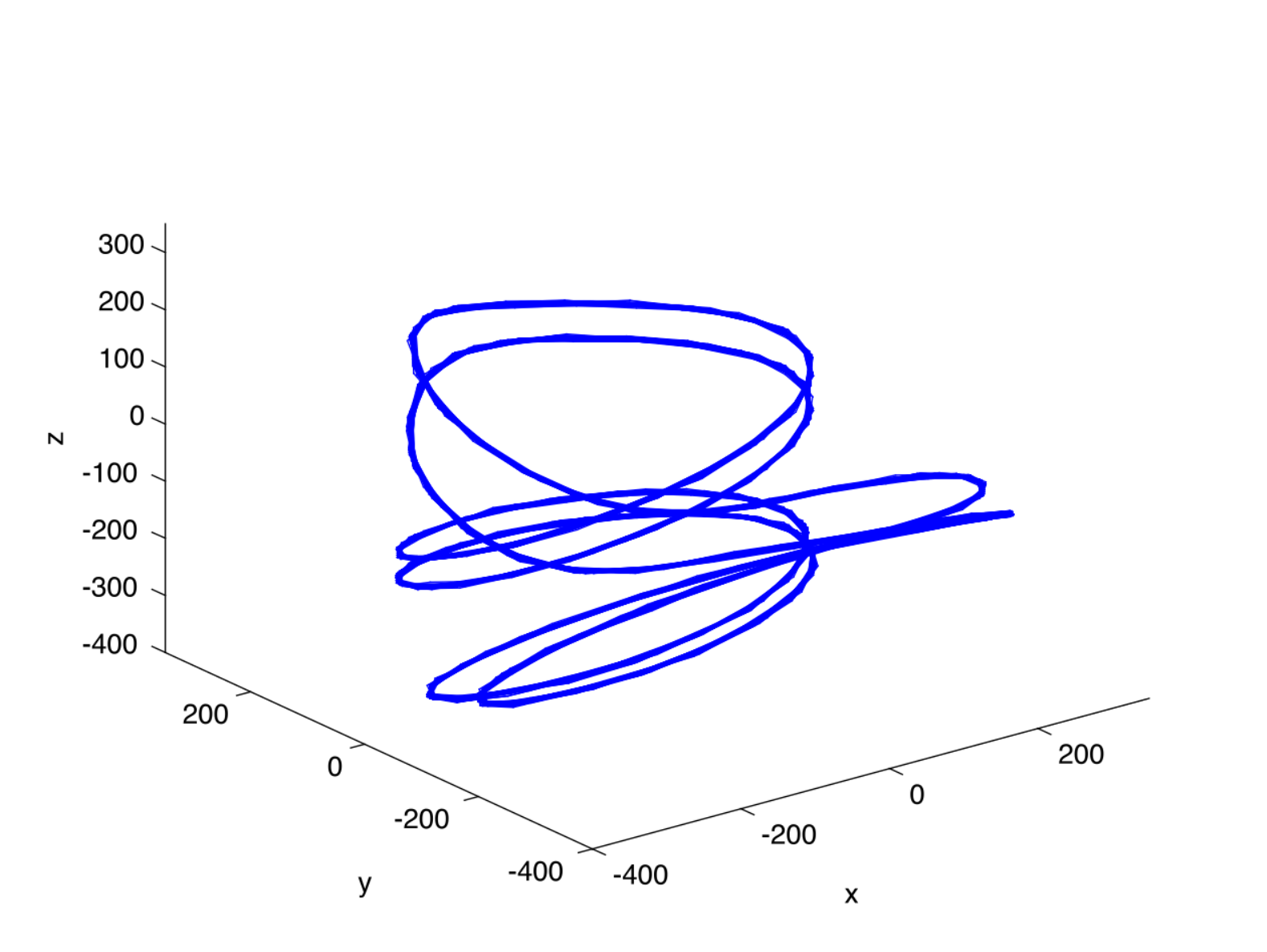}
		\subcaption{}
		\label{fig:Diode0350}
	\end{subfigure}
	\begin{subfigure} {0.24\textwidth}
		\centering
		\includegraphics [width={1.15\textwidth}] {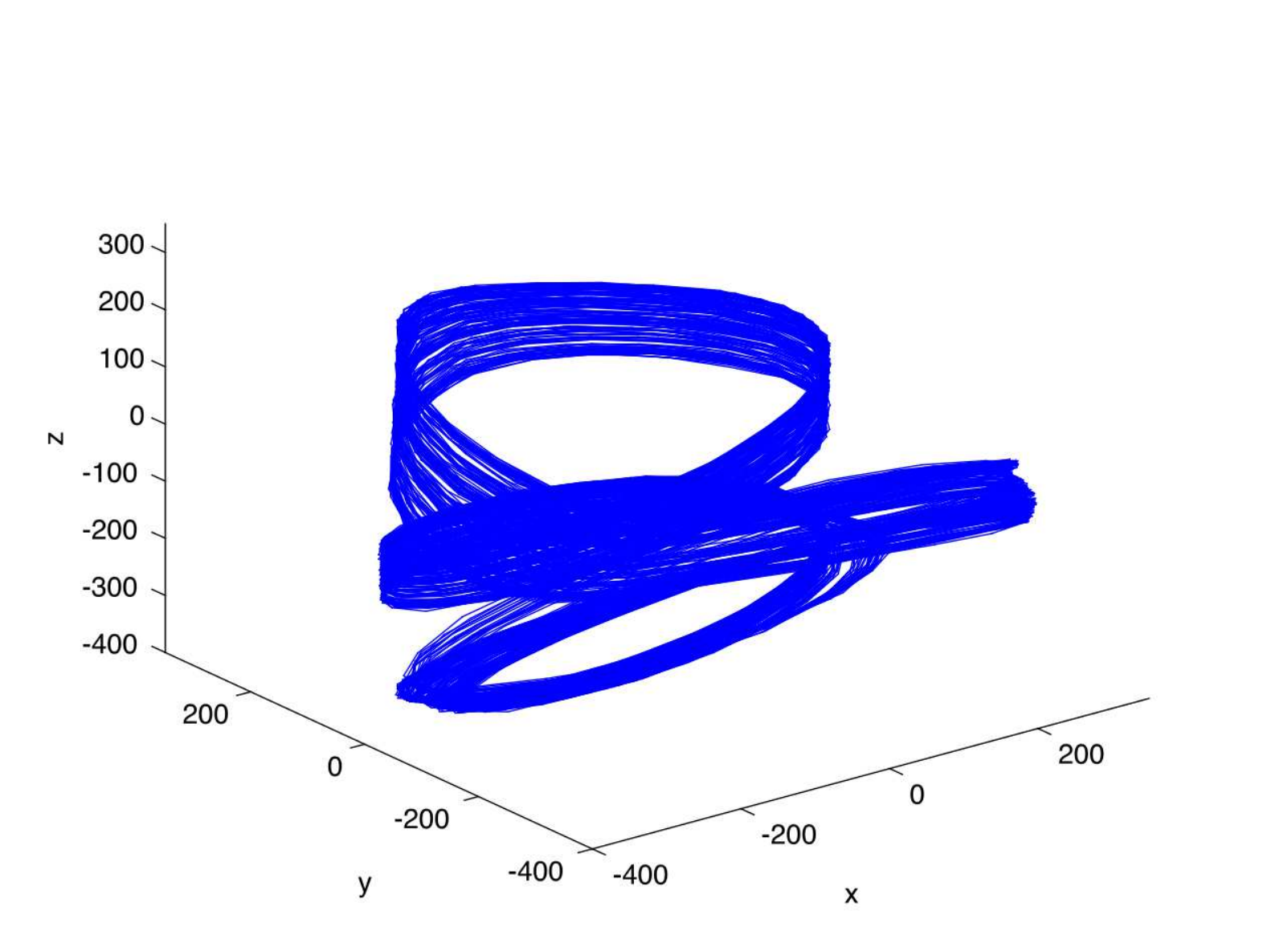}
		\subcaption{}
		\label{fig:Diode0500}
	\end{subfigure}
	\begin{subfigure} {0.24\textwidth}
		\centering
		\includegraphics [width={1.15\textwidth}] {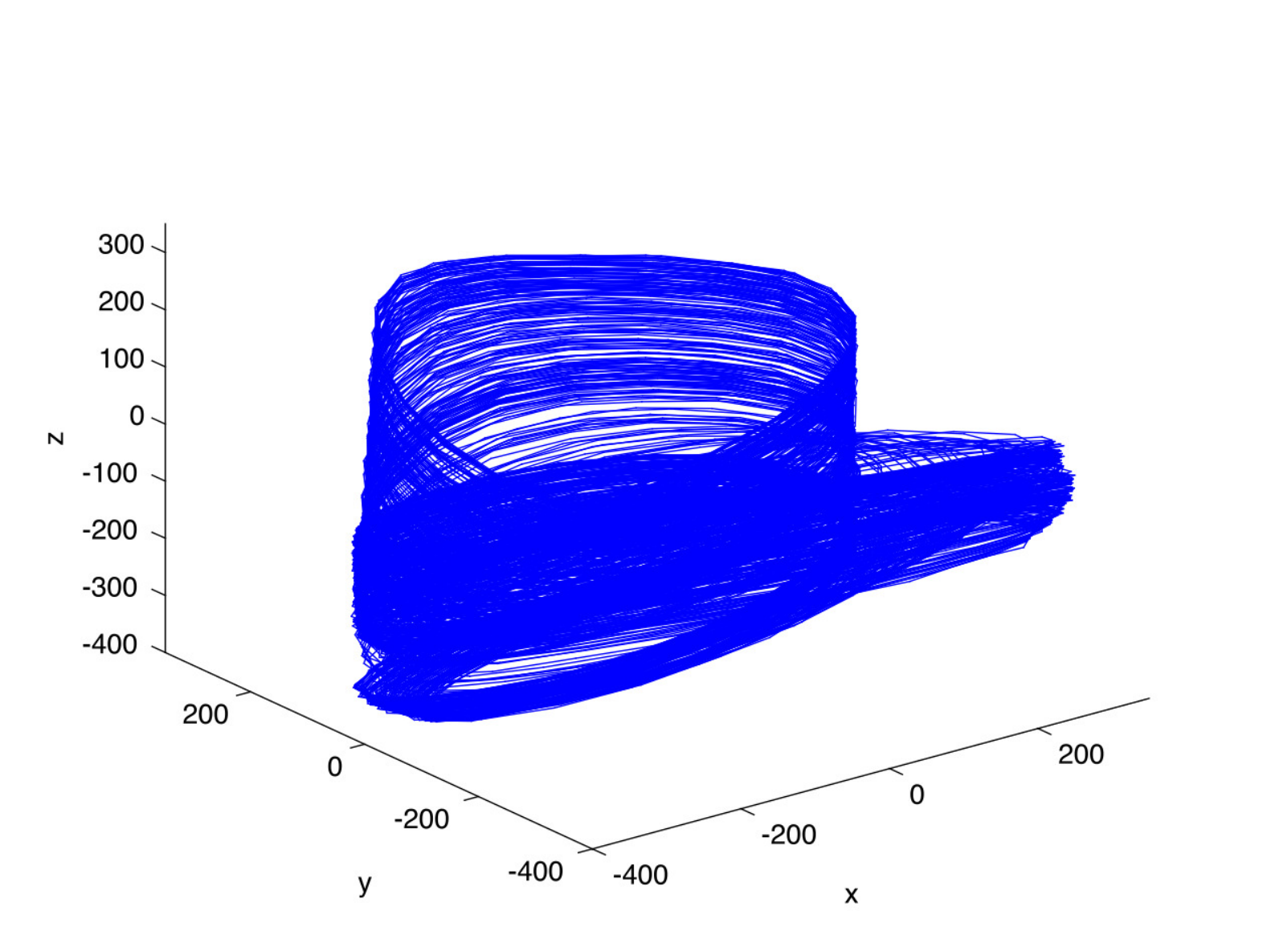}
		\subcaption{}
		\label{fig:Diode0750}
	\end{subfigure}
		\caption{Attractors from time delay embeddings of time series generated by the experimental diode resonator circuit for different values of the driving voltage amplitude \(U_0\). Each time series was embedded with delay \(\tau=8\). This figure shows the system in \subref{fig:Diode0000} period-3 oscillation, \(U_0=3V\); \subref{fig:Diode0350} period-6 oscillation, \(U_0=3.7V\); \subref{fig:Diode0500} multiband chaos, \(U_0=4V\); and \subref{fig:Diode0750} chaos-chaos intermittency after the interior crisis, \(U_0=4.5V\).}
	\label{fig:DiodeAtr}
\end{figure}

\begin{figure}[]
	\centering
		\centering
	\begin{subfigure} {0.24\textwidth}
		\centering
		\includegraphics [width={1\textwidth}] {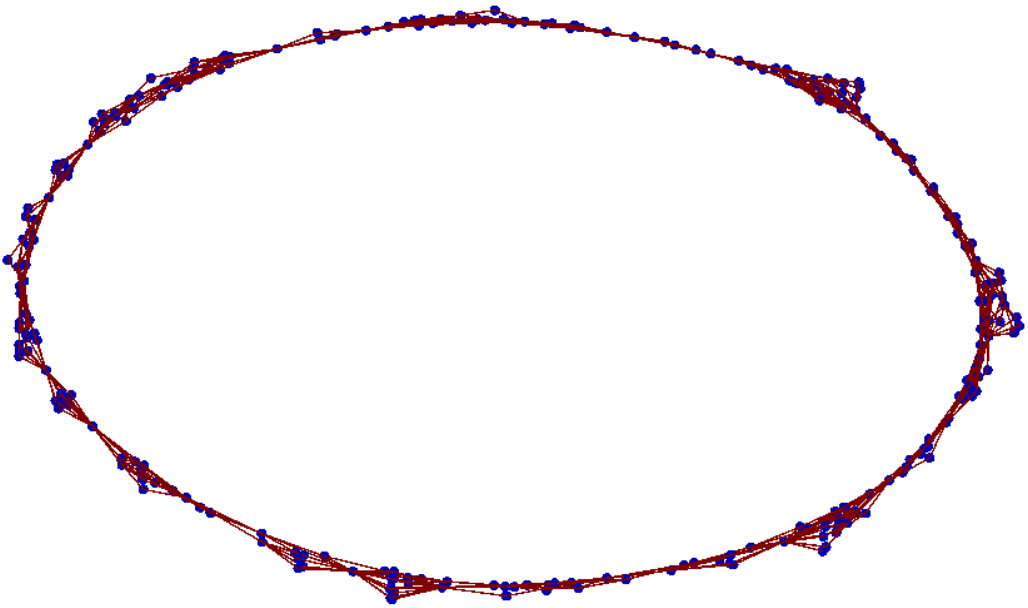}
		\subcaption{}
		\label{fig:DiodeNET0000}
	\end{subfigure}
	\begin{subfigure} {0.24\textwidth}
		\centering
		\includegraphics [width={1\textwidth}] {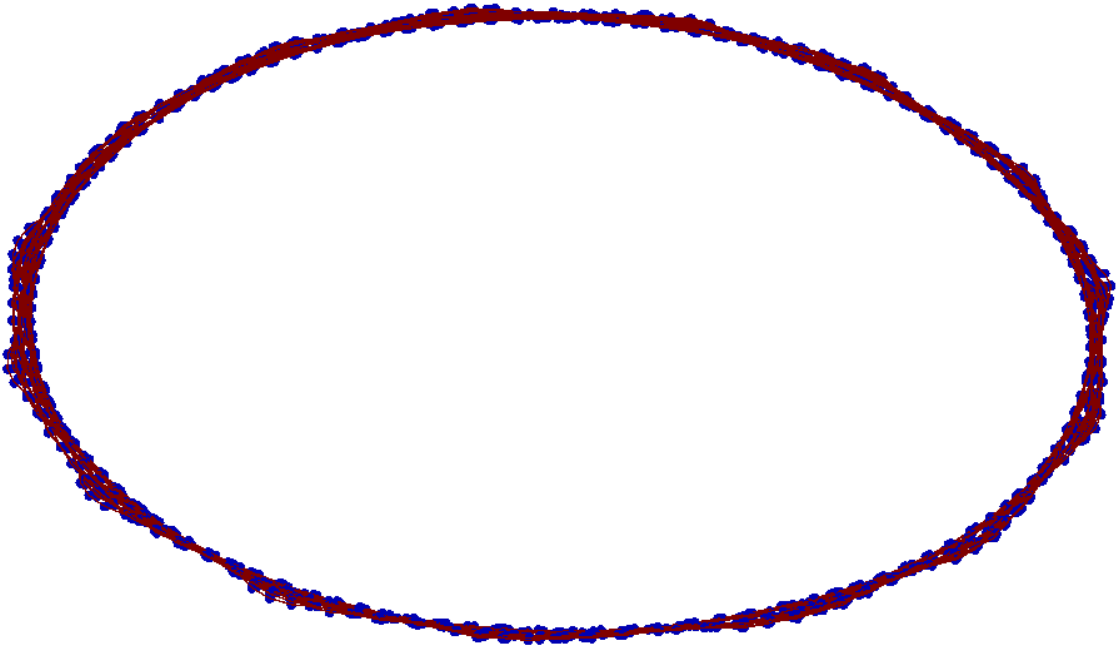}
		\subcaption{}
		\label{fig:DiodeNET0350}
	\end{subfigure}
	\begin{subfigure} {0.24\textwidth}
		\centering
		\includegraphics [width={1\textwidth}] {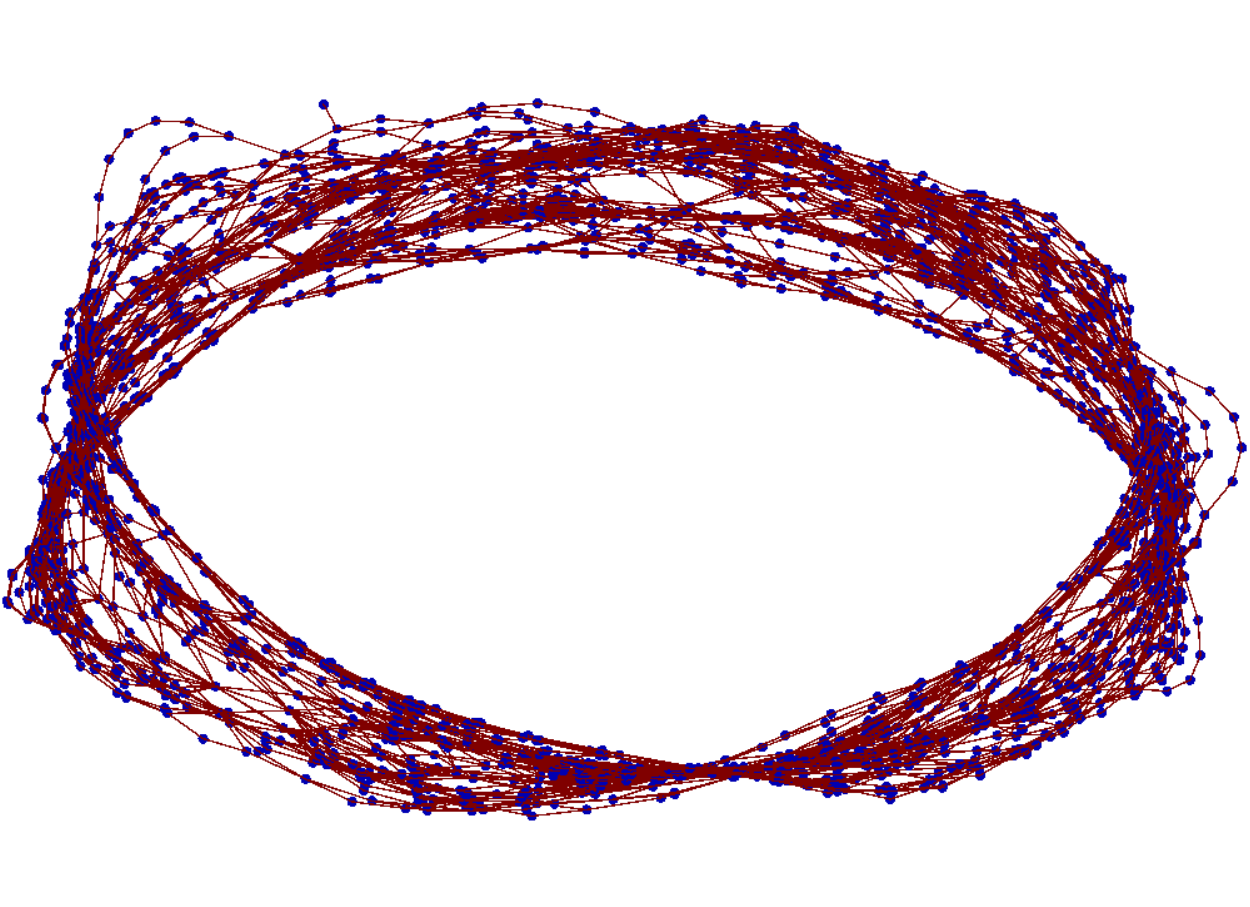}
		\subcaption{}
		\label{fig:DiodeNET0500}
	\end{subfigure}
	\begin{subfigure} {0.16\textwidth}
		\centering
		\includegraphics [width={1\textwidth}] {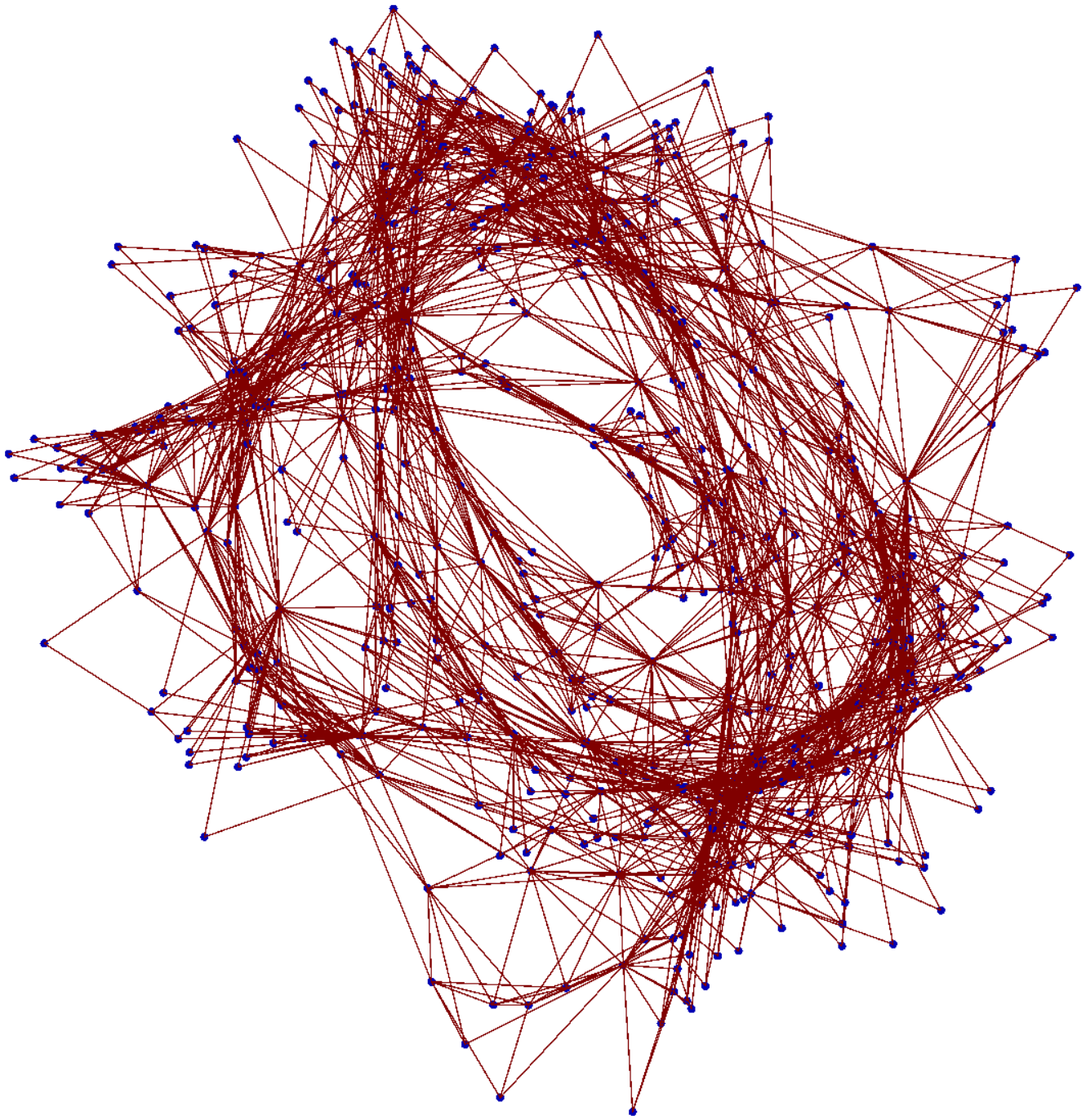}
		\subcaption{}
		\label{fig:DiodeNET0750T}
	\end{subfigure}
		\begin{subfigure} {0.08\textwidth}
		\centering
		\includegraphics [width={1.8\textwidth}] {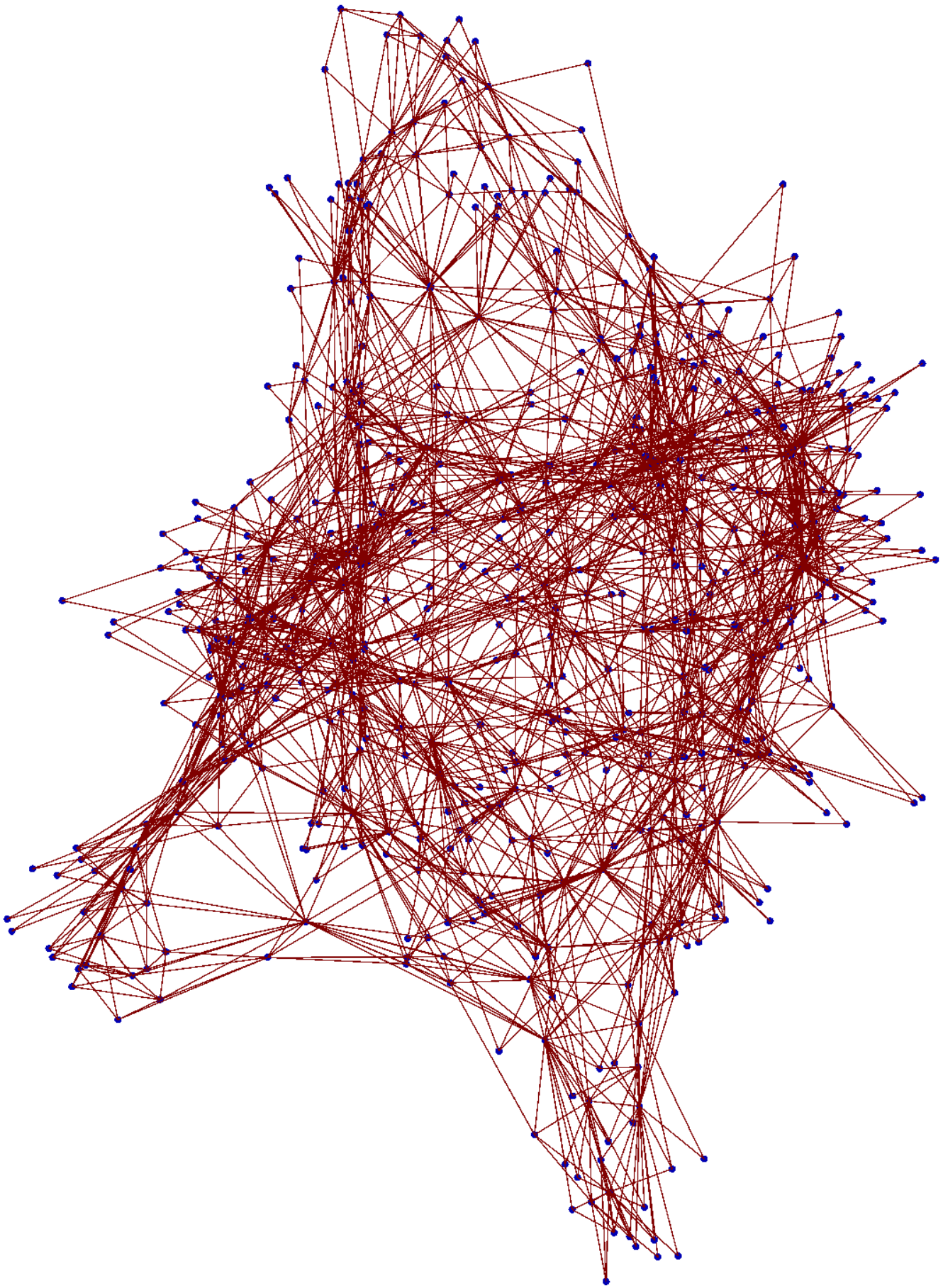}
		\subcaption{}
		\label{fig:DiodeNET0750S}
	\end{subfigure}
	\caption{Networks corresponding to the time series shown in Figure \ref{fig:DiodeAtr}. The embedding dimensions used for the period-3, period-6, multiband chaotic and chaos-chaos intermittent time series were \subref{fig:DiodeNET0000} \(D=10\), \subref{fig:DiodeNET0350} \(D=16\), \subref{fig:DiodeNET0500} \(D=14\), and \subref{fig:DiodeNET0750T}/\subref{fig:DiodeNET0750S} \(D=7\) respectively. Two views are given for the network from the chaos-chaos intermittent time series where \subref{fig:DiodeNET0750T} is a top down view and \subref{fig:DiodeNET0750S} is a side view of the network's tube like structure.}
	\label{fig:DiodeNET}
\end{figure}

We now present results of the method as applied to experimental time series data from an externally driven diode resonator circuit (see Figure \ref{fig:Circuit}) as previously reported in~\cite{jungling_noise-free_2008}. Each time series comprises 65536 observations of the voltage \(U_R\). Time series were recorded for 1000 evenly spaced values of the control parameter --- the amplitude of the driving sinusoidal voltage \(U_0\) --- in the range \(3V\leq U_0 < 5V\). The system begins in period-3 oscillation (Figure \ref{fig:Diode0000}) and undergoes a period doubling bifurcation into period-6 (Figure \ref{fig:Diode0350}). This is followed by a period doubling cascade into multiband chaos (Figure \ref{fig:Diode0500}) and then an interior crisis at \(U_0 \approx 4.05\) into chaos-chaos intermittency~\cite{grebogi_critical_1986} (Figure \ref{fig:Diode0750}). Networks for these different dynamical regimes are shown in Figure \ref{fig:DiodeNET}. Embedding lag is set to \(\tau=8\). The period-3 and period-6 time series map to ring structures. In the period-6 case it was necessary to select a large value for the embedding dimension, \(D=16\), to unfold the ring structure (i.e. there are no false transitional edges in the network) because of the closeness of trajectories in phase space (see Figure \ref{fig:Diode0350}). The multiband chaotic time series exhibits a narrow flat band structure with a clear pinch point that likely corresponds to the folding region on the attractor. Following the interior crisis the attractor (Figure \ref{fig:Diode0750}) has a corresponding tube like network structure, consistent with the results from the R\"ossler system.

\begin{figure}[]
	\centering
		\centering
	\begin{subfigure} {0.32\textwidth}
		\centering
		\includegraphics [width={1.1\textwidth}] {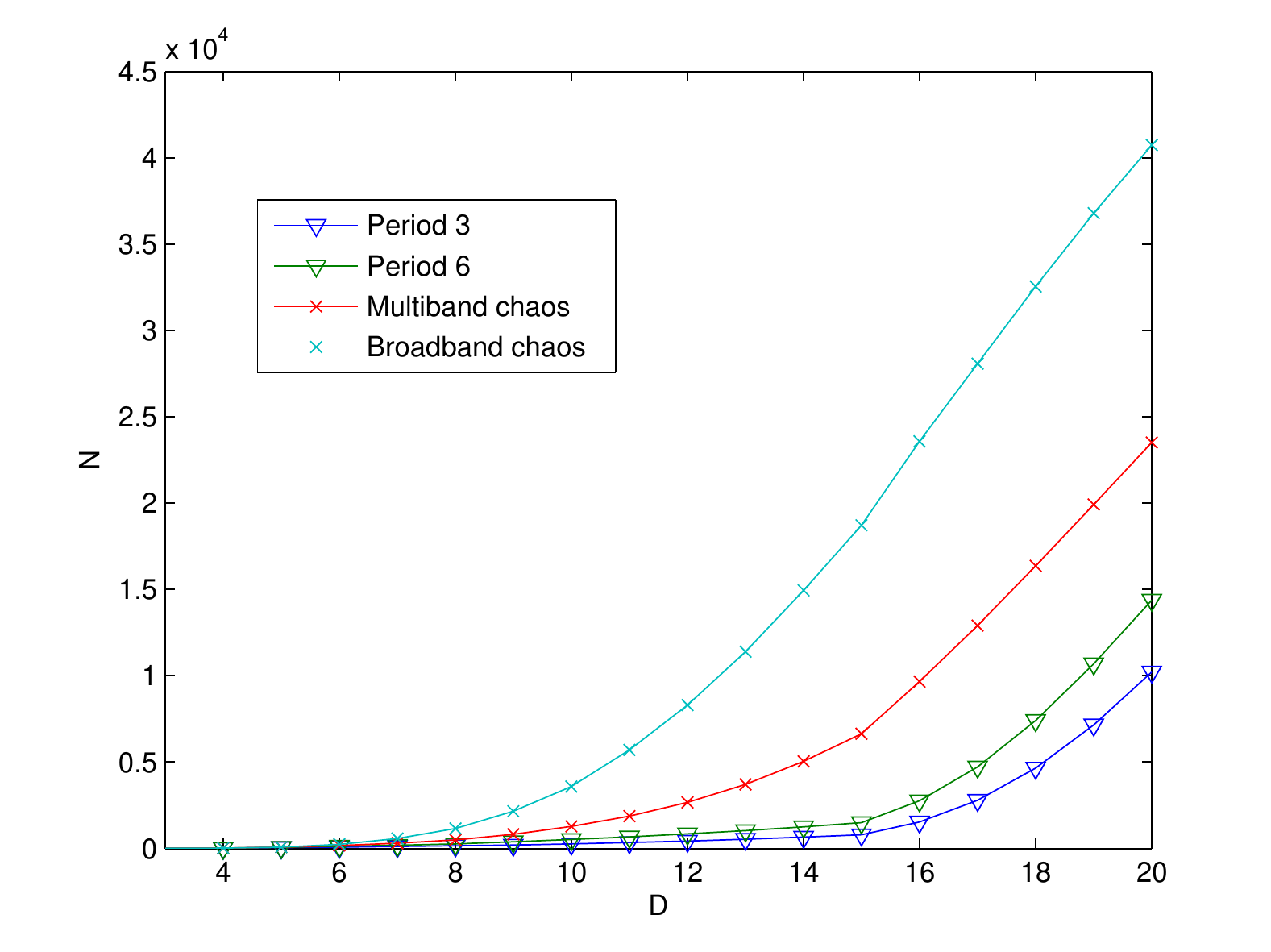}
		\subcaption{}
		\label{fig:Diode-NVsD}
	\end{subfigure}
	\begin{subfigure} {0.32\textwidth}
		\centering
		\includegraphics [width={1.1\textwidth}] {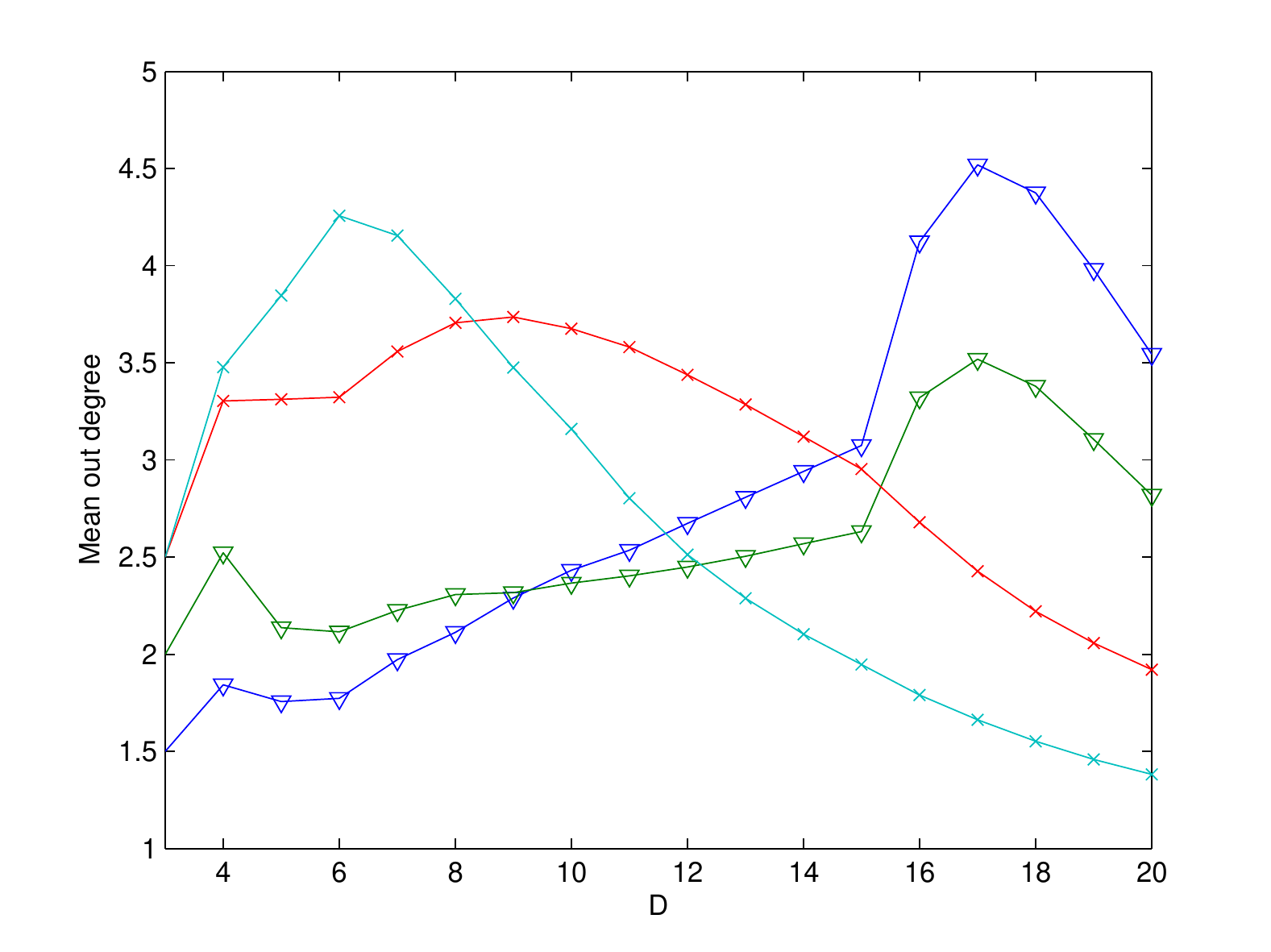}
		\subcaption{}
		\label{fig:Diode-MeanDegVsD}
	\end{subfigure}
		\begin{subfigure} {0.32\textwidth}
		\centering
		\includegraphics [width={1.1\textwidth}] {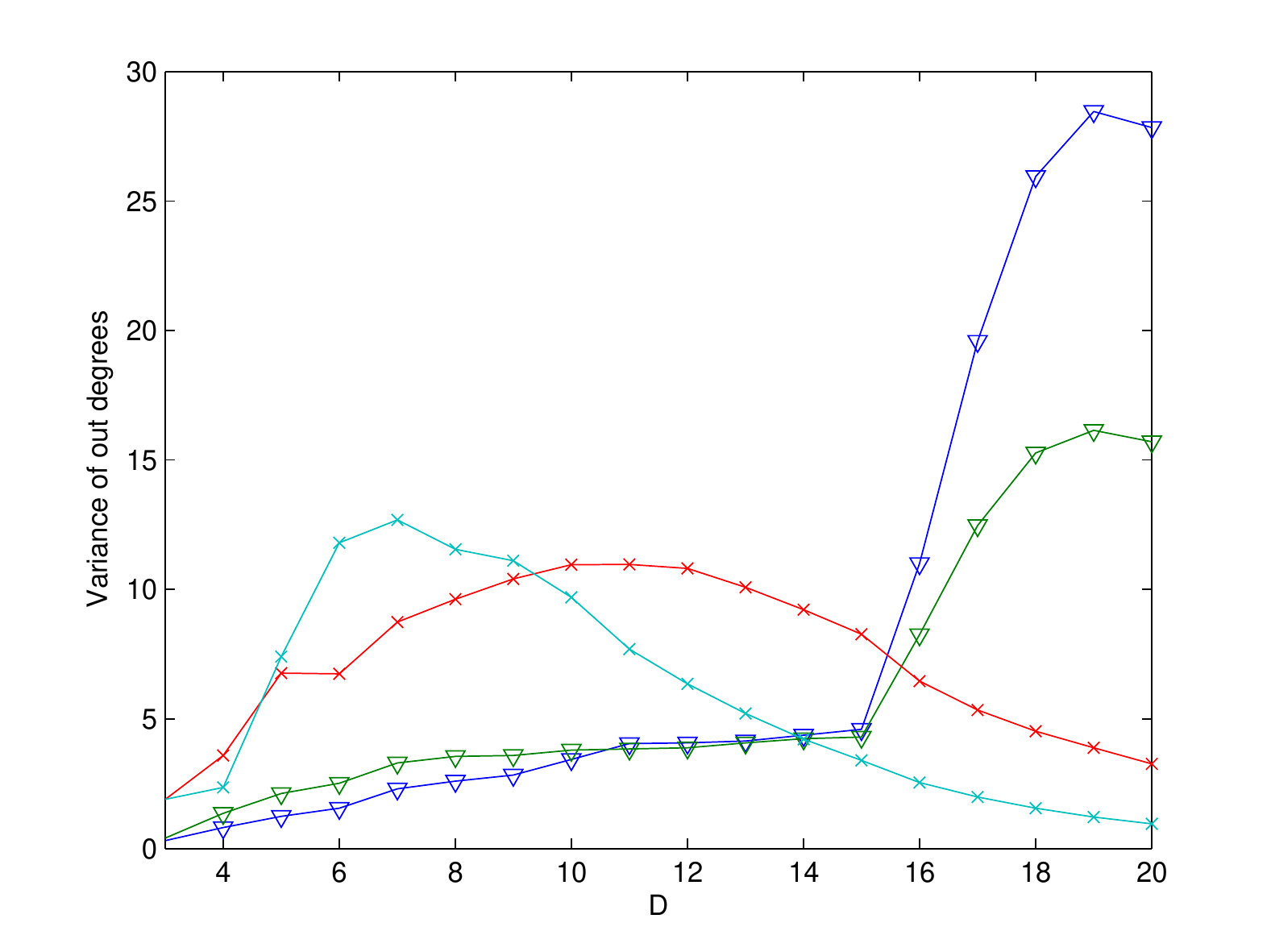}
		\subcaption{}
		\label{fig:Diode-DegVarVsD}
	\end{subfigure}
	\caption{Network measures plotted against the embedding dimension for the diode resonator time series data in different dynamical regimes: period 3, \(U_0=3V\) (blue line); period 6, \(U_0=3.7V\) (green line); multiband chaos, \(U_0=4V\) (red line); and chaos-chaos intermittency, \(U_0=4.5V\) (cyan line). Network measures shown are \subref{fig:Diode-NVsD} number of nodes, \subref{fig:Diode-MeanDegVsD} mean out degree, and \subref{fig:Diode-DegVarVsD} variance of out degrees.}
	\label{fig:DiodeDim}
\end{figure}

The size of the networks grow with \(D\) in the same manner as the R\"ossler system with the exception that the periodic networks begin to grow at a similar rate to the chaotic networks after \(D=15\) (Figure \ref{fig:Diode-NVsD}). This anomaly is reflected in the plots for \(\langle k_{out} \rangle\) and \(\sigma\) against \(D\), shown in Figures \ref{fig:Diode-DegVarVsD} and \ref{fig:Diode-MeanDegVsD} respectively, which both have have peaks for the periodic regimes after this point but otherwise exhibit similar trends to the periodic R\"ossler time series. These later peaks are likely due to low levels of noise in the experimental time series. When \(D\) becomes large, the states will effectively span a region of the phase space that is smaller than the cross section of the noisy period trajectory therefore resulting in a rapid increase to a peak in \(\langle k_{out} \rangle\) and \(\sigma\), just as occurs for the chaotic time series at lower values of \(D\).

\begin{figure}[]
	\centering
		\centering
	\begin{subfigure} {0.32\textwidth}
		\centering
		\includegraphics [width={1.1\textwidth}] {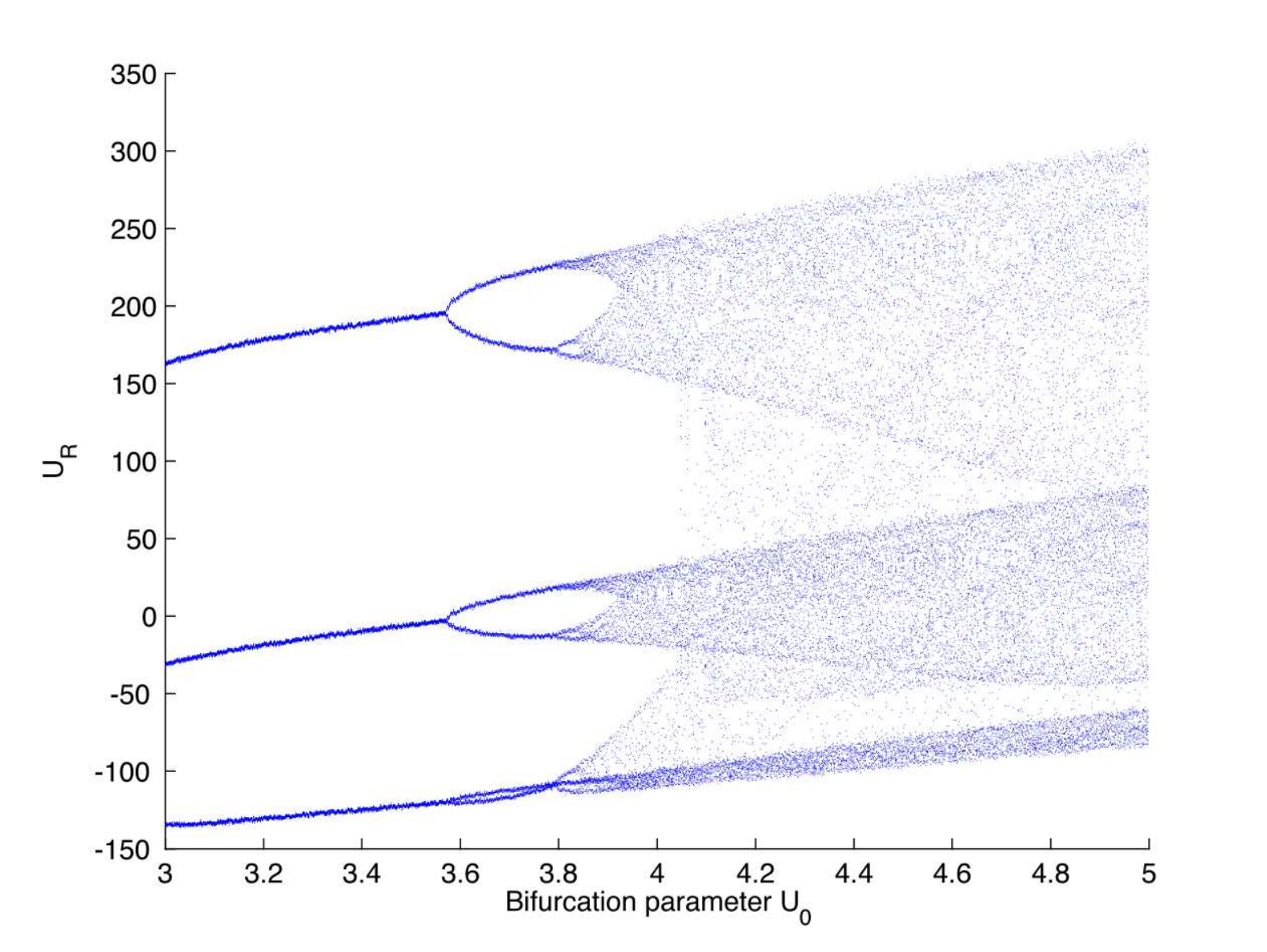}
		\subcaption{}
		\label{fig:Diode-Bif}
	\end{subfigure}
	\begin{subfigure} {0.32\textwidth}
		\centering
		\includegraphics [width={1.1\textwidth}] {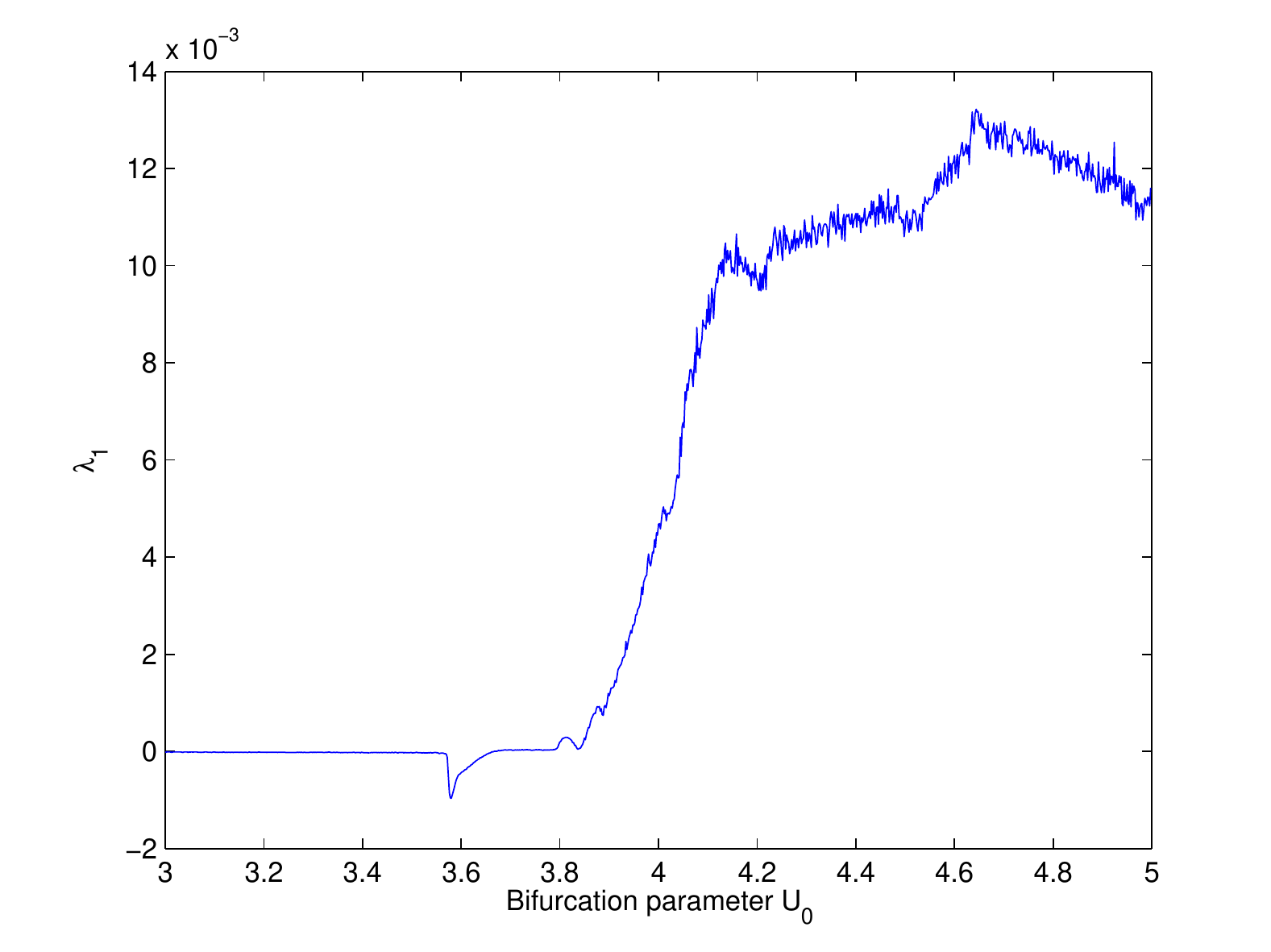}
		\subcaption{}
		\label{fig:Diode-Lyap}
	\end{subfigure}
	\begin{subfigure} {0.32\textwidth}
		\centering
		\includegraphics [width={1.1\textwidth}] {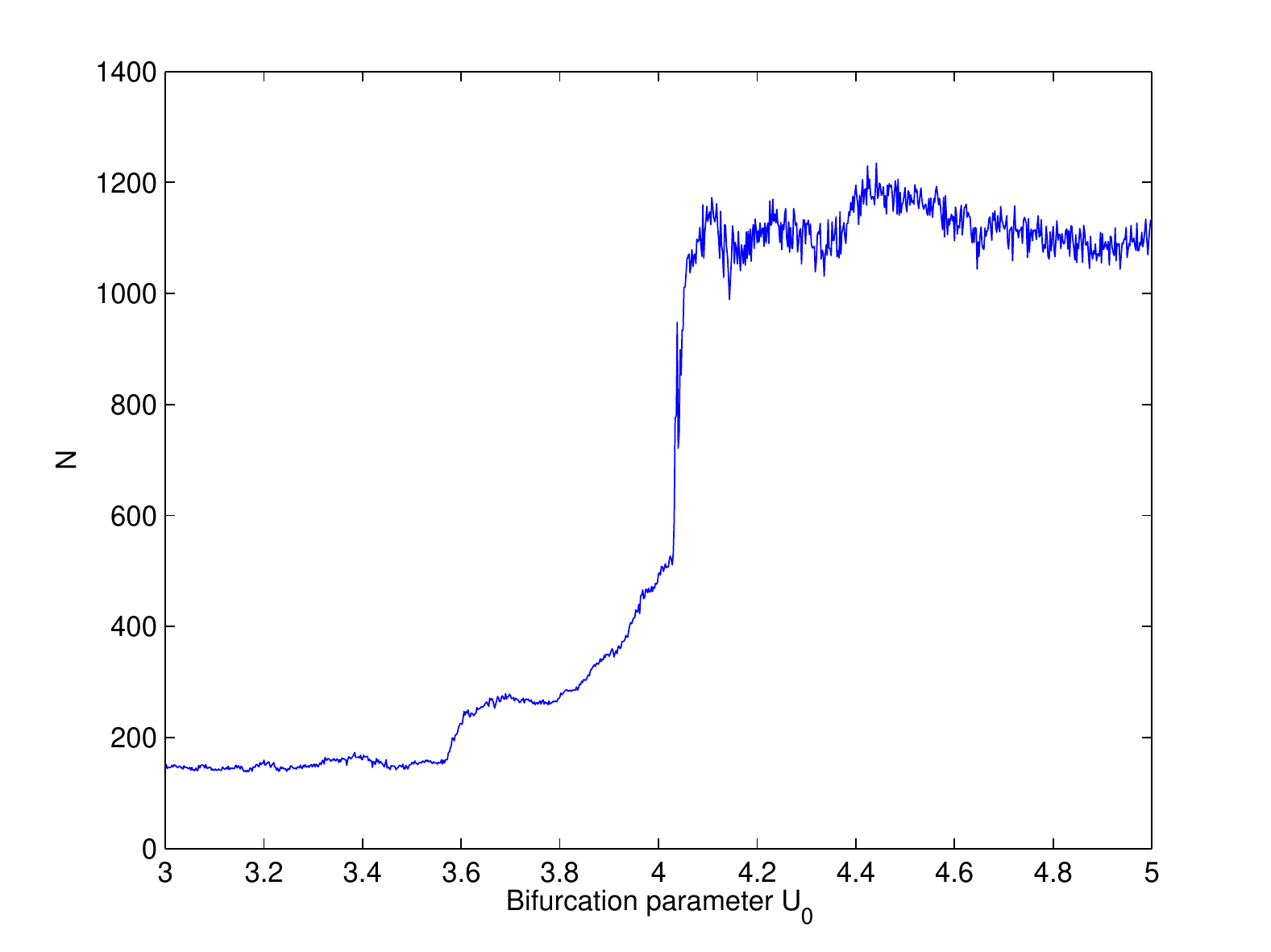}
		\subcaption{}
		\label{fig:Diode-NVsBif}
	\end{subfigure}
	\begin{subfigure} {0.32\textwidth}
		\centering
		\includegraphics [width={1.1\textwidth}] {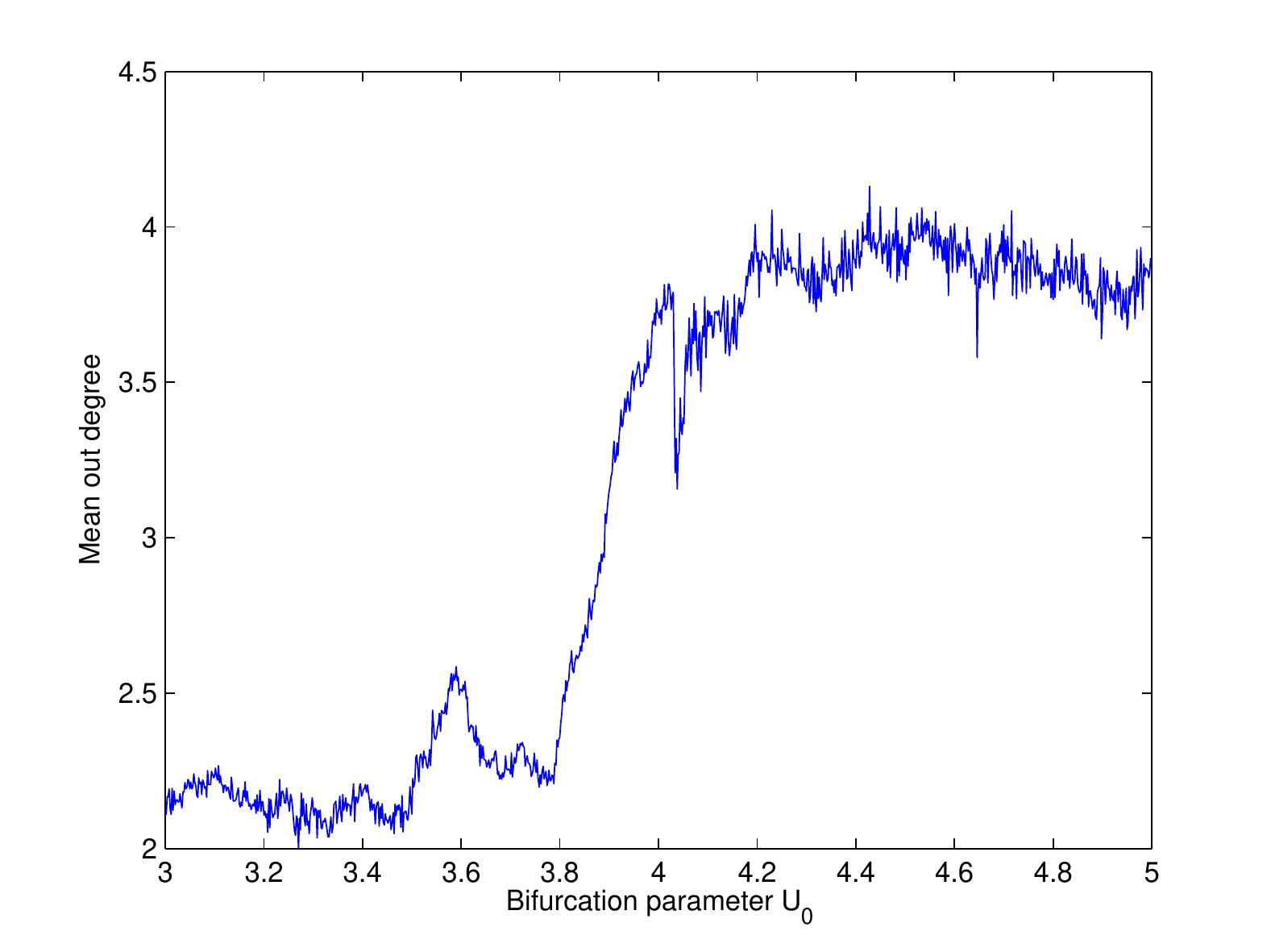}
		\subcaption{}
		\label{fig:Diode-MeanDegVsBif}
	\end{subfigure}
	\begin{subfigure} {0.32\textwidth}
		\centering
		\includegraphics [width={1.1\textwidth}] {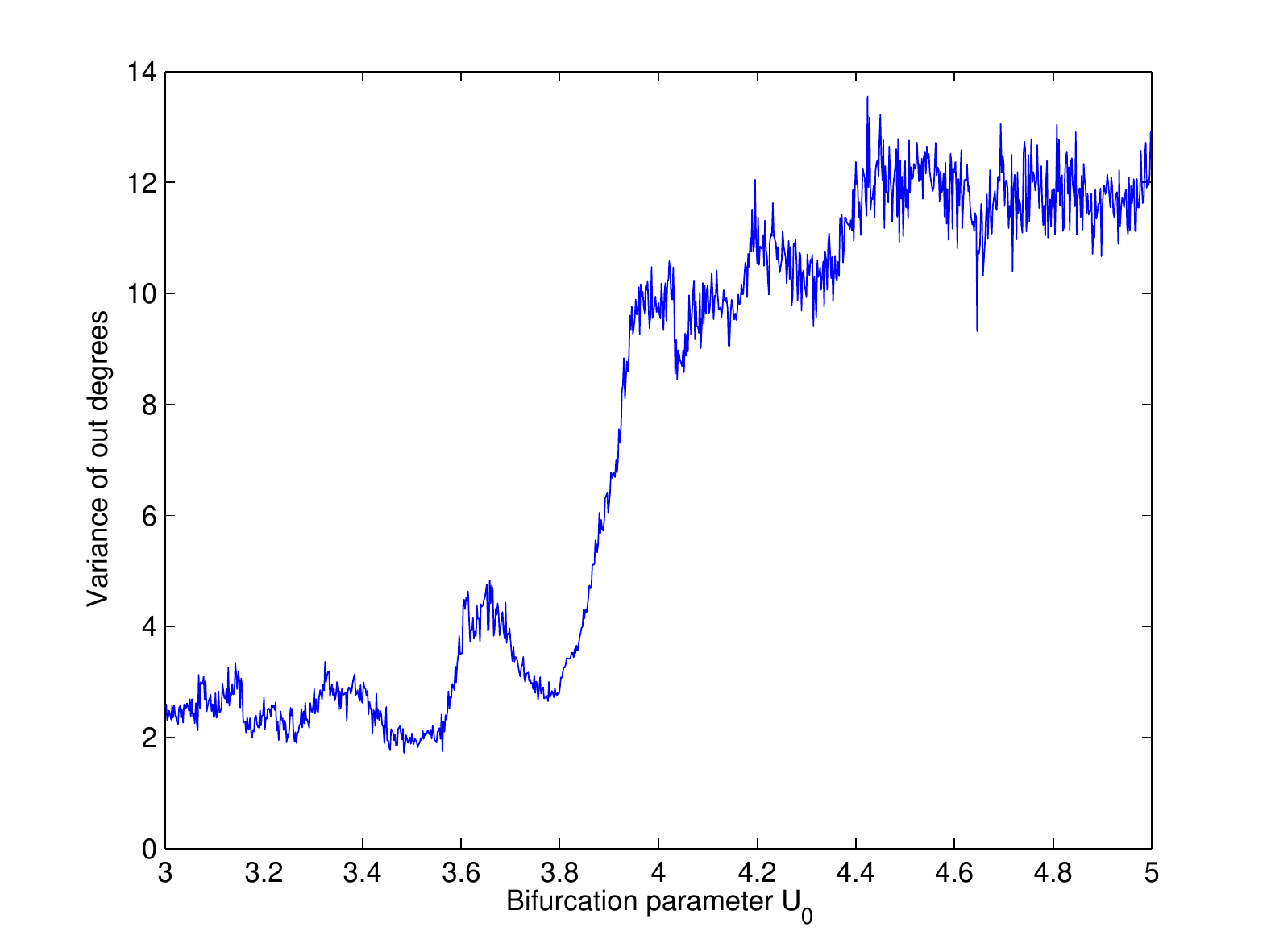}
		\subcaption{}
		\label{fig:Diode-DegVarVsBif}
	\end{subfigure}
	\begin{subfigure} {0.32\textwidth}
		\centering
		\includegraphics [width={1.1\textwidth}] {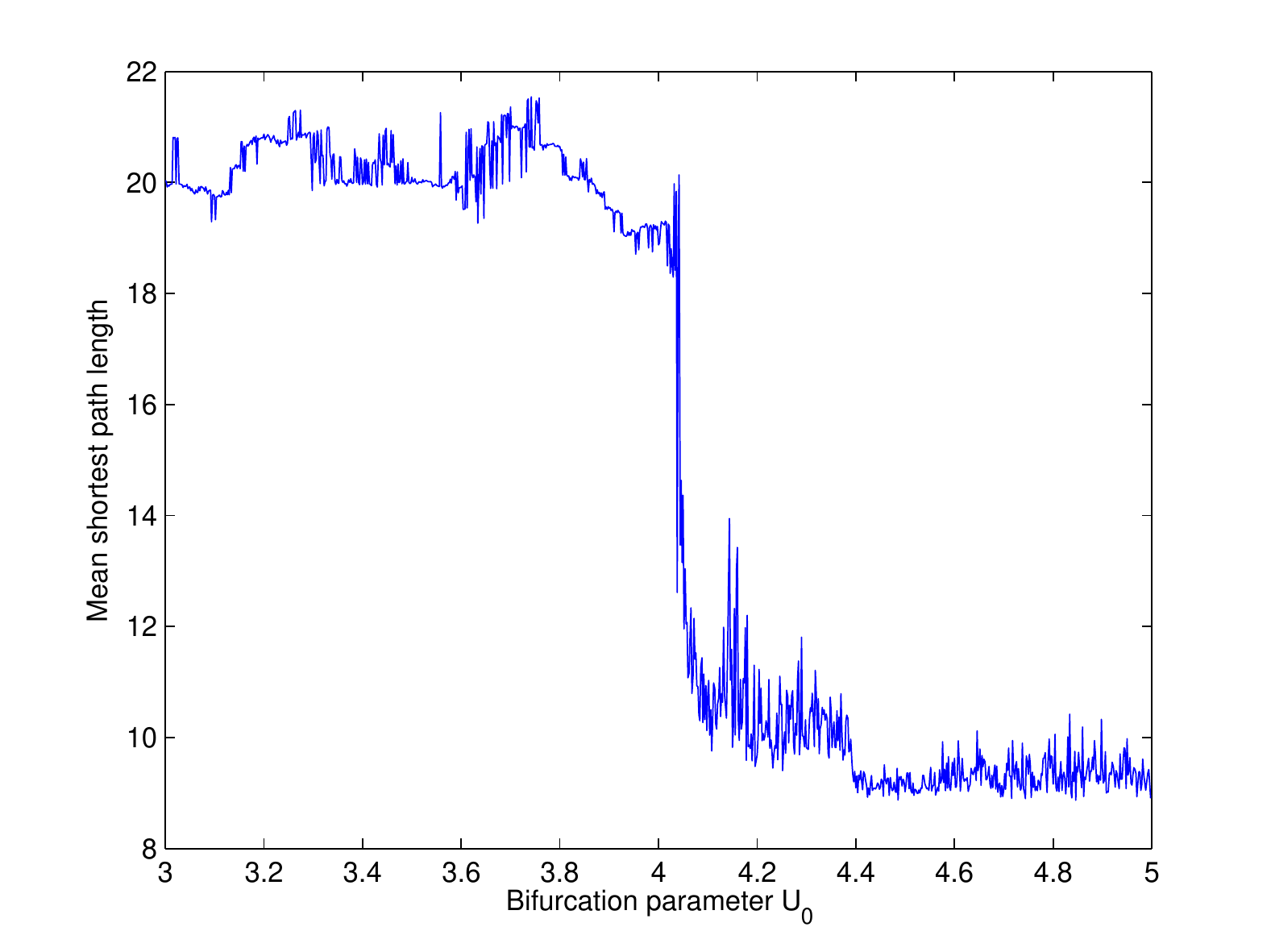}
		\subcaption{}
		\label{fig:Diode-MeanSPLVsBif}
	\end{subfigure}
	\begin{subfigure} {0.32\textwidth}
		\centering
		\includegraphics [width={1.1\textwidth}] {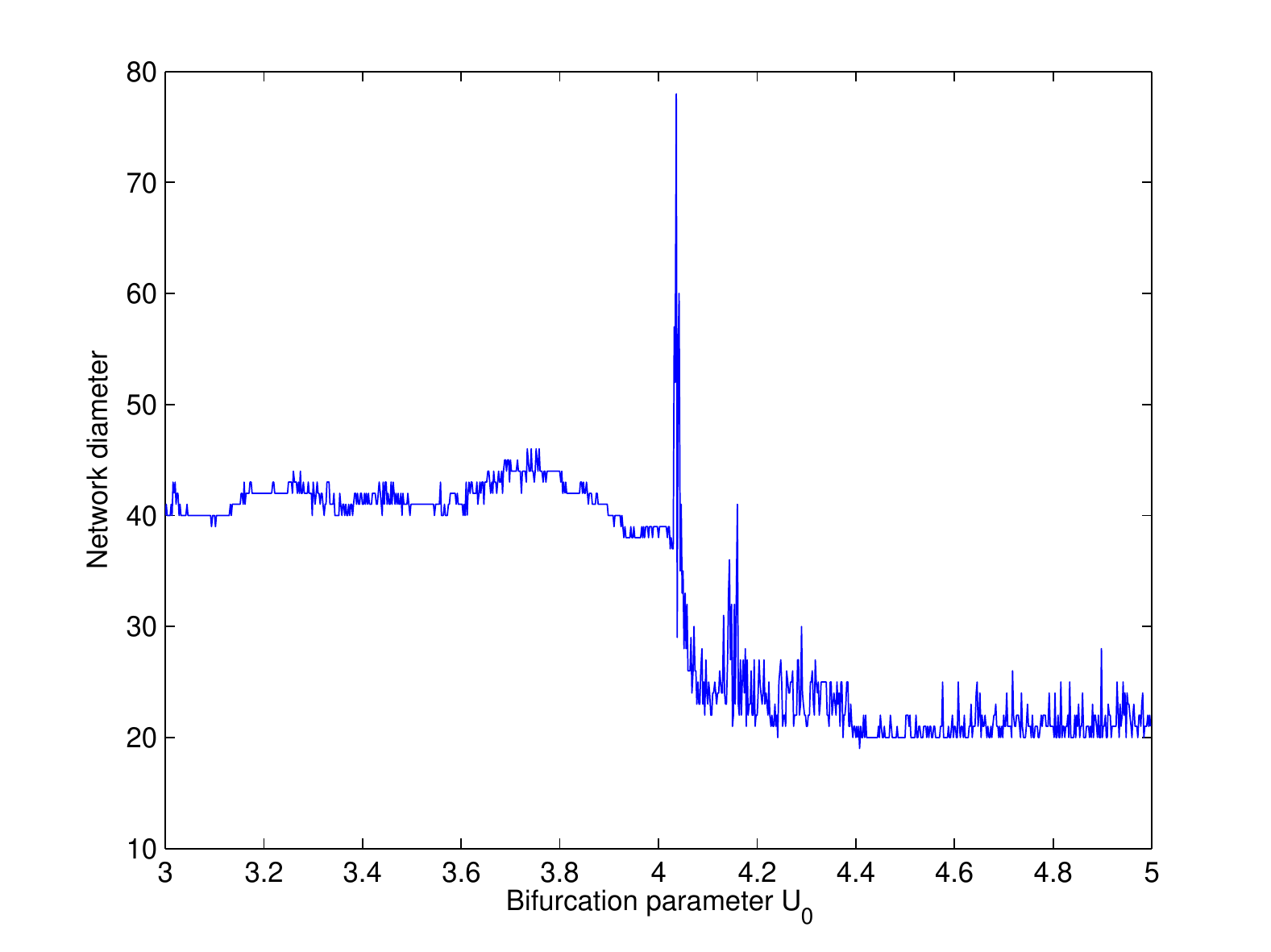}
		\subcaption{}
		\label{fig:Diode-DiamVsBif}
	\end{subfigure}
	\caption{\subref{fig:Diode-Bif} The bifurcation diagram constructed from the diode resonator dataset and \subref{fig:Diode-Lyap} the largest Lyapunov exponent for the range \(3V\leq U_0 \leq 5V\). Networks were generated for each time series with \(\tau=8\) and \(D=8\). Also shown are plots of the \subref{fig:Diode-NVsBif} number of nodes, \subref{fig:Diode-MeanDegVsBif} the mean out degree, \subref{fig:Diode-DegVarVsBif} variance of out degrees, \subref{fig:Diode-MeanSPLVsBif} mean shortest path length, and \subref{fig:Diode-DiamVsBif} network diameter over the same range.}
	\label{fig:DiodeMeas}
\end{figure}

The full bifurcation spectrum of the data set is plotted in Figure \ref{fig:Diode-Bif} using the \textit{extrema} function from the TISEAN software package. The largest Lyapunov exponent \(\lambda_1\), computed using the \textit{lyapk} function is shown in Figure \ref{fig:Diode-Lyap}. The brief range of the bifurcation parameter for which \(\lambda_1\) becomes negative after the first period doubling bifurcation is an error due to poor parameter selection for those particular corresponding time series (the exponent was computed with fixed parameters over the entire dataset for practical reasons). We then generated networks for each time series with \(\tau=8\) and \(D=8\). Network measures are plotted against the bifurcation spectrum in Figures \ref{fig:Diode-NVsBif}-\ref{fig:Diode-DiamVsBif}. The size of the network exhibits sensitivity to both the period doubling bifurcation at \(U_0\approx 3.6\), the period doubling cascade to chaos for approximately \(0.38\leq U_0 \leq 0.405\), and undergoes a step change at the interior crisis, reflecting the filling of the attractor. As was the case in the model system, \(\langle k_{out} \rangle\) and \(\sigma\) provide robust tracking of dynamical change similar to \(\lambda_1\), and also appear sensitive to the period doubling bifurcation and the interior crisis. The mean shortest path length and network diameter both undergo a clearly discernible step change at the interior crisis, with the latter also exhibiting a peak value at the change point. Both of these results are easily understood in terms of the relationship between the networks and phase space as follows: additional nodes and edges are created immediately after the crisis, corresponding to the intermittent chaotic trajectories that begin to fill the space between the bands of the pre-crisis attractor in phase space. These new nodes and edges become shortcuts in the network. The spike in diameter corresponds to the small number of time series which have only a limited number of trajectories in between the bands of the pre-crisis attractor because they are in the immediate vicinity of the crisis and we are dealing with finite non-stationary data. These trajectories will form new strands in the network which are only connected to the main structure where they leave and rejoin the bands of the pre-crisis attractor, and hence these trajectories will have a significant impact on the network diameter. Moreover, these strands or subgraphs will have a far lower degree and degree variance than the remainder of the network, hence why the value for mean out degree and degree variance also dips at the interior crisis.

\section{Discussion and Conclusions}
\label{sec:Conc}
We have defined a generalised form of the ordinal partition time series to network transformation algorithm by specifying partitions with time lagged elements. By selecting the time lag using traditional time series embedding techniques, each ordinal symbol represents a region of the embedded phase space, and the networks are therefore a Markov chain representation of the dynamics.

We applied the new method to chaotic flows, the R\"ossler system and experimental data from a diode resonator circuit, and found that dynamical information was successfully embedded in the network structure. Due to the nature of the transformation, the network will contain both topological and temporal information. Moreover, the mean out degree and the variance of out degrees appear to track the relative change in the largest Lyapunov exponent while also displaying sensitivity to period doubling bifurcations. As described in Section \ref{sec:Rossler}, the reason for this tracking is that node degree reflects transitional probability in the partitioned phase space and is therefore sensitive to the exponential divergence of trajectories in chaotic systems. Furthermore, the degree variance of ordinal partition networks derived from chaotic systems will be large due to the stretching and folding processes in the attractor. Unfortunately both of these simple measures are easily skewed by noise because they are local node properties in a transition network and therefore are merely global averages of temporal behaviour for a single time step.

However, graphical evidence showing relative consistency in network structure despite significant additive noise suggests that there should exist measures of network structure that are robust to noise. In~\cite{padberg_local_2009} Padberg \textit{et al.} proposed analytical links between the Lyapunov exponents and network structure, approaching the problem by evenly partitioning phase space and comparing the divergence of trajectories to subgraph expansion rates. Whilst the ordinal partition method does not partition phase space into evenly sized and bounded regions, as was the case in Padberg's paper, measuring the growth of a subgraph over a timespan of intermediate length should provide a more robust indication of the divergence of trajectories in the presence of noise. We note here that the possibility of a link between the Lyapunov exponents and network properties was also proposed by Nicolis \textit{et al.} in~\cite{nicolis_dynamical_2005} with respect to transition networks built from chaotic maps, and by Donner \textit{et al.} in~\cite{donner_recurrence_2010} with respect to recurrence networks, but we are not aware of any subsequent work which investigates these ideas in detail.

We also found that mean shortest path length and network diameter track dynamical changes and undergo a step change for an attractor merging crisis in an experimental data set. In this work we have restricted our attention to network measures that do not depend on edge weight. Edge weights will be considered in future investigations.

To conclude, we propose that the ordinal partition method, which explicitly captures temporal information, is complimentary to existing proximity network time series analysis methods, which are better suited for analysis of attractor topology. However, unlike proximity networks and visibility graphs, whose size is dependent solely on the length of the time series, the size of an ordinal partition network is dependent on attractor topology. Furthermore, the parametric simplicity of the generalised version of the ordinal partition algorithm presents an advantage over proximity network algorithms and the dual symbol ordinal partition method.

\newpage



\begin{thebibliography}{10}
\providecommand{\url}[1]{#1}
\csname url@samestyle\endcsname
\providecommand{\newblock}{\relax}
\providecommand{\bibinfo}[2]{#2}
\providecommand{\BIBentrySTDinterwordspacing}{\spaceskip=0pt\relax}
\providecommand{\BIBentryALTinterwordstretchfactor}{4}
\providecommand{\BIBentryALTinterwordspacing}{\spaceskip=\fontdimen2\font plus
\BIBentryALTinterwordstretchfactor\fontdimen3\font minus
 \fontdimen4\font\relax}
\providecommand{\BIBforeignlanguage}[2]{{%
\expandafter\ifx\csname l@#1\endcsname\relax
\typeout{** WARNING: IEEEtran.bst: No hyphenation pattern has been}%
\typeout{** loaded for the language `#1'. Using the pattern for}%
\typeout{** the default language instead.}%
\else
\language=\csname l@#1\endcsname
\fi
#2}}
\providecommand{\BIBdecl}{\relax}
\BIBdecl

\bibitem{donner_recurrence-based_2011}
\BIBentryALTinterwordspacing
R.~V. Donner, M.~Small, J.~F. Donges, N.~Marwan, Y.~Zou, R.~Xiang, and
  J.~Kurths, ``Recurrence-based time series analysis by means of complex
  network methods,'' \emph{International Journal of Bifurcation and Chaos},
  vol.~21, no.~04, pp. 1019--1046, 2011. [Online]. Available:
  \url{http://www.worldscientific.comdoiabs10.1142S0218127411029021}
\BIBentrySTDinterwordspacing

\bibitem{lacasa_time_2008}
\BIBentryALTinterwordspacing
L.~Lacasa, B.~Luque, F.~Ballesteros, J.~Luque, and J.~C. Nu\~no, ``From time
  series to complex networks: The visibility graph,'' \emph{Proceedings of the
  National Academy of Sciences}, vol. 105, no.~13, pp. 4972--4975, 2008.
  [Online]. Available: \url{http://www.pnas.org/content/105/13/4972.abstract}
\BIBentrySTDinterwordspacing

\bibitem{nicolis_dynamical_2005}
\BIBentryALTinterwordspacing
G.~Nicolis, A.~G. Cantú, and C.~Nicolis, ``Dynamical aspects of interaction
  networks,'' \emph{International Journal of Bifurcation and Chaos}, vol.~15,
  no.~11, pp. 3467--3480, 2005. [Online]. Available:
  \url{http://www.worldscientific.com/doi/abs/10.1142/S0218127405014167}
\BIBentrySTDinterwordspacing

\bibitem{padberg_local_2009}
\BIBentryALTinterwordspacing
K.~Padberg, B.~Thiere, R.~Preis, and M.~Dellnitz, ``Local expansion concepts
  for detecting transport barriers in dynamical systems,'' \emph{Communications
  in Nonlinear Science and Numerical Simulation}, vol.~14, no.~12, pp. 4176 --
  4190, 2009. [Online]. Available:
  \url{http://www.sciencedirect.com/science/article/pii/S1007570409001518}
\BIBentrySTDinterwordspacing

\bibitem{campanharo_duality_2011}
\BIBentryALTinterwordspacing
A.~S. L.~O. Campanharo, M.~I. Sirer, R.~D. Malmgren, F.~M. Ramos, and L.~A.~N.
  Amaral, ``Duality between time series and networks,'' \emph{{PLoS} {ONE}},
  vol.~6, no.~8, p. e23378, 2011. [Online]. Available:
  \url{http://dx.doi.org/10.1371\%2Fjournal.pone.0023378}
\BIBentrySTDinterwordspacing

\bibitem{small_complex_2013}
M.~Small, ``Complex networks from time series: Capturing dynamics,'' in
  \emph{Circuits and Systems ({ISCAS}), 2013 {IEEE} International Symposium
  on}, May 2013, pp. 2509--2512.

\bibitem{bandt_permutation_2002}
\BIBentryALTinterwordspacing
C.~Bandt and B.~Pompe, ``Permutation entropy: A natural complexity measure for
  time series,'' \emph{Phys. Rev. Lett.}, vol.~88, no.~17, p. 174102, Apr.
  2002. [Online]. Available:
  \url{http://link.aps.org/doi/10.1103/PhysRevLett.88.174102}
\BIBentrySTDinterwordspacing

\bibitem{amigo_true_2007}
\BIBentryALTinterwordspacing
J.~M. Amig\'o, S.~Zambrano, and M.~A.~F. Sanjuán, ``True and false forbidden
  patterns in deterministic and random dynamics,'' \emph{{EPL} (Europhysics
  Letters)}, vol.~79, no.~5, p. 50001, 2007. [Online]. Available:
  \url{http://stacks.iop.org/0295-5075/79/i=5/a=50001}
\BIBentrySTDinterwordspacing

\bibitem{zhang_complex_2006}
\BIBentryALTinterwordspacing
J.~Zhang and M.~Small, ``Complex network from pseudoperiodic time series:
  Topology versus dynamics,'' \emph{Phys. Rev. Lett.}, vol.~96, no.~23, p.
  238701, Jun. 2006. [Online]. Available:
  \url{http://link.aps.org/doi/10.1103/PhysRevLett.96.238701}
\BIBentrySTDinterwordspacing

\bibitem{yang_complex_2008}
\BIBentryALTinterwordspacing
Y.~Yang and H.~Yang, ``Complex network-based time series analysis,''
  \emph{Physica A: Statistical Mechanics and its Applications}, vol. 387, no.
  5–6, pp. 1381 -- 1386, 2008. [Online]. Available:
  \url{http://www.sciencedirect.com/science/article/pii/S0378437107011235}
\BIBentrySTDinterwordspacing

\bibitem{marwan_complex_2009}
\BIBentryALTinterwordspacing
N.~Marwan, J.~F. Donges, Y.~Zou, R.~V. Donner, and J.~Kurths, ``Complex network
  approach for recurrence analysis of time series,'' \emph{Physics Letters A},
  vol. 373, no.~46, pp. 4246 -- 4254, 2009. [Online]. Available:
  \url{http://www.sciencedirect.com/science/article/pii/S0375960109011852}
\BIBentrySTDinterwordspacing

\bibitem{xu_superfamily_2008}
\BIBentryALTinterwordspacing
X.~Xu, J.~Zhang, and M.~Small, ``Superfamily phenomena and motifs of networks
  induced from time series,'' \emph{Proceedings of the National Academy of
  Sciences}, vol. 105, no.~50, pp. 19\,601--19\,605, 2008. [Online]. Available:
  \url{http://www.pnas.org/content/105/50/19601.abstract}
\BIBentrySTDinterwordspacing

\bibitem{zhao_geometrical_2014}
\BIBentryALTinterwordspacing
Y.~Zhao, T.~Weng, and S.~Ye, ``Geometrical invariability of transformation
  between a time series and a complex network,'' \emph{Phys. Rev. E}, vol.~90,
  no.~1, p. 012804, Jul. 2014. [Online]. Available:
  \url{http://link.aps.org/doi/10.1103/PhysRevE.90.012804}
\BIBentrySTDinterwordspacing

\bibitem{xiang_multiscale_2012}
\BIBentryALTinterwordspacing
R.~Xiang, J.~Zhang, X.-K. Xu, and M.~Small, ``Multiscale characterization of
  recurrence-based phase space networks constructed from time series,''
  \emph{Chaos: An Interdisciplinary Journal of Nonlinear Science}, vol.~22,
  no.~1, pp.~--, 2012. [Online]. Available:
  \url{http://scitation.aip.org/content/aip/journal/chaos/22/1/10.1063/1.3673789}
\BIBentrySTDinterwordspacing

\bibitem{donner_recurrence_2010}
\BIBentryALTinterwordspacing
R.~V. Donner, Y.~Zou, J.~F. Donges, N.~Marwan, and J.~Kurths, ``Recurrence
  networks-—a novel paradigm for nonlinear time series analysis,'' \emph{New
  Journal of Physics}, vol.~12, no.~3, p. 033025, 2010. [Online]. Available:
  \url{http://stacks.iop.org/1367-2630/12/i=3/a=033025}
\BIBentrySTDinterwordspacing

\bibitem{iwayama_change-point_2013}
K.~Iwayama, Y.~Hirata, H.~Suzuki, and K.~Aihara, ``Change-point detection with
  recurrence networks,'' \emph{Nonlinear Theory and Its Applications, {IEICE}},
  vol.~4, no.~2, pp. 160--171, 2013.

\bibitem{zou_power-laws_2012}
\BIBentryALTinterwordspacing
Y.~Zou, J.~Heitzig, R.~V. Donner, J.~F. Donges, J.~D. Farmer, R.~Meucci,
  S.~Euzzor, N.~Marwan, and J.~Kurths, ``Power-laws in recurrence networks from
  dynamical systems,'' \emph{Europhysics Letters}, vol.~98, no.~4, p. 48001,
  2012. [Online]. Available:
  \url{http://stacks.iop.org/0295-5075/98/i=4/a=48001}
\BIBentrySTDinterwordspacing

\bibitem{liu_topological_2014}
\BIBentryALTinterwordspacing
J.-L. Liu, Z.-G. Yu, and V.~Anh, ``Topological properties and fractal analysis
  of a recurrence network constructed from fractional brownian motions,''
  \emph{Phys. Rev. E}, vol.~89, no.~3, p. 032814, Mar. 2014. [Online].
  Available: \url{http://link.aps.org/doi/10.1103/PhysRevE.89.032814}
\BIBentrySTDinterwordspacing

\bibitem{lacasa_visibility_2009}
\BIBentryALTinterwordspacing
L.~Lacasa, B.~Luque, J.~Luque, and J.~C. Nu\~no, ``The visibility graph: A new
  method for estimating the hurst exponent of fractional brownian motion,''
  \emph{Europhysics Letters}, vol.~86, no.~3, p. 30001, 2009. [Online].
  Available: \url{http://stacks.iop.org/0295-5075/86/i=3/a=30001}
\BIBentrySTDinterwordspacing

\bibitem{bezsudnov_time_2014}
\BIBentryALTinterwordspacing
I.~V. Bezsudnov and A.~A. Snarskii, ``From the time series to the complex
  networks: The parametric natural visibility graph,'' \emph{Physica A:
  Statistical Mechanics and its Applications}, vol. 414, no.~0, pp. 53 -- 60,
  2014. [Online]. Available:
  \url{http://www.sciencedirect.com/science/article/pii/S0378437114005676}
\BIBentrySTDinterwordspacing

\bibitem{ahmadlou_visibility_2012}
\BIBentryALTinterwordspacing
M.~Ahmadlou and H.~Adeli, ``Visibility graph similarity: A new measure of
  generalized synchronization in coupled dynamic systems,'' \emph{Physica D:
  Nonlinear Phenomena}, vol. 241, no.~4, pp. 326 -- 332, 2012. [Online].
  Available:
  \url{http://www.sciencedirect.com/science/article/pii/S0167278911002491}
\BIBentrySTDinterwordspacing

\bibitem{donges_testing_2013}
\BIBentryALTinterwordspacing
J.~F. Donges, R.~V. Donner, and J.~Kurths, ``Testing time series
  irreversibility using complex network methods,'' \emph{Europhysics Letters},
  vol. 102, no.~1, p. 10004, 2013. [Online]. Available:
  \url{http://stacks.iop.org/0295-5075/102/i=1/a=10004}
\BIBentrySTDinterwordspacing

\bibitem{cao_detecting_2004}
\BIBentryALTinterwordspacing
Y.~Cao, W.~Tung, J.~B. Gao, V.~A. Protopopescu, and L.~M. Hively, ``Detecting
  dynamical changes in time series using the permutation entropy,'' \emph{Phys.
  Rev. E}, vol.~70, no.~4, p. 046217, Oct. 2004. [Online]. Available:
  \url{http://link.aps.org/doi/10.1103/PhysRevE.70.046217}
\BIBentrySTDinterwordspacing

\bibitem{zunino_distinguishing_2012}
\BIBentryALTinterwordspacing
L.~Zunino, M.~C. Soriano, and O.~A. Rosso, ``Distinguishing chaotic and
  stochastic dynamics from time series by using a multiscale symbolic
  approach,'' \emph{Phys. Rev. E}, vol.~86, no.~4, p. 046210, Oct. 2012.
  [Online]. Available: \url{httplink.aps.orgdoi10.1103PhysRevE.86.046210}
\BIBentrySTDinterwordspacing

\bibitem{kulp_discriminating_2014}
\BIBentryALTinterwordspacing
C.~W. Kulp and L.~Zunino, ``Discriminating chaotic and stochastic dynamics
  through the permutation spectrum test,'' \emph{Chaos: An Interdisciplinary
  Journal of Nonlinear Science}, vol.~24, no.~3, 2014. [Online]. Available:
  \url{http://scitation.aip.org/content/aip/journal/chaos/24/3/10.1063/1.4891179}
\BIBentrySTDinterwordspacing

\bibitem{sun_characterizing_2014}
\BIBentryALTinterwordspacing
X.~Sun, M.~Small, Y.~Zhao, and X.~Xue, ``Characterizing system dynamics with a
  weighted and directed network constructed from time series data,''
  \emph{Chaos: An Interdisciplinary Journal of Nonlinear Science}, vol.~24,
  no.~2, p. 024402, 2014. [Online]. Available:
  \url{http://scitation.aip.org/content/aip/journal/chaos/24/2/10.1063/1.4868261}
\BIBentrySTDinterwordspacing

\bibitem{kennel_estimating_2003}
\BIBentryALTinterwordspacing
M.~B. Kennel and M.~Buhl, ``Estimating good discrete partitions from observed
  data: Symbolic false nearest neighbors,'' \emph{Phys. Rev. Lett.}, vol.~91,
  p. 084102, Aug 2003. [Online]. Available:
  \url{http://link.aps.org/doi/10.1103/PhysRevLett.91.084102}
\BIBentrySTDinterwordspacing

\bibitem{hirata_characterizing_2005}
\BIBentryALTinterwordspacing
Y.~Hirata, K.~Judd, and K.~Aihara, ``Characterizing chaotic response of a squid
  axon through generating partitions,'' \emph{Physics Letters A}, vol. 346, no.
  1–3, pp. 141 -- 147, 2005. [Online]. Available:
  \url{http://www.sciencedirect.com/science/article/pii/S0375960105011965}
\BIBentrySTDinterwordspacing

\bibitem{wolfram_documentation_2014}
\BIBentryALTinterwordspacing
{Wolfram Langauge and System: Documentation Center}. (2014) {GraphPlot3D}.
  [Online]. Available:
  \url{http://reference.wolfram.com/language/ref/GraphPlot3D.html}
\BIBentrySTDinterwordspacing

\bibitem{hegger_practical_1999}
\BIBentryALTinterwordspacing
R.~Hegger, H.~Kantz, and T.~Schreiber, ``Practical implementation of nonlinear
  time series methods: The {TISEAN} package,'' \emph{Chaos: An
  Interdisciplinary Journal of Nonlinear Science}, vol.~9, no.~2, pp. 413--435,
  1999. [Online]. Available:
  \url{http://scitation.aip.org/content/aip/journal/chaos/9/2/10.1063/1.166424}
\BIBentrySTDinterwordspacing

\bibitem{jungling_noise-free_2008}
\BIBentryALTinterwordspacing
T.~J\"ungling, H.~Benner, T.~Stemler, and W.~Just, ``Noise-free stochastic
  resonance at an interior crisis,'' \emph{Phys. Rev. E}, vol.~77, no.~3, p.
  036216, Mar. 2008. [Online]. Available:
  \url{http://link.aps.org/doi/10.1103/PhysRevE.77.036216}
\BIBentrySTDinterwordspacing

\bibitem{grebogi_critical_1986}
\BIBentryALTinterwordspacing
C.~Grebogi, E.~Ott, and J.~A. Yorke, ``Critical exponent of chaotic transients
  in nonlinear dynamical systems,'' \emph{Phys. Rev. Lett.}, vol.~57, pp.
  1284--1287, Sep 1986. [Online]. Available:
  \url{http://link.aps.org/doi/10.1103/PhysRevLett.57.1284}
\BIBentrySTDinterwordspacing

\end{thebibliography}
\end{document}